\documentclass[journal,onecolumn,romanappendices]{IEEEtran}
\usepackage{cite}
\ifCLASSINFOpdf
\usepackage[pdftex]{graphicx}
\graphicspath{./figures}
\DeclareGraphicsExtensions{.pdf,.jpeg,.png}
\else
\fi
\usepackage[cmex10]{amsmath}
\usepackage{array}
\usepackage{url}
\usepackage{color}
\usepackage[utf8]{inputenc}  
\usepackage{citesort}
\usepackage[T1]{fontenc} 
\usepackage{mathtools}
\usepackage{amsfonts}
\usepackage{amssymb}
\usepackage[english]{babel}
\usepackage{algorithm}
\usepackage[noend]{algpseudocode}
\usepackage{float}
\usepackage{multirow}
\usepackage{bbm}

\newtheorem{claim}{Claim}
\newtheorem{assumption}{\textit{\textbf{Assumption}}}

\algrenewcommand{\algorithmiccomment}[1]{\hfill$\triangleright$ #1}
\newcolumntype{P}[1]{>{\centering\arraybackslash}p{#1}}
\newcolumntype{M}[1]{>{\centering\arraybackslash}m{#1}}
\newcolumntype{N}{@{}m{0pt}@{}}

\DeclareUnicodeCharacter{00A0}{~}

\def \({\left(}
\def \){\right)}
\def \[{\left[}
\def \]{\right]}
\newcommand{\tbf}[1]{{\textbf{#1}}}

\newcommand{\txt}[1]{\text{#1}}
\newcommand{\defeq}{\vcentcolon=}

\newcommand{\br}{{\textbf {r}}}
\newcommand{\bF}{{\textbf {F}}}

\newcommand{\bX}{{\textbf {X}}}
\newcommand{\bx}{{\textbf {x}}}

\newcommand{\by}{{\textbf {y}}}
\newcommand{\bz}{{\textbf {z}}}
\newcommand{\bZ}{{\textbf {Z}}}

\newcommand{\bs}{{\textbf {s}}}
\newcommand{\bS}{{\textbf {S}}}

\newcommand{\bR}{{\textbf {R}}}

\newcommand{\bxi}{{\boldsymbol{\xi}}}

\newcommand{\be}{\begin{equation}}
\newcommand{\ee}{\end{equation}}
\newcommand{\bea}{\begin{eqnarray}}
\newcommand{\eea}{\end{eqnarray}}

\newtheorem{theorem}{Theorem}[section]
\newtheorem{lemma}[theorem]{\textbf{Lemma}}
\newtheorem{thm}[theorem]{\textbf{Theorem}}

\newtheorem{corollary}[theorem]{\textbf{Corollary}}
\newtheorem{definition}[theorem]{\textbf{Definition}}

\begin{document}

\title{Universal Sparse Superposition Codes \\with Spatial Coupling and GAMP Decoding}
\author{Jean~Barbier$^{\dagger\star}$, Mohamad~Dia$^{\dagger}$, and~Nicolas~Macris$^{\dagger}$
\thanks{\!\!\!\!\!\!$\dagger$ School of Computer and Communication Sciences, Ecole Polytechnique F\'ed\'erale de Lausanne, Switzerland.\newline
\noindent $\star$ International Center for Theoretical Physics, Trieste, Italy.\newline
emails: firstname.lastname@epfl.ch.}% <-this % stops a space
}
%\thanks{Manuscript submitted xxxx xx, 2017.}}
%Manuscript received xxxx xx, 2017; revised xxxx xx, 2017.}}

%\markboth{Submitted TO IEEE TRANSACTIONS ON INFORMATION THEORY}%
%{Barbier \MakeLowercase{\textit{et al.}}: Universal Sparse Superposition Codes with Spatial Coupling and GAMP Decoding}

\maketitle

\begin{abstract}
Sparse superposition codes, or sparse regression codes, constitute a new class of codes which was first introduced for communication over the additive white Gaussian noise (AWGN) channel. It has been shown that such codes are capacity-achieving over the AWGN channel under optimal maximum-likelihood decoding as well as under various efficient iterative decoding schemes equipped with power allocation or spatially coupled constructions. Here, we generalize the analysis of these codes to a much broader setting that includes all memoryless channels. We show, for a large class of memoryless channels, that spatial coupling allows an efficient decoder, based on the generalized approximate message-passing (GAMP) algorithm, to reach the potential (or Bayes optimal) threshold of the underlying (or uncoupled) code ensemble. Moreover, we argue that spatially coupled sparse superposition codes universally achieve capacity under GAMP decoding by showing, through analytical computations, that the error floor vanishes and the potential threshold tends to capacity as one of the code parameter goes to infinity. Furthermore, we provide a closed form formula for the algorithmic threshold of the underlying code ensemble in terms of a Fisher information. Relating an algorithmic threshold to a Fisher information has theoretical as well as practical importance. Our proof relies on the state evolution analysis and uses the potential method developed in the theory of low-density parity-check (LDPC) codes and compressed sensing.
\end{abstract}


% \fi
\IEEEpeerreviewmaketitle
\tableofcontents

\section{Introduction}
\IEEEPARstart{S}{parse} superposition (SS) codes, or sparse regression codes, were first introduced by Barron and Joseph \cite{barron2010sparse} for reliable communication over the additive white Gaussian noise (AWGN) channel. The SS codes were then proven to be capacity-achieving under adaptive successive decoding along with power allocation \cite{JosephB14,barron2012high}. Later on, the connection between SS codes and compressed sensing was made in \cite{barbier2014replica}. The decoding of SS codes can be interpreted as an estimation of a sparse signal, with structured prior distribution, based on a relatively small number of noisy observations. Hence, the approximate message-passing (AMP) algorithm, originally developed for compressed sensing, was adapted in \cite{barbier2014replica} to decode SS codes where it exhibited better finite-length performance than adaptive successive decoding. SS codes, with appropriate power allocation on the transmitted signal, were then proven to achieve capacity under AMP decoding \cite{rush2015capacity}. Furthermore, the extension of the state evolution (SE) equations, originally developed to track the performance of AMP for compressed sensing \cite{BayatiMontanari10}, was proven to be exact for SS codes in \cite{rush2015capacity}.

The idea of \emph{spatial coupling} was originaly introduced for low-density parity-check (LDPC) codes under the name of LDPC 
convolutional codes \cite{FelstromZigangirov99,LentmaierZigangirov05}. Spatial coupling has been then successfully applied 
to various problems including error correcting codes \cite{KRU11}, code division multiple access (CDMA) \cite{TTK11,SchlegelTruhachev11}, 
satisfiability \cite{HamedHassani2013}, and compressed sensing \cite{KudekarPfister10,KMSSZ12,6283053}; where it has been shown to boost 
the performance under iterative algorithms. Recently, spatial coupling was applied to SS codes in \cite{barbierSchulkeKrzakala,BarbierK15}. 
The construction of coding matrices for SS codes with local coupling and a proper termination was shown to considerably improve the
performance. Moreover, practical Hadamard-based operators were used in \cite{barbierSchulkeKrzakala} to encode SS codes, where they showed better finite-length performance than random operators under AMP decoding. The spatially 
coupled construction used in \cite{barbierSchulkeKrzakala,BarbierK15} has many similarities with that introduced in the context of compressed sensing \cite{KrzakalaMezard12,CaltagironeZ14,Montanari-Javanmard}. Empirical evidence shows that spatially coupled SS codes perform much better than power allocated ones and that they achieve capacity under AMP decoding without any need for power allocation. This motivated the initiation of their rigorous study \cite{barbierDiaMacris_isit2016} using the \emph{potential method}, originally developed for the spatially coupled Curie-Weiss model \cite{Hassani10,Hassani12} and LDPC codes \cite{YedlaJian12,PfisterMacrisBMS,6887298}. The phenomenon of \emph{threshold saturation} for AWGN channels was shown in \cite{barbierDiaMacris_isit2016}, i.e. the \emph{potential threshold} that characterizes the performance of SS codes under the Bayes optimal minimum mean-square error (MMSE) decoder can be reached using spatial coupling and AMP decoding. Moreover, the potential threshold itself was shown to achieve capacity in the large input alphabet size limit.

Threshold saturation was first established 
in the context of spatially coupled LDPC codes for general binary input memoryless symmetric channels 
in \cite{PfisterMacrisBMS,Kudekar-Urbanke-Richardson-2013}, and is recognized as the mechanism underpinning the excellent performance of 
such codes \cite{Zigangirov-Costello-2010}. It is interesting that essentially the same phenomenon can be established for 
a coding system operating on a channel with {\it continuous inputs}. 
%Since in the large 
%"section" size limit the potential threshold approaches capacity, 
This result
was a stepping-stone towards establishing that spatially coupled SS codes achieve capacity on the AWGN channel under 
AMP decoding \cite{barbierDiaMacris_isit2016}. Note that a similar (but different) potential to the one used in \cite{barbierDiaMacris_isit2016} has been introduced in the context of scalar compressed sensing \cite{BayatiMontanari10,6887298}. It is interesting that the potential method 
goes through for the present system involving a dense coding matrix and a fairly wide class of spatial couplings. Related results on the optimality of spatial coupling in compressed sensing \cite{6283053} and on the threshold saturation of systems characterized by a $1$-dimensional state evolution \cite{6887298,7115123} have been obtained by different approaches.

In the classical noisy compressed sensing problem, the AMP algorithm and the SE recursion tracking the algorithmic performance were derived for the AWGN channel \cite{BayatiMontanari10,DMM09}. The extension of AMP to general memoryless (possibly non-linear) channels with arbitrary input and output distributions was introduced in \cite{rangan2011generalized} via the generalized approximate message-passing (GAMP) algorithm. Moreover, an extension of SE describing the exact behavior of GAMP was also provided in \cite{rangan2011generalized}. Later on, a full rigorous analysis proving the tractability of GAMP via SE was given in \cite{Montanari-Javanmard} (for the case of fully factorized prior). These encouraging results naturally led us to generalize the analysis of SS codes in \cite{barbierDiaMacris_isit2016} to a much broader setting that includes all memoryless channels and potentially any input signal model that factorizes over B-dimensional sections \cite{barbierDiaMacris_itw2016,BBD_ISIT2017}. Moreover, SS codes under GAMP decoding were recently proposed for an inverse source coding problem \cite{Dia_ISIT2018,DiaThesis}.

In this work we prove that threshold saturation is a universal phenomenon for SS codes; i.e. we show that, for any memoryless channel, spatial coupling allows 
GAMP decoding to reach the potential threshold of the code ensemble (Section \ref{sec:proofsketch} Theorem \ref{th:mainTheorem} and Corollary \ref{cor:maincorollary}). Moreover, we argue, through non-rigorous analytical computations, that spatially coupled SS codes universally achieve capacity under GAMP decoding by showing that the error floor vanishes and the potential threshold tends to capacity as one of the code's parameters (the section size, or input alphabet size, $B$) goes to infinity. Note that a fully rigorous statement about the capacity achieving property of SS codes still requires the following: $i)$ a rigorous asymptotic analysis in the large section size limit $B\to\infty$ (see Section \ref{sec:larg_B}), $ii)$ the proof that state evolution tracks the performance of GAMP over general memoryless channels when the prior factorizes over $B$-dimensional sections (as opposed to the fully factorized case treated in \cite{Montanari-Javanmard}). Furthermore, we give a simple expression of the GAMP algorithmic threshold of the underlying code ensemble in terms of a Fisher information (Section \ref{sec:larg_B}). Although we focus on coding for the sake of coherence with our previous results, the framework and methods are very general and hold for a wide class of non-linear estimation problems with random linear mixing.

Our proof strategy uses a potential function, which is inspired from the statistical physics replica method. However, we stress that the proof {\it does not} rely on the 
replica method (which is not rigorous). Recently, it has been shown that the replica prediction is exact for generalized random linear estimation problems including compressed sensing and SS codes on general channels \cite{BDMK_alerton2016,BMDK_2017,ReevesPfister_isit16,ReevesPfister_trans,pmlr-v75-barbier18a}. Hence, the potential threshold can be rigorously interpreted as the optimal threshold under MMSE decoding.

The paper is organized as follows.
The code construction of the underlying and coupled ensembles are described in Section \ref{sec:codeens}. Section \ref{sec:GAMP} reviews the GAMP algorithm, while Section \ref{sec:stateandpot} presents the SE equations and potential function adapted to the present context. The
GAMP thresholds of the underlying and coupled ensembles as well as the potential threshold are then given precise definitions. The essential steps for the proof of threshold saturation are presented in Section \ref{sec:proofsketch}. The connection between the potential threshold at infinite input alphabet size $B\to\infty$ and Shannon's capacity, as well as the closed form expression of the algorithmic threshold in terms of a Fisher information, are given in Section \ref{sec:larg_B}. Four different channel models are used to illustrate the results. Section \ref{sec:openChallenges} is dedicated to conclusion and open challenges.

\section{Code ensembles}\label{sec:codeens}
We first define the underlying and spatially coupled ensembles of SS codes for transmission over a generic memoryless channel. 
In the rest of the paper a subscript ``$\text{un}$'' indicates a quantity related to the underlying ensemble and a subscript ``$\text{co}$''
a quantity related to the spatially coupled ensemble. 
The probability law of a Gaussian random variable $X$ with mean $m$ and variance $\sigma^2$ is denoted $X \sim \mathcal{N}(m, \sigma^2)$ and the corresponding probability density function as $\mathcal{N}(x\vert m, \sigma^2)$. 
\subsection{The underlying ensemble} \label{sec:underlyingEns}
In the framework of SS codes, the \emph{information word} or \emph{message} is a vector made of $L$ \emph{sections}, $\bs = [\bs_1, \dots, \bs_L]$. Each section $\bs_l$, $l\in\{1, \dots, L\}$,  is a $B$-dimensional vector with a single component
equal to $1$ and $B-1$ components equal to $0$. The non-zero component of each section can be set differently especially when 
schemes with power allocation are considered \cite{JosephB14,barron2012high}. However, we will restrict ourselves to the binary case in this work where spatial 
coupling is used to achieve capacity instead of power allocation. We call $B$ the \emph{section size} (or alphabet size usually chosen to be a power of 2) and set $N=LB$. 
The message $\bs$ can be seen as a one-to-one mapping from an original message $\tbf{u} \in \{0,1\}^{L\log_2(B)}$, where the position of the non-zero component in $\bs_l$ is 
specified by the binary representation of $\tbf{u}_l$ (i.e. $\bs$ is obtained from $\tbf{u}$ using a simple position modulation (PM) scheme). For example if $B=4$ and $L=5$, a 
valid message is $\tbf s = [0001,0010,1000,0100,0010]$ which corresponds to $\tbf u = [00,01,11,10,01]$ .  
One can think of the information words as being defined for a $B$-ary alphabet with a constant power allocation 
for each symbol. 

We consider random codes generated by a fixed \emph{coding matrix} $\bF\in \mathbb{R}^{M \times N}$ drawn from the ensemble of random matrices with i.i.d real 
Gaussian entries distributed as $\mathcal{N}(0, 1/L)$. The variance of the coding matrix entries is such that the \emph{codeword} $\bF\bs\in \mathbb{R}^{M}$ has a normalized average 
power $\mathbb{E}[||\bF \bs||_{2}^{2}]/M = 1$. Note that the cardinality of this code is $B^L$ and the length of the codeword is $M$. Hence, the (design) rate is defined as
\be \label{eq:designRate}
R = \frac{L\log_2 B}{M} = \frac{N\log_2 B}{M B}.
\ee
The code is thus specified by $(M, R, B)$ where $R$ is the code rate, $M$ the block length, $B$ the section size. 
\begin{figure}[!t]
\centering
\includegraphics[draft=false,width=.7\textwidth, height = 175pt]{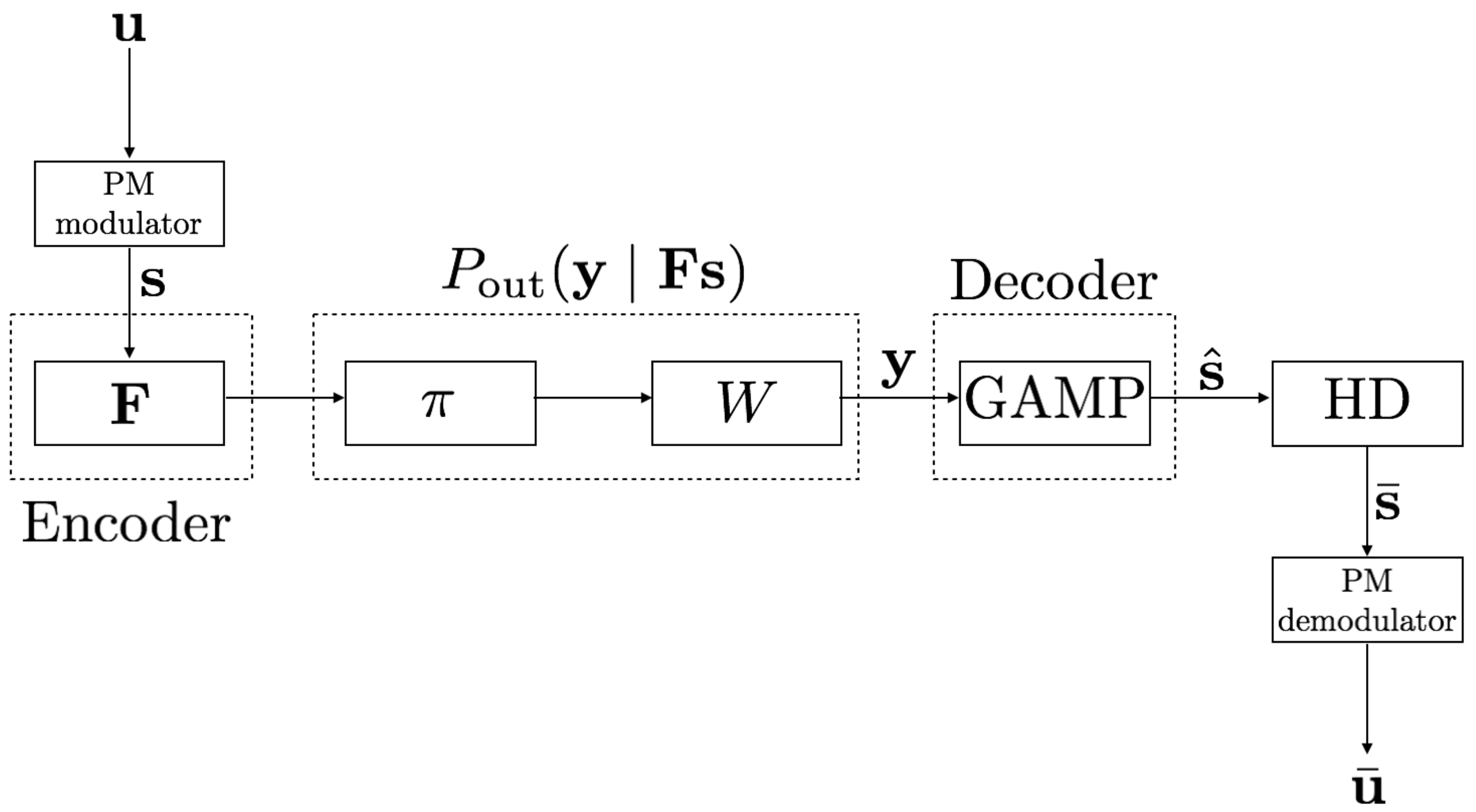}
\caption{The encoder/decoder block diagram of the SS codes under GAMP decoding over any memoryless channel $W$. The map $\pi$ is needed when the capacity achieving input distribution of $W$ is not Gaussian. The GAMP algorithm provides soft valued estimate $\hat{\bs}$ of $\bs$ in the MMSE sense. A simple hard decision (HD) mechanism is used to provide the binary decoded message $\bar{\bs}$ by setting the most biased component in each section of $\hat{\bs}$ to $1$ and the others to $0$. The original message $\tbf u$ and its decoded version $\bar{\tbf{u}}$ can be easily recovered from $\bs$ and $\bar{\bs}$ respectively using PM modulator and demodulator as illustrated in Section~\ref{sec:underlyingEns}.}
\label{fig:blockDiag}
\end{figure}

Codewords are transmitted through a known memoryless channel $W$. This requires to 
map the codeword components $[\bF\bs]_\mu\in \mathbb{R}$, $\mu \in \{1, \dots, M\}$, 
onto the input alphabet of $W$. We call $\pi$ this map and refer to Section~\ref{sec:larg_B} for various examples. 
The concatenation of $\pi$ and $W$ can be seen as an \emph{effective memoryless channel} $P_{\text{out}}$, such that
\begin{equation}
P_{\text{out}}(\by|\bF\bs) = \prod_{\mu = 1}^M P_{\text{out}}(y_\mu|[\bF\bs]_\mu) \defeq \prod_{\mu = 1}^M W(y_\mu|\pi([\bF\bs]_\mu)).
\end{equation}
Note that one can look equivalently at $\pi$ as a part of the channel model or as a part of the encoder. In the present framework, it is more convenient to work with the effective memoryless channel
from which the receiver obtains the noisy channel observation $\by$. However in the analysis of Section~\ref{sec:larg_B}, the capacity of $W$ is considered. 

The decoding task is to recover $\bs$ from channel observations $\by$ as depicted in Fig.~\ref{fig:blockDiag}. This can be interpreted as a compressed sensing problem with structured sparsity, due to the sectionwise 
structure of $\bs$, where $\by$ would be the compressed measurements.
The rate $R$ can be linked to the ``measurement rate'' $\alpha$, used in the compressed sensing literature, by
\be \label{eq:measurementRate}
\alpha = \frac{M}{N} = \frac{\log_2 B}{BR}.
\ee
Thus, the same algorithms and analysis used in compressed 
sensing theory like the GAMP algorithm and SE can be used in the present context. See \cite{BarbierK15} for more details on this interconnection.

\subsection{The spatially coupled ensemble}\label{subsec:SC_SS}
\begin{figure}[!t]
\centering
\includegraphics[draft=false,width=.6\textwidth, height = 140pt]{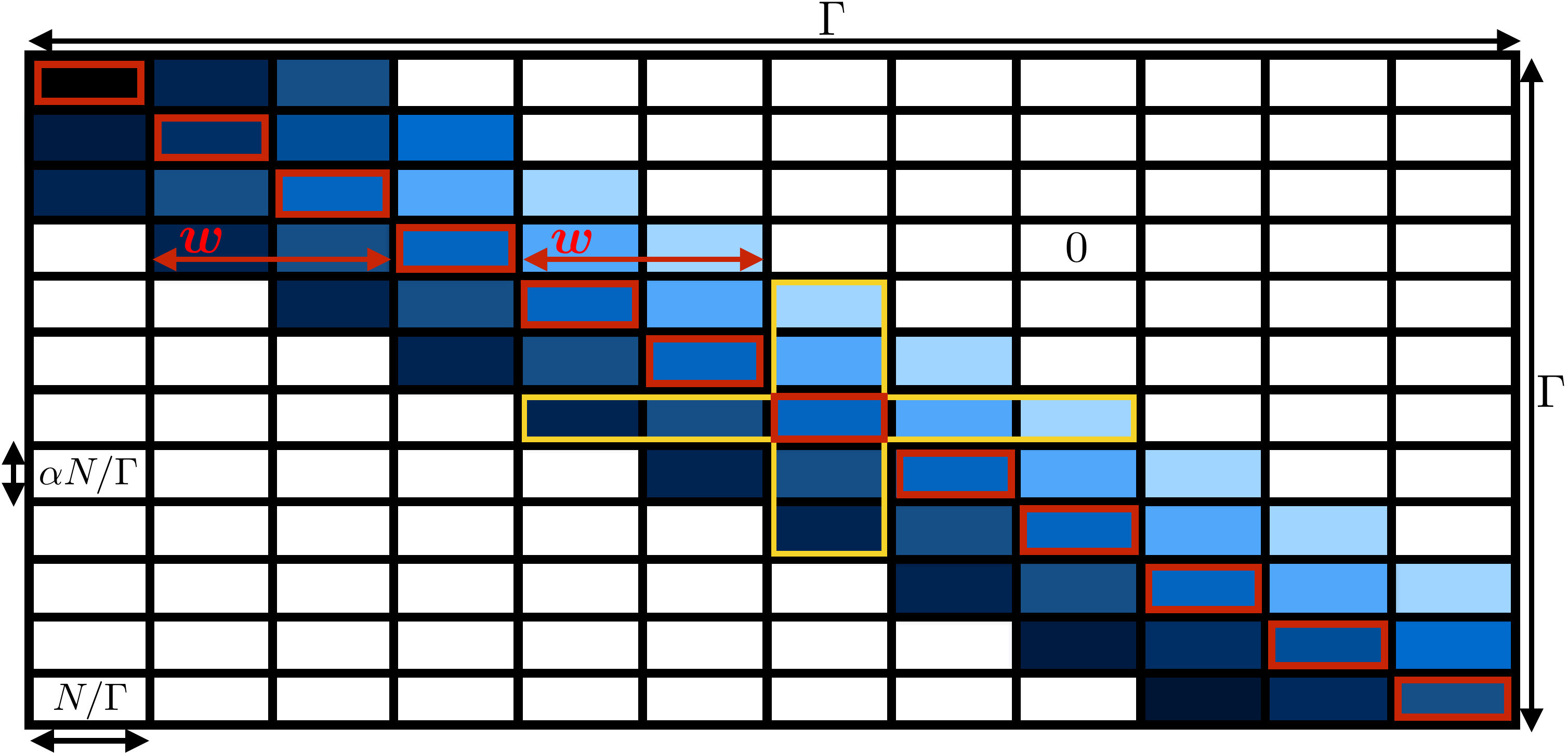}
\caption{A spatially coupled coding matrix $\bF^{\text{co}} \in\mathbb{R}^{M\times N}$ made of $\Gamma\times \Gamma$ blocks indexed by $(r,c)$, each with $N/\Gamma$ 
columns and $M/\Gamma =\alpha N/\Gamma$ rows where $\alpha= (\log_2B)/BR$. The i.i.d elements in block $(r,c)$ are distributed as $\mathcal{N}(0,J_{r,c}\Gamma/L)$. Away 
from the boundaries, in addition to the diagonal (in red), there are $w$ forward and $w$ backward coupling blocks. In this example, the design function $g_w$ enforces a stronger backward coupling where the non-uniform variance across blocks is illustrated by the level of shading. Blocks are darker at the boundaries because the variances are larger so as to enforce the \emph{variance normalization} $\sum_{c=1}^\Gamma J_{r,c} = 1 \ \forall \ r$. The yellow shape emphasizes \emph{variance symmetry}.}
\label{fig:opSpCoupling}
\end{figure}
We consider spatially coupled codes based on coding 
matrices $\bF^{\text{co}} \in \mathbb{R}^{M\times N}$ as depicted in Fig.~\ref{fig:opSpCoupling}. A spatially coupled coding matrix $\bF^{\text{co}}$ is made of $\Gamma\times \Gamma$ \emph{blocks} 
indexed by $(r,c)$, each with $N/\Gamma$ columns and $M/\Gamma =\alpha N/\Gamma$ rows. The structure of $\bF^{\text{co}}$ induces a natural decomposition of the 
message into $\Gamma$ blocks, $\bs = [\bs_1, \dots, \bs_\Gamma]$, where each block is made of $L/\Gamma$ sections.\footnote{Of course $N, M, L, \Gamma$ can always 
be chosen s.t $N/\Gamma, M/\Gamma, L/\Gamma$ are integers.} $\bF^{\text{co}}$ is constructed such 
that each block is coupled (except at the boundaries) with $w$ forward blocks and $w$ backward blocks, where $w$ is the \emph{coupling window}. 
The strength of the coupling is specified by the variance $J_{r,c}$ of each block $(r,c)$. The entries inside each block $(r,c)$ of $\bF^{\text{co}}$ are i.i.d. distributed 
as $\mathcal{N}(0,J_{r,c}\Gamma/L)$.\footnote{In the uncoupled construction the variance scales as the inverse number of sections.
In the coupled construction the variances within a block scales as the inverse number of sections within a block.} 
In order to impose homogeneous power over all the components of $\bF^{\text{co}} \bs$, we tune the (unscaled) block \emph{variances} $J_{r,c}$ such 
that the following \emph{variance normalization} condition holds for all $r \in \{1, \dots,\Gamma\}$
\be\label{eq:varianceNormalization}
\sum_{c=1}^{\Gamma}J_{r,c}= 1.
\ee
This normalization induces homogeneous average power over all codeword components, i.e. $M^{-1}||\bF^{\text{co}} \bs||_{2}^{2} = 1$. 
There are various ways to construct the variance matrix $J$ of the spatially coupled matrix such that \eqref{eq:varianceNormalization} holds. For instance, one can pick $J_{r,c}$'s such that the coupling strength is uniform over the window. However, we will consider a more general construction in this work by using a \emph{design function} $g_w$. The design function satisfies
\be \label{eq:designFunction}
\begin{cases}
g_w(x) = 0 \quad \, \, &\text{if} \quad \,\, |x|>1, \\
\underline{g} \leq g_w(x)\leq \bar{g} \quad &\text{if} \quad\,\, |x| \le 1,
\end{cases}
\ee
where $\bar{g}$, $\underline{g}$ are strictly positive constants independent of $w$. Moreover, $g_w$ is assumed to be Lipschitz continuous on $|x|<1$ with Lipschitz constant $g_*$ independent of $w$. In particular
\be\label{eq:Lipschitz_g}
\big\vert g_w\big(\frac{k}{w}\big) - g_w\big(\frac{k^{\prime}}{w}\big)\big\vert \le \frac{g_*}{w} \vert k - k^{\prime}\vert,
\ee
for $k, k^{\prime} \in \{-w,\dots,w\}$.
Furtheremore, we impose the following normalization
\be \label{eq:normalization_g}
\frac{1}{2w+1}\sum_{k=-w}^{w} g_w(\frac{k}{w}) = 1.
\ee
The design function is then used to construct the variances such that \eqref{eq:varianceNormalization} and \eqref{eq:normalization_g} are satisfied. Hence, we choose
\be \label{eq:blockVariance}
J_{r,c} = \gamma_r  \frac{g_w((c-r)/w)}{2w+1} = 
\frac{g_w((c-r)/w)/(2w+1)}{\sum_{c=1}^\Gamma g_w((c-r)/w)/(2w+1)},
\ee
where $\gamma_r$ is tuned to enforce \eqref{eq:varianceNormalization}. Note that, away from the 
boundaries, $\gamma_r$ is a trivial term equal to $1$. However, $\gamma_r$ changes at the boundaries to compensate for the lower number 
of blocks being coupled (see Fig.~\ref{fig:opSpCoupling} where darker colors were used at the boundaries to stress on this point). The following remarks will be used in the analysis.
We always have $1\leq \gamma_r \leq \underline{g}^{-1}$ and
\be\label{useful-small-bound}
J_{r,c} \leq ({\bar g}/{\underline g})(2w+1)^{-1}.
\ee
In 
the bulk (i.e. away from the boundaries), the following \emph{variance symmetry} property holds for $k\in \{2w+1, \dots,\Gamma-2w\}$
\be \label{eq:varianceSymmetry}
\sum_{r=1}^\Gamma J_{r,k}=\sum_{c=1}^\Gamma J_{k,c}=1.
\ee

%
%It implies $\gamma_r = (2w+1)\big[\sum_{c=\dot c}^{\ddot c} g((r-c)/w)\big]^{-1}$, where $\dot c \defeq {\rm max}(r-w,1)$ and $\ddot c \defeq {\rm min}(r+w,\Gamma)$. Thus, $\gamma_r=1 \ \forall \ r\in\{w+1:\Gamma-w\}$, $\gamma_r > 1$ else. 
%

The ensemble of spatially coupled matrices is then parametrized by $(M,R,B,\Gamma,w,g_w)$. Note that the coupling induced by $g_w$ is not necessarily symmetric, hence the present construction generalizes the ones in \cite{YedlaJian12,6887298,7115123} which all require $g_w(-x)=g_w(x)$, while we do not. This relaxation may strongly improve the perfomances in practice \cite{CaltagironeZ14}.
 
One key element of spatially coupled codes is the \emph{seed} introduced at the boundaries. 
We assume the sections in the first $4w$ and 
last $4w$ blocks of the message ${\tbf s}$ to be known by the decoder (the choice of $4w$ blocks is convenient for the proofs and will become clear in Section \ref{sec:proofsketch}). This boundary 
condition can be interpreted as perfect side information that propagates inwards and boosts the performance. Note that one could also impose the seed differently by constructing a coding matrix with lower communication rate (higher measurement rate) at the boundaries \cite{barbierSchulkeKrzakala,BarbierK15,KrzakalaMezard12,CaltagironeZ14,Montanari-Javanmard}. The seed induces a rate loss in the \emph{effective rate} of the code
\be \label{eq:effectiverate}
R_{\text{eff}} = R \Big(1-\frac{8w}{\Gamma}\Big).
\ee
However, this loss vanishes as $L \rightarrow \infty$ and then $\Gamma \rightarrow \infty$ for any fixed $R$.
As already mentioned, in addition to lower decoding error, the main advantage of coupled SS codes w.r.t power allocated ones is that they allow communication at high rate with a small section size $B$, while power allocated codes require a much larger $B$, which prevents communication of messages of practically relevant sizes \cite{BarbierK15}. Recently, the power allocated SS codes have been optimized in order to achieve better finite size performance \cite{Greig18}.
%
% We naturally denote $\bF_{\mu l}\in \mathbb{R}^{B}$ the vector of components of $\bF$ located at the row $\mu$ and acting on $\bs_l$. Using that $\bF$ has i.i.d entries of zero mean together with the central limit theorem, we obtain for a bulk component: $\tilde y_\mu^2 \to (L/\Gamma)(2w+1)J_0/L = 1$ as $L \to \infty$, indeed verified thanks to the normalization $\sum_{c=1}^{\Gamma} J_{r,c}=1 \ \forall \ r \Rightarrow J_0=\Gamma/(2w+1)$. We obtain similarly $J_*=\Gamma/(w+1)$, and . The rate of a spatially coupled coding matrix is $R = \Gamma R_{\text{bulk}} R_{\text{seed}}/[(\Gamma-w-1) R_{\text{seed}} + (w+1)R_{\text{bulk}}]$, where the \emph{local rate} of the $r^{th}$ block-row is $R_r=\log_2(B)/(\alpha_r B)$ with $\sum_{r=1}^{\Gamma} \alpha_r/\Gamma=M/N$. As $\Gamma\to \infty$, the boundary influence vanishes and $R\to R_{\text{bulk}}$.
%

\section{Generalized approximate message-passing algorithm}\label{sec:GAMP}
\begin{figure}[!t]
\centering
\includegraphics[draft=false,width=.95\textwidth, height = 140pt,trim={5pt 160 5 160},clip]{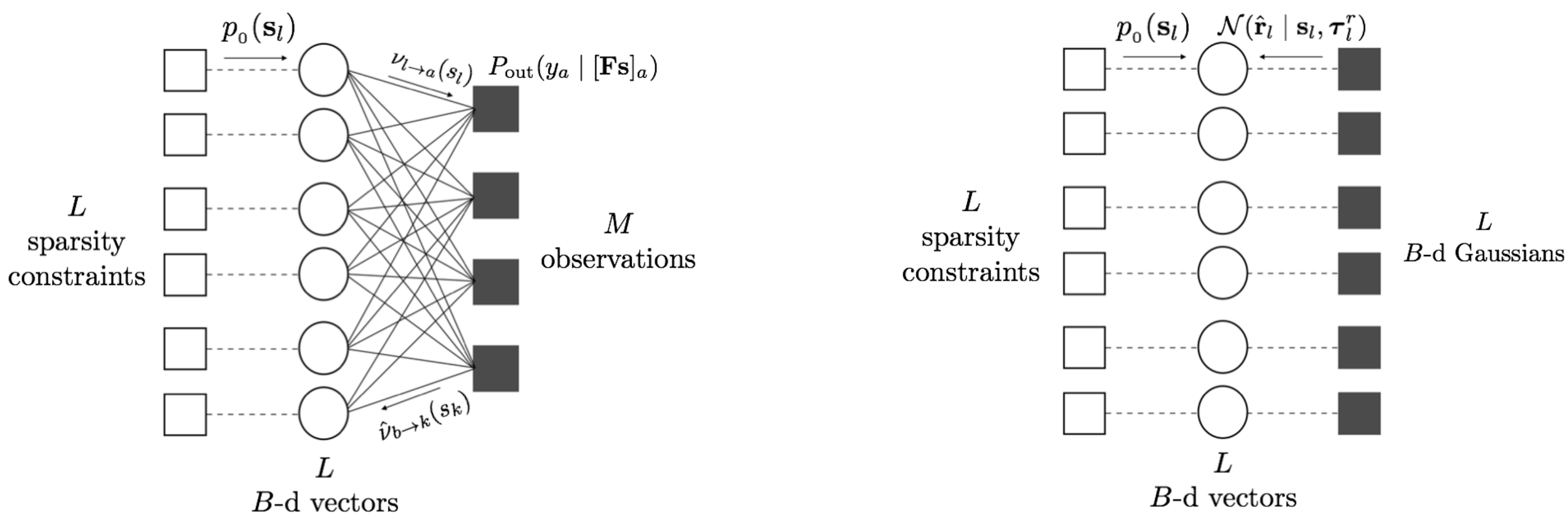}
\caption{Left: Factor graph of the underlying ensemble showing the statistical relationships between the $B$-dimensional sections (circles) of the information word $\bs$ given the known prior $p_0(\bs)$ (plain squares), the coding matrix $\bF$ and the channel observation $\by$ (colored squares). The BP algorithm estimates $\bs$ via iterative exchange of messages, along edges, between circle-nodes and square-nodes. Right: The GAMP algorithm simplifies the BP operations to a sequence of estimation problems from Gaussian noise. At the $l^{th}$ section, $\hat{\bf r}_l$ is the output of an effective Gaussian channel of zero mean and covariance matrix $\text{diag}(\boldsymbol{\tau}^r_l)$.}
\label{fig:factorGraph}
\end{figure}

The posterior distribution describing the statistical relationships in the decoding task is given by (in the following discussion $\bF$ denotes a generic coding matrix)
\begin{equation}\label{eq:posteriorDistribution}
P(\bs|\by,\bF) = \frac{\prod_{l = 1}^L p_0(\bs_l) \prod_{\mu = 1}^M P_{\text{out}}(y_\mu|[\bF\bs]_\mu)}{\int d\bs\prod_{l = 1}^L p_0(\bs_l) \prod_{\mu = 1}^M P_{\text{out}}(y_\mu|[\bF\bs]_\mu)}.
\end{equation}
In the SS codes setting, the sections of the information word are uniformly distributed over all the possible $B$-dimensional vectors with a single non-zero component equal to 1. Hence, the prior of each section reads
\be \label{eq:sectionPrior}
p_0(\bs_l)=\frac{1}{B}\sum_{i=1}^B \delta_{s_{li},1}\prod_{j\neq i}^{B-1} \delta_{s_{lj}, 0},
\ee
where $s_{li}$ is the $i^{th}$ component of the $l^{th}$ section (here $i \in \{1,\dots,B\}$ and $l\in \{1,\dots,L\}$). 
The posterior distribution \eqref{eq:posteriorDistribution} can be represented via a graphical model as shown in the l.h.s of Fig.~\ref{fig:factorGraph}. Therefore, it is natural 
to consider an iterative message-passing algorithm to perform the decoding. For a dense graphical model Belief Propagation (BP) is computationally prohibitive but 
can be simplified down to the AMP algorithm which has been successfully used in many applications, mainly in compressed sensing \cite{BayatiMontanari10,DMM09}. 
The AMP algorithm uses efficient Gaussian (or quadratic) approximations of BP that ``decouple'' the vector-valued estimation problem into a sequence 
of scalar estimation problems under an \emph{effective Gaussian noise} (r.h.s of Fig.~\ref{fig:factorGraph}). The sum-product version 
of AMP (originally used to perform MMSE estimation in compressed sensing with AWGN channel) was adapted in \cite{barbier2014replica,BarbierK15} to SS 
codes by incorporating the structured $B$-dimensional prior distribution (\ref{eq:sectionPrior}). The GAMP algorithm extends the approximations made 
in AMP to any memoryless channel \cite{rangan2011generalized}. Interestingly, the same Gaussian approximations on a dense graph remain valid under GAMP, even for a 
non-Gaussian channel, and the only difference appears in the computation of the effective Gaussian noise levels.
%%
%\begin{figure}[!t]
%\centering
%\includegraphics[draft=false,width=.95\textwidth, height = 155pt]{./figures/factorGraph12.png}
%\caption{Left: Factor graph of the underlying ensemble showing the statistical relationships between the $B$-dimensional sections (circles) of the information word $\bs$ given the known prior $p_0(\bs)$ (light green squares), the coding matrix $\bF$ and the channel observation $\by$ (dark blue squares). The BP algorithm estimates $\bs$ via iterative exchange of messages, along edges, between circle-nodes and square-nodes. Right: The GAMP algorithm simplifies the BP operations to a sequence of estimation problems from Gaussian noise. At the $l^{th}$ section, $\hat{\bf r}_l$ is the output of an effective Gaussian channel of zero mean and covariance matrix $\text{diag}(\boldsymbol{\tau}^r_l)$.}
%\label{fig:factorGraph}
%\end{figure}
%%
%

The GAMP algorithm was originally introduced to estimate signals with i.i.d components \cite{rangan2011generalized}. In the present context the message components are 
correlated through $p_0(\bs_l)$, therefore we adapt GAMP to cover this vectorial setting. The steps of GAMP are shown in Algorithm \ref{alg:gamp} below. 
The ``${\circ2}$'' and ``${\circ-1}$'' symbols mean that the square and inverse operations are taken componentwise: $({\bf F}^{\circ2})_{\mu i} = F_{\mu i}^2$ and 
$({\bf F}^{\circ-1})_{\mu i} = F_{\mu i}^{-1}$. 
All the derivatives in Algorithm \ref{alg:gamp} are also taken componentwise. The sum-product GAMP algorithm produces a sequence of the estimated posterior mean $\hat{\bs}^{(t)}$ and the corresponding estimated posterior variance ${\boldsymbol{\tau}^s}^{(t)}$. The dimensions of the various estimated vectors $\hat{\bs}, \hat{\br}, \dots$ and their corresponding variances $\boldsymbol{\tau}^s, \boldsymbol{\tau}^r, \dots$ are given in Algorithm 1.

In this generalization to the vectorial setting of
SS codes, only steps $12$ and $13$ of Algorithm \ref{alg:gamp} differ from the canonical GAMP algorithm in \cite{rangan2011generalized}. The function $g_\text{in}$ depends on the input prior distibution and it is adapted from \cite{rangan2011generalized} to act on $B$-dimensional vectors. Due to the code construction, $g_{\text{in}}(\hat{\br}_l,\text{diag}(\boldsymbol{\tau}^r_l))$ can be interpreted as the MMSE estimator, or \emph{denoiser}, of a given $B$-dimensional section $\bs_l$ sent through an effective Gaussian channel of zero mean and covariance matrix $\text{diag}(\boldsymbol{\tau}^r_l)$ where
\begin{align} \label{eq:gaussianObs}
\hat{\br}_l = \bs_l + \bxi, \quad \bxi \sim \mathcal{N}(\mathbf{0}, \text{diag}(\boldsymbol{\tau}^r_l)).
\end{align}
\begin{definition}[Denoiser] \label{def:gin}
Formally, we define the denoiser acting \emph{sectionwise} on each $B$-dimensional section of the message as follows
\begin{align}
g_{\text{in}}(\hat{\br}_l,\text{diag}(\boldsymbol{\tau}^r_l)) \defeq  \mathbb{E}[\bS_l \mid \hat{\bR}_l = \hat{\br}_l] = \frac{\int d{\bs_l}\, p_0(\bs_l) \mathcal{N}(\hat{\br}_l\vert\bs_l, \text{diag}(\boldsymbol{\tau}^r_l)) \bs_l}{ \int d{\bs_l} \,p_0(\bs_l) \mathcal{N}(\hat{\br}_l\vert\bs_l, \text{diag}(\boldsymbol{\tau}^r_l)) },
\end{align}
where $\bS_l \sim p_0({\bs_l})$. Plugging \eqref{eq:sectionPrior} yields the componentwise expression of the denoiser used in the GAMP algorithm for SS codes
\begin{align*}
	[g_{\text{in}}(\hat{\br}_l,\text{diag}(\boldsymbol{\tau}^r_l))]_i & = \frac{\exp((2\hat{r}_{li}-1)/(2 \tau^r_{li}))}{\sum_{j=1}^{B}\exp((2\hat{r}_{lj}-1)/(2 \tau^r_{lj}))}
	\nonumber \\ &
	=
	\Big[ 1+ \sum_{j\neq i}^B \exp\Big((2\hat{r}_{lj}-1)/(2 \tau^r_{lj}) -  (2\hat{r}_{li}-1)/(2 \tau^r_{li}) \Big)\Big]^{-1},
\end{align*}
where $i \in \{1,\dots,B\}$.
\end{definition}

Moreover, the componentwise product $\boldsymbol{\tau}^r_l \circ \frac{\partial}{\partial \hat{\textbf{r}}_l} g_\text{in}$ is the estimate of the posterior variance, which quantifies
how "confident" GAMP is in its current iteration, and is given by
\begin{align}\label{eq:dgin_gamp}
\boldsymbol{\tau}^r_l \circ \frac{\partial}{\partial \hat{\textbf{r}}_l} g_\text{in}(\hat{\textbf{r}}_l, \text{diag}(\boldsymbol{\tau}^r_l)) 
& 
\defeq \text{var}(\bS_l \mid \hat{\bR}_l 
= \hat{\br}_l) 
\nonumber \\ &
= \mathbb{E}[\bS_l^{\circ2} \mid \hat{\bR}_l = \hat{\br}_l] - (\mathbb{E}[\bS_l \mid \hat{\bR}_l = \hat{\br}_l])^{\circ2},
\end{align}
where the expectation and the variance are induced from \eqref{eq:gaussianObs}. As the message $\bs$ 
in SS codes consists of only $0$'s and $1$'s, we have that $\mathbb{E}[\bS_l^{\circ2} \mid \hat{\bR}_l = \hat{\br}_l] = \mathbb{E}[\bS_l \mid \hat{\bR}_l = \hat{\br}_l]$. 
Hence, the calculation of $\text{var}(\bS_l \mid \hat{\bR}_l = \hat{\br}_l)$ is immediate using \eqref{eq:sectionPrior} which yields the following componentwise expression
\begin{align*}
	[\boldsymbol{\tau}^r_l \circ \frac{\partial}{\partial \hat{\textbf{r}}_l} g_\text{in}(\hat{\textbf{r}}_l, \text{diag}(\boldsymbol{\tau}^r_l))]_i =[g_{\text{in}}(\hat{\br}_l,\text{diag}(\boldsymbol{\tau}^r_l))]_i - ( [g_{\text{in}}(\hat{\br}_l,\text{diag}(\boldsymbol{\tau}^r_l))]_i)^2.
\end{align*}
The function $g_\text{out}$ of GAMP (see Algorithm \ref{alg:gamp}) is acting componentwise and depends solely on the physical channel $P_\text{out}$. 
The general expression of $g_\text{out}$ is given in Appendix \ref{sec:App_GAMP} as well as examples for different communication channels. The function $g_\text{out}$ can be interpreted as a \emph{score function} of the parameter $\hat{p}$ in the distribution of the random variable $Y \sim P_{\text{out}}(y\mid z)$ with $Z \sim \mathcal{N}(\hat{p},\tau^p)$.

Note that the functions $g_\text{in}$ and $g_\text{out}$ can be seen heuristically as Gaussian (or quadratic) approximations of the sum-product loopy BP updates used in the MMSE estimation. The detailed interpretation of these functions, as well as that of the various parameters of the GAMP algorithm, is given in \cite{rangan2011generalized}, which we omit here since it is lengthy and beyond the scope of this work.

The computational complexity of GAMP is dominated by the $\mathcal{O}(MN) = \mathcal{O}(L^2B\ln(B))$ matrix-vector multiplication. It can be reduced, 
for practical implementations, by using structured operators such as Fourier and Hadamard matrices \cite{barbierSchulkeKrzakala,BBD_ISIT2017}. Fast 
Hadamard-based operators constructed as in \cite{barbierSchulkeKrzakala}, with random sub-sampled modes of the full Hadamard operator, 
allow to achieve a lower $\mathcal{O}(L\ln(B) \ln(BL))$ decoding complexity and strongly reduce the memory 
need \cite{BarbierK15,condo_practical}. Besides practical advantages, using structured operators can lead to a more robust 
finite-length performance \cite{barbierSchulkeKrzakala}. However, random operators are mathematically more tractable and easier to analyse. Hence, we restrict ourselves in this work to random operators.  

\begin{algorithm}[H]
\caption{GAMP ($\by,\bF,B,\rm nIter$)}\label{alg:gamp}
\begin{algorithmic}[1]
\State $\hat{\bs}^{(0)} \quad \gets \, \, \mathbf{0}_{N,1}$%\Comment{Initializations}
\State ${\boldsymbol{\tau}^s}^{(0)} \, \,\gets \, \, (1/B) \mathbf{1}_{N,1}$
\State $\hat{\bz}^{(-1)} \, \,\gets \, \, \mathbf{0}_{M,1}$
\State $t \, \qquad \gets \, \, 0$%\Comment{Iteration counter}
\While{$t<\rm nIter$}
\State ${\boldsymbol{\tau}^p}^{(t)} \quad \, \,\gets \, \, \textbf{F}^{\circ2}{\boldsymbol{\tau}^s}^{(t)} \qquad \in \mathbb{R}^{M}$%\Comment{Output linear step}
\State ${\hat{\textbf{p}}}^{(t)} \qquad \gets \, \, \textbf{F}\hat{\bs}^{(t)} - {\boldsymbol{\tau}^p}^{(t)} \circ \hat{\bz}^{(t-1)}\qquad  \in \mathbb{R}^{M}$
\State $\hat{\bz}^{(t)} \qquad \gets \, \, g_\text{out}(\hat{\textbf{p}}^{(t)}, \by, {\boldsymbol{\tau}^p}^{(t)}) \qquad \in \mathbb{R}^{M}$%\Comment{Output nonlinear step}
\State ${\boldsymbol{\tau}^z}^{(t)} \quad \, \,\gets \, \, -\frac{\partial}{\partial \hat{\textbf{p}}^{(t)}} g_\text{out}(\hat{\textbf{p}}^{(t)}, \by, {\boldsymbol{\tau}^p}^{(t)}) \qquad \in \mathbb{R}^{M}$
\State ${\boldsymbol{\tau}^r}^{(t)} \quad \, \,\gets \, \, {((( {\boldsymbol{\tau}^z}^{(t)})^{\intercal}\textbf{F}^{\circ2})^{\intercal})}^{\circ-1} \qquad \in \mathbb{R}^{N}$%\Comment{Input linear step}
\State $\hat{\textbf{r}}^{(t)} \qquad \gets \, \, \hat{\bs}^{(t)} + {\boldsymbol{\tau}^r}^{(t)} \circ ((\hat{\bz}^{(t)})^{\intercal}\textbf{F})^{\intercal} \qquad \in \mathbb{R}^{N}$
\State $\hat{\bs}^{(t+1)}\quad \gets \, \, g_\text{in}(\hat{\textbf{r}}^{(t)}, \text{diag}({\boldsymbol{\tau}^r}^{(t)})) \qquad \in \mathbb{R}^{N}$%\Comment{Input nonlinear step}
\State ${\boldsymbol{\tau}^s}^{(t+1)} \, \,\gets \, \, {\boldsymbol{\tau}^r}^{(t)} \circ \frac{\partial}{\partial \hat{\textbf{r}}^{(t)}} g_\text{in}(\hat{\textbf{r}}^{(t)}, \text{diag}({\boldsymbol{\tau}^r}^{(t)})) \qquad \in \mathbb{R}^{N}$
\State $t \qquad \quad \gets \, \, t+1$
\EndWhile\label{gampwhile}
%\State \textbf{return} $\hat{\bs}$%\Comment{The prediction scores for each bit are in $\hat{\bs}$}
\end{algorithmic}
\end{algorithm}
Decoding SS codes using iterative message-passing algorithm, such as GAMP, leads asymptotically in $L$ to a sharp \emph{phase transition} below Shannon's capacity. 
The decoder is therefore blocked at a certain threshold separating the ``decodable'' and ``non-decodable'' regions. Moreover, SS codes 
under message-passing decoding may exhibit, asymptotically in $L$ and for any fixed alphabet size $B$, a non-negligible \emph{error floor}\footnote{In fact, the existence of an error floor depends on the communication channel being used. For example there is no error floor for the BEC and BSC (or any binary input channel) when $L\to\infty$ and any fixed $B$ (see \cite{BBD_ISIT2017} for a proof in the BEC case)  but there is one for the AWGN channel as long as $B$ remains finite.} in the decodable region (similarly to low-density generator-matrix codes \cite{PfisterMacrisBMS}). Whenever the error floor exists, it can be made arbitrarily small by increasing $B$ \cite{barbierSchulkeKrzakala,BarbierK15}.

\section{State evolution and potential formulation}\label{sec:stateandpot}
The asymptotic behavior of the AMP algorithm operating on dense graphs can be tracked by a simple recursion called state evolution (SE), similar to the density evolution (DE) for sparse graphs. The rigorous proof showing that SE tracks exactly the asymptotic performance of AMP and GAMP was given in \cite{BayatiMontanari10,Montanari-Javanmard}. Moreover, the extension of the SE equation of AMP to SS code settings, with $B$-dimensional structured prior distribution and power allocation, was proven to be exact in \cite{rush2015capacity}. We believe that the methods of \cite{rush2015capacity} and \cite{Montanari-Javanmard} can be extended to the present setting of spatially coupled SS codes and GAMP algorithm. This would prove that SE correctly tracks GAMP, a conjecture which is firmly supported by numerical simulations \cite{BBD_ISIT2017}.
\subsection{State evolution of the underlying system}
SE tracks the performance of GAMP by computing the average asymptotic mean-square error (MSE) of the GAMP estimate $\hat{\bs}^{(t)}$ at each iteration $t$ 
\be
%\tilde E^{(t)} \defeq \mathbb{E}_{\bs, \by}[\frac{1}{L}\sum_{l=1}^L \|\hat{\bs}_l^{(t)} - \bs_l\|_2^2].
\tilde E^{(t)} \defeq \lim_{L \rightarrow \infty} \frac{1}{L}\sum_{l=1}^L \|\hat{\bs}_l^{(t)} - \bs_l\|_2^2.
\ee
It turns out that tracking the GAMP algorithm is equivalent to running a simple recursion that iteratively computes the MMSE of a single section sent through an \emph{equivalent AWGN channel}. This equivalent channel is induced by the code construction and has an \emph{effective} noise variance that depends solely on 
the physical channel $P_{\rm out}(y|x)$. 
 In order to formalize this, we first need some definitions.
\begin{definition}[Effective noise] \label{def:effNoise}
The effective noise variance $\Sigma^{2}(E)$, parametrized by $E\in [0,1]$, is defined via the following relation
\begin{align*}
\Sigma^{-2}(E) \defeq \frac{\mathbb{E}_{p\vert E} [\mathcal{F}(p|E)]}{R},
\end{align*}
where the expectation $\mathbb{E}_{p\vert E}$ is w.r.t $\mathcal{N}(p|0,1-E)$ and 
\begin{align*}
\mathcal{F}(p|E) \defeq \int dy f(y|p,E) (\partial_p \ln f(y|p,E))^2 
\end{align*}
is the Fisher information of the parameter $p$ associated with the probability distribution of the random variable $Y$ with density
\begin{align*}
f(y|p,E) \defeq \int dx P_{\text{out} }(y|x) \mathcal{N}(x|p,E).
\end{align*}
See Appendix \ref{sec:App_GAMP} for explicit expressions for various communication channels. To get an intuition about this Fisher information, observe that in the AWGN channel, for example, the effective noise variance is directly related to the channel noise parameter with $\Sigma^2(E)=R(E+\rm{snr}^{-1})$, where $\rm{snr}$ is the signal-to-noise ratio.  Intuitively speaking, the effective noise variance is tracking the denoising variance $\tau^r$ of the Algorithm \ref{alg:gamp}. 
\end{definition}

We will need some regularity properties for the function $\Sigma(E)$ which boils down to mild assumptions on the channel transition probability
$P_{\rm out}(y|x)$.
\begin{assumption}[Continuity and boundedness of $\Sigma(E)$]\label{continuity-assumption}
 The channel transition probability $P_{\rm out}(y|x)$ is such that $\Sigma(E)$ is a continuous and twice differentiable function of $E\in [0,1]$. 
 \end{assumption}
\begin{assumption}[Scaling of $\Sigma^{-2}(E)$ as $E\to 0$]\label{scaling-assumption}
The channel transition probability $P_{\rm out}(y|x)$ is such that $\Sigma^{-2}(E)$ and its first two derivatives are bounded by a polynomial in $E^{-1}$. 
Formally, for a given channel there exist two constants $C>0$ and $\beta >0$ such that 
\begin{align}\label{upper-scale}
\max\Big(\Sigma^{-2}(E),\big|\frac{\partial \Sigma^{-2}(E)}{\partial E}\big|, \big|\frac{\partial^2 \Sigma^{-2}(E)}{\partial E^2}\big|\Big) \leq  \frac{C}{RE^{\beta}} \equiv \lambda(E)
\end{align}
for all $E\in [0, 1]$.
\end{assumption}
\begin{figure}[!t]
\centering
\includegraphics[draft=false,width=0.38\textwidth, height=150pt, trim={0pt 0 1 0},clip]{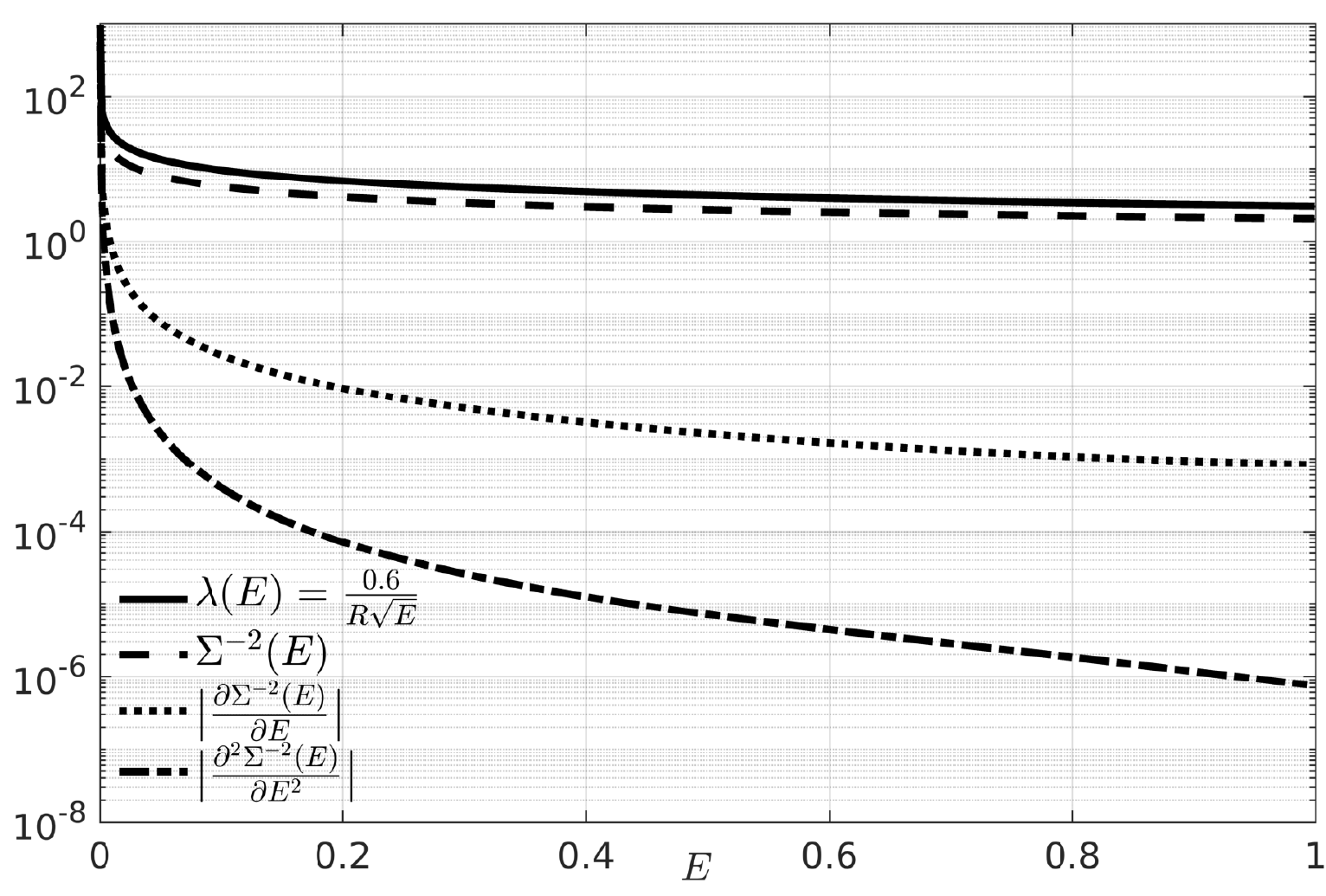}
\centering
\includegraphics[draft=false,width=0.38\textwidth, height=150pt, trim={0pt 0 1 0},clip]{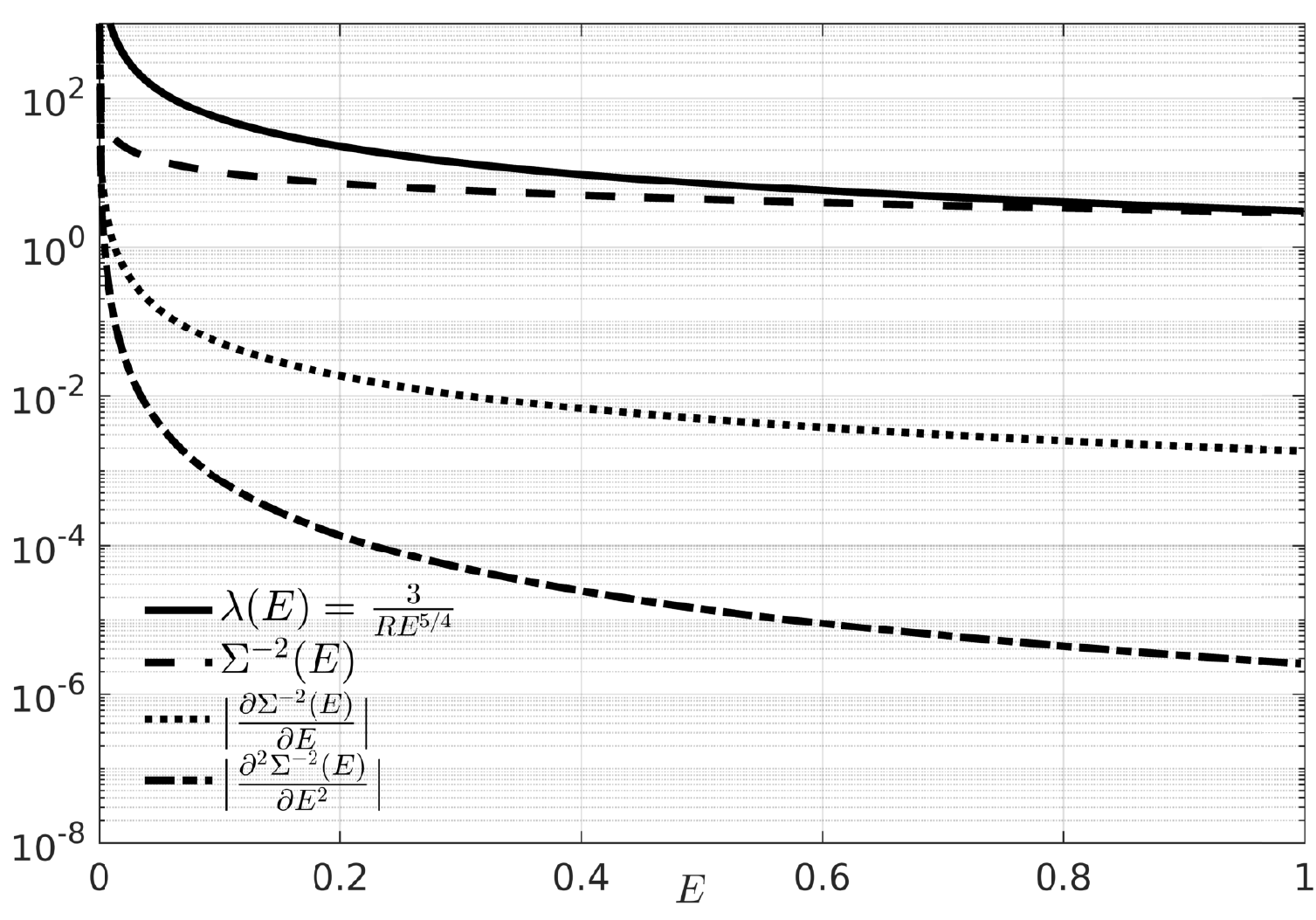}
\caption{$\Sigma^{-2}(E)$ and its first two derivatives in a semi-log scale for the BSC (left) and the BEC (right)  with flip and erasure probabilities
$\epsilon = 0.1$ and $R=0.2$. Assumption \ref{scaling-assumption} is satisfied with exponents $\beta =1/2$ and $5/4$. 
Furthermore, the effective noise variance of both channels is bounded with  $\Sigma^2(E)< 1/2$. Note that the mapping $\pi([\bF\bs]_\mu) = {\rm sign}([\bF\bs]_\mu)$ was used here.}
\label{fig:Sigma_BSC}
\end{figure}

These assumptions will be needed in the proof of threshold saturation in Section \ref{sec:proofsketch}. 
In practice they can be checked on a case by case basis for each channel at hand. For the AWGN channel we have the analytic simple expression 
$\Sigma^2(E) = R(E+{\rm snr}^{-1})$ so the assumptions are obviously satisfied. One can also check them for the
binary symmetric channel (BSC), binary erasure channel (BEC) and Z channel (ZC), using the tedious
expressions for the Fisher information given in Table \ref{table:gout} in Appendix \ref{sec:App_GAMP}.
Fig. \ref{fig:Sigma_BSC} illustrates $\Sigma^{-2}(E)$ and its derivatives for the BSC and BEC. 

The following lemma (which is independent from 
the assumptions) will also be needed.

\begin{lemma} \label{lemma:SigmaIncreases}
$\Sigma^{2}(E)$ is non-negative and increasing with $E$. In particular $\Sigma^2(E) \leq \Sigma^2(1) < +\infty$.
\end{lemma}
\begin{IEEEproof}
Positivity of the Fisher information implies $\Sigma^{2}(E) \ge 0$. The proof that it is increasing is a straightforward application of the data 
processing inequality for Fisher information (e.g. Corollary 6 in \cite{fisherInfoProperties}).
\end{IEEEproof}

From now on, $\bS\sim p_0(\bs)$ and $\mathbf{Z}\sim \mathcal{N}(\mathbf{0},\tbf{I}_B)$ are $B$-dimensional random vectors with corresponding expectations 
denoted $\mathbb{E}_{\bS,\bZ}$, and $Z\sim\mathcal{N}(0,1)$ with expectation denoted $\mathbb{E}_{Z}$.
\begin{definition}[SE of the underlying system] \label{def:SE}
The SE operator of the underlying system is the average MMSE of the equivalent channel 
\begin{align*}
T_{\text{un}}(E) \defeq \text{mmse}\big(\Sigma(E)\big) =\mathbb{E}_{\bS, \bZ}\bigg[\sum_{i=1}^B \Big( \Big[g_{\text{in}}\Big(\bS +\frac{\mathbf{Z} \Sigma(E)}{\sqrt{\log_2 B}},\frac{\tbf{I}_B\,\Sigma^{2}(E)}{\log_2 B}\Big)\Big]_{i} - S_i\Big)^2\bigg],
\end{align*}
%
% \begin{align}\label{equ:stateevopunderlying}
% T_{\rm un}(E) \defeq \mathbb{E}_{\bs, \bz}\biggl[\sum_{i=1}^B \big(g_{\text{in},i}(\bs,\bz,\Sigma(E)) - s_i\big)^2\biggr].
% \end{align}
% %
where $g_{\text{in}}$ is the denoiser given in Definition \ref{def:gin}
\begin{align}\label{denoiser-simplified}
\Big[g_{\text{in}}\Big(\bs +\frac{\bz\Sigma}{\sqrt{\log_2 B}},\frac{\tbf{I}_B\,\Sigma^{2}}{\log_2 B}\Big)\Big]_{i} =
\Bigg[1+\sum_{k\neq i}^B 
e^{(s_k- s_i)\log_2 B/\Sigma^{2} + (z_k - z_i)\sqrt{\log_2 B}/\Sigma}\Bigg]^{-1}.
\end{align}
The SE iteration tracking the performance of the GAMP decoder for the underlying system can be expressed as
\begin{align*}
\tilde E^{(t+1)} = T_\text{un}(\tilde E^{(t)}), \qquad t\geq 0,
\end{align*}
with the initialization $\tilde E^{(0)}=1$.
\end{definition}

Note for further use that \eqref{denoiser-simplified} is a well defined continuous function of $\Sigma >0$ (all other arguments being fixed). At $\Sigma=0$ we define the function by its continuous extension which is obviously finite. Thus we will consider that $g_{\text{in}}$ is continuous for $\Sigma\geq 0$.

After $t$ iterations of the GAMP algorithm, the MSE tracked by SE is denoted by $T_{\text{un}}^{(t)}(\tilde{E}^{(0)})$. The monotonicity properties and the continuity of the SE operator, discussed in Section~\ref{sec:propCoupledSyst}, ensure that eventually all initial conditions converge to a fixed point. More specifically, the following limit exists
\be
\lim_{t \rightarrow \infty} T_{\text{un}}^{(t)}(\tilde{E}^{(0)}) \defeq T_{\text{un}}^{(\infty)}(\tilde{E}^{(0)}),
\ee
for all $\tilde{E}^{(0)}\in [0,1]$ and satisfies
\be
T_{\text{un}} (T_{\text{un}}^{(\infty)}(\tilde{E}^{(0)})) = T_{\text{un}}^{(\infty)}(\tilde{E}^{(0)}).
\ee

Having introduced the SE iteration, the following definitions can be properly stated.
\begin{definition}[MSE Floor]\label{def:MSEfloor} The MSE floor $E_{\rm f}$ is the fixed point reached from the initial condition of zero error,
\begin{align*}
E_{\rm f} = T_{\text{un}}^{(\infty)}(0). 
\end{align*}

Note that for the channels where $E=0$ is {\it not} a trivial fixed point of the SE at a finite section size $B$, the MSE floor $E_{\rm f}$ is {\it strictly positive}. For example, this is the case for the AWGN channel \cite{barbier2014replica,BarbierK15}. However, one can show that for certain channels $W$ there exists a {\it trivial} fixed point $E=0$ of  SE leading to {\it vanishing} MSE floor even at finite $B$. This is typically the case for binary input channels and has been proved explicitly 
for the BEC, BSC and Z channels \cite{BBD_ISIT2017}. For generality, we will always denote the MSE floor as $E_{\rm f}$ whether it is zero or not.
\end{definition}
\begin{definition}[Basin of attraction]
For a fixed channel, the basin of attraction $\mathcal{V}_0$ to the MSE floor $E_{\rm f}$ is defined as
\begin{align*}
\mathcal{V}_0 \defeq \big\{ E \in [0,1] \ \! |\ \! T_{\text{un}}^{(\infty )}(E) = E_{\rm f} \big\}.
\end{align*}
Note that for a given channel, the basin of attraction is a function of the rate as the $T_{\text{un}}$ operator varies with the rate.
\end{definition}
\begin{definition} [Threshold of underlying ensemble]
The GAMP threshold of the underlying ensemble is defined as
\begin{align*}
R_{\text{un}} \defeq {\rm sup}\{R>0\ \! |\ \! T_{\text{un}}^{(\infty)}(1) = E_{\rm f}\}.
\end{align*}
 \end{definition}

For the present system, one can show that the only two possible fixed points are $T_{\text{un}}^{(\infty)}(0)$ and $T_{\text{un}}^{(\infty)}(1)$. For $R<R_{\text{un}}$, there is only one fixed point, namely the ``good'' one
$T_{\text{un}}^{(\infty)}(1)=E_{\rm f}$. Whenever $E_{\rm f}$ is non-zero, it will vanish as the section size $B$ increases (see Section \ref{sec:larg_B}). Instead if $R>R_{\text{un}}$, the GAMP decoder is blocked by the ``bad'' fixed point $T_{\text{un}}^{(\infty)}(1)> E_{\rm f}$. The ``bad'' fixed point does not vanish as B increases.

The GAMP algorithm ``tries'' to minimize the MSE. 
Thus the natural quantity being tracked by SE is the MSE. But one can also assess the performance of GAMP by looking at the section error rate (SER)
% \be
% \rm{SER}^{(t)} = \frac{1}{L}\sum_{l=1}^{L} \mathbbm{1}({\bs_l \neq {\hat{\bs}_l}^{(t)}}),
% \ee
(which is more natural for coding problems) after applying a hard decision (HD) thresholding on the decoder's output.
The analytical relationship between MSE and the SER has been discussed in \cite{barbier2014replica,BarbierK15} and one verifies that an MSE going to 
zero implies a SER going to zero.

\subsection{State evolution of the coupled system}
For a spatially coupled system, the performance of GAMP at each iteration $t$ is described by
an average {\it MSE vector} $[\tilde E_c^{(t)} \, \ |\, \ c \in \{1,\dots, \Gamma\}]$ along the ``spatial dimension'' indexed by the blocks of the message with
\be
%\tilde E_c^{(t)} \defeq \mathbb{E}_{\bs, \by}[\frac{\Gamma}{L}\sum_{l\in c} \|\hat{\tbf s}_l^{(t)} - {\tbf s}_l\|_2^2],\quad c \in \{4w+1,\dots,\Gamma-4w\},
\tilde E_c^{(t)} \defeq \lim_{L\rightarrow \infty}\frac{\Gamma}{L}\sum_{l\in c} \|\hat{\tbf s}_l^{(t)} - {\tbf s}_l\|_2^2,\quad c \in \{4w+1,\dots,\Gamma-4w\},
\ee
where the sum $l\in c$ is over the set of indices of the $L/\Gamma$ sections composing the $c$-th block of $\bs$. To reflect the seeding at the boundaries, we enforce the following \emph{pinning condition} for all $c\in\{1,\dots, 4w\}\cup\{\Gamma-4w+1,\dots,\Gamma\}$
\be\label{eq:pinningConcition}
\tilde E_c^{(t)}=0, \qquad t \ge 0,
\ee
where the message at these positions is assumed to be known to the decoder at all times.

It turns out that the following change of variables
\be
E_r^{(t)} \defeq  \sum_{c=1}^{\Gamma} J_{r,c} \tilde E_c^{(t)},
\ee
where ${\tbf E}= [E_r \ \! | \ \! r \in \{1,\dots,\Gamma\}]$ is called the \emph{profile}, makes the problem mathematically more tractable for spatially coupled codes. The pinning condition implies
\be\label{eq:pinningConcition1}
E_r^{(t)}=0, \qquad t\ge 0,
\ee 
for all $r\in\mathcal{R}\defeq\{1,\dots,3w\}\cup\{\Gamma-3w+1,\dots, \Gamma\}$.

An important concept is that of \emph{degradation} because it allows to compare different profiles. 
\begin{definition}[Degradation] \label{def:degradation}
A profile ${\tbf{E}}$ is degraded (resp. strictly degraded) with respect to another one ${\tbf{G}}$, 
denoted as ${\tbf{E}} \succeq {\tbf{G}}$ (resp. ${\tbf{E}} \succ {\tbf{G}}$), if $E_r \ge  G_r \ \forall \ r$ 
(resp. there exists some $r$ such that the inequality is strict).
\end{definition}

In order to define the SE of the spatially coupled system, we need first the following definition.

\begin{definition}[Per-block effective noise]\label{def:sigmac}
The per-block effective noise variance $\Sigma_{c}^{2}({\tbf E})$ is defined, for all $\,c \in\{1,\dots, \Gamma\}$, by
\begin{align*}
\Sigma_{c}^{-2}({\tbf E}) \defeq  \sum_{r=1}^{\Gamma} \frac{J_{r,c}}{\Sigma^{2}(E_r)}=\sum_{r=1}^{\Gamma} \frac{J_{r,c}}{R} \mathbb{E}_{p\vert E_r} [\mathcal{F}(p|E_r)].
\end{align*}
\end{definition}

\begin{definition}[SE of the coupled system] \label{def:SEc}
The vector valued coupled SE operator is defined componentwise for $t\ge0$ as
\begin{equation*}
E_r^{(t+1)} = [T_{\text{co}}({\tbf E}^{(t)})]_r =
\begin{cases} 
\sum_{c=1}^{\Gamma} J_{r, c}\mathbb{E}_{\bS, \bZ}\Big[\sum_{i=1}^B \Big(g_{\text{in},i}\Big(\bS +\frac{\mathbf{Z} \Sigma_{c}({\tbf E}^{(t)})}{\sqrt{\log_2 B}},\frac{\Sigma_{c}^2({\tbf E}^{(t)})}{\log_2 B}\Big) - S_i\Big)^2\Big] \qquad &r \notin \mathcal{R},\\
0 \quad    &r \in \mathcal{R}.
\end{cases}
\end{equation*}
Note that for $r\in \mathcal{R}$, the pinning condition $E_r^{(t)} =0$ is enforced at all times. SE is initialized with $E_r^{(0)}=1$ for $r\notin \mathcal{R}$.
%(note that 
%for $r\in \mathcal{R}$ we may also initialise with $E_r^{(0)}=1$ since the pinning condition is enforced at all times). 
%
\end{definition}
\begin{definition} [Threshold of coupled ensemble]\label{def:AMPcoupled}
The GAMP threshold of the spatially coupled system is defined as 
\begin{align*}
R_{\text{co}} \defeq {\liminf}_{w\to \infty}{\liminf}_{\Gamma\to \infty} {\rm sup}\{R>0\ \! |\ \! T_{\text{co}}^{(\infty)}(\boldsymbol{1}) \preceq \tbf E_{\rm f}\}
\end{align*}
where $\boldsymbol{1}$ is the all ones vector and ${\tbf E}_{\rm f} \defeq [E_r=E_{\rm f} \ \!|\ \! r \in \{1,\dots,\Gamma\}]$ is the \emph{MSE floor profile} (recall $E_{\rm f}$ 
in Definition \ref{def:MSEfloor}). The existence of the limit $T_{\text{co}}^{(\infty)}(\boldsymbol{1})$ is verified 
in Section \ref{sec:propCoupledSyst}. Note that the degradation $\preceq$ holds with equality for the cases where $E_{\rm f} =0$.
\end{definition}

For the noisy compressed sensing problem, the rigorous proof that SE tracks the performance of GAMP, on both the underlying and spatially coupled models, was already done in \cite{Montanari-Javanmard} by generalizing the work of \cite{BayatiMontanari10}. For the SS codes, we assume that the same results hold.
\begin{assumption}[Accuracy of state evolution]\label{SE-assumption}
We assume that, at least for the channels under Assumption \ref{continuity-assumption} and Assumption \ref{scaling-assumption}, the state evolution equation tracks the performance of GAMP on both the underlying and spatially coupled SS codes.
\end{assumption}

\begin{figure}[!t]
\centering
\includegraphics[draft=false,width=0.45\textwidth, height=145pt, trim={0pt 2 1 0},clip]{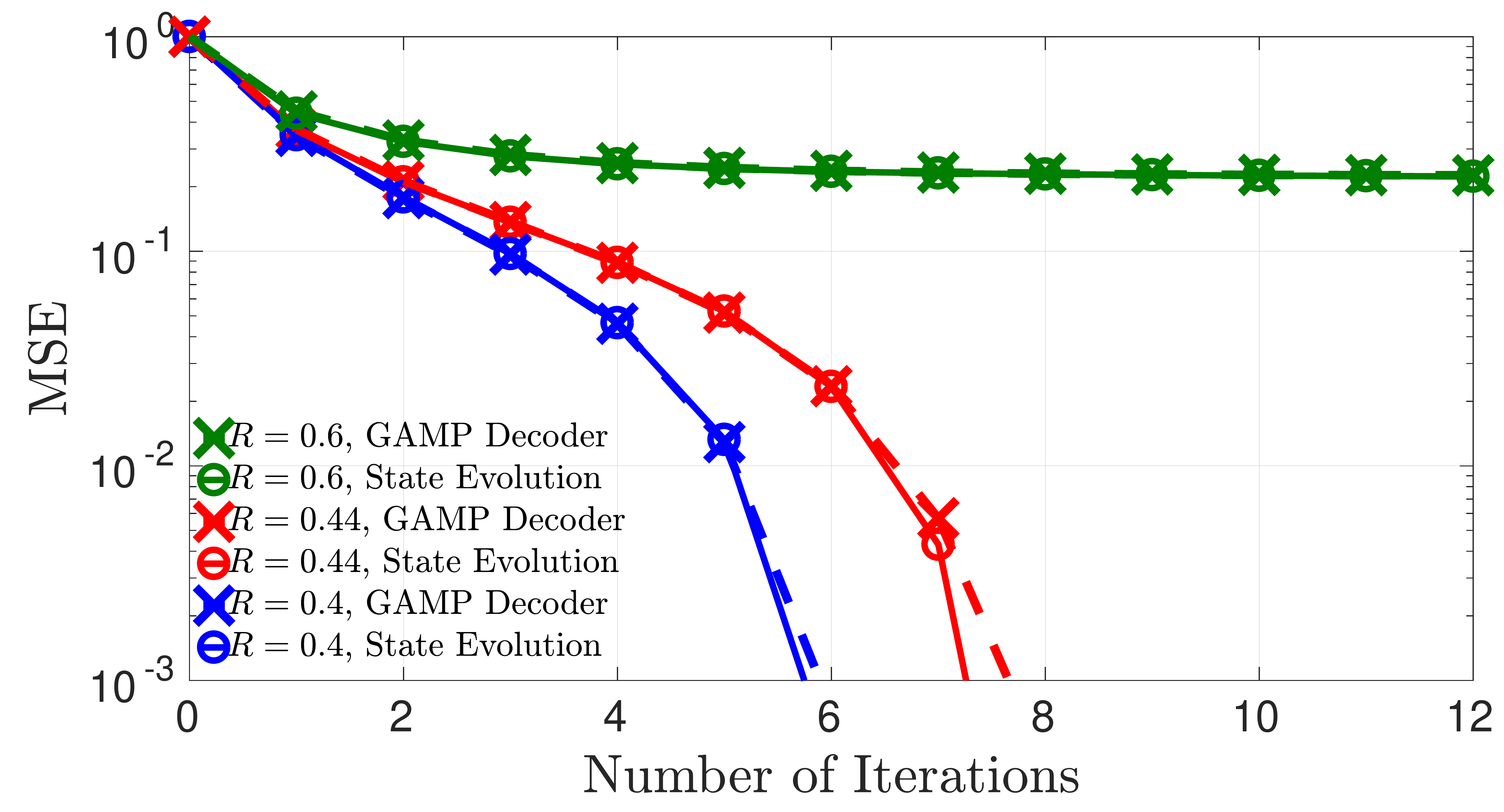}
\centering
\includegraphics[draft=false,width=0.45\textwidth, height=145pt, trim={0pt 2 1 0},clip]{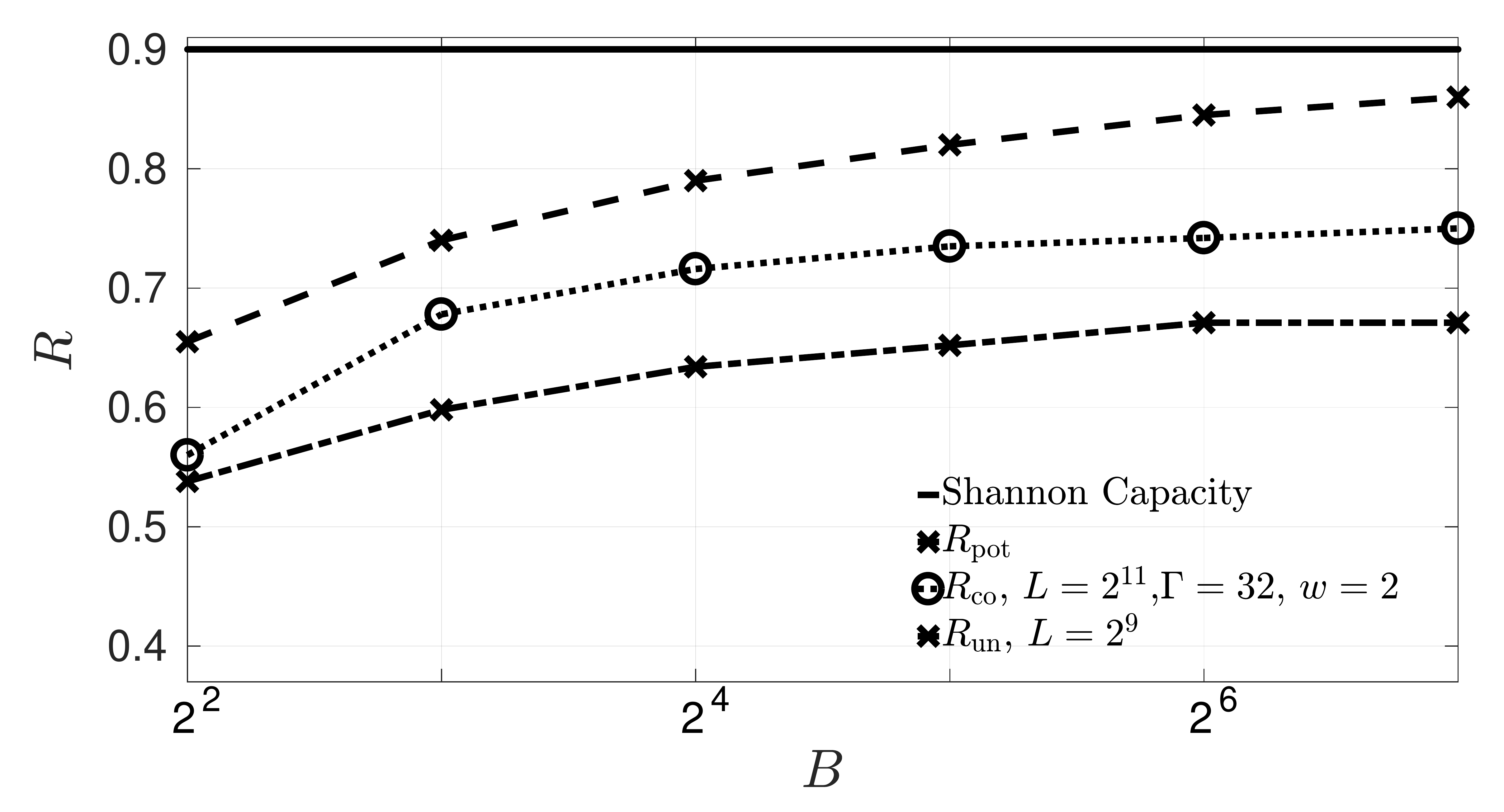}
\caption{The GAMP performance over the binary erasure channel (BEC) with erasure probability $\epsilon = 0.1$. Left: SE tracking the GAMP decoder (MSE performance) at each iteration for three different rates with $L=2^{11}$ and $B=4$. Right: The potential threshold (Definition \ref{def:potThresh}) as well as the GAMP thresholds of the underlying and coupled ensembles are shown as a function of $B$. $R_{\rm{co}}$ saturates $R_{\rm pot}$ as $L$, $\Gamma$ and $w$ go to infinity for all values of $B$. However, $R_{\text{un}}$ maintains a gap to $R_{\rm pot}$, and hence to channel capacity, for all values of $B$ when $L\rightarrow \infty$. See \cite{BBD_ISIT2017} for further numerical simulations on various channels.}
\label{fig:SE_tracking}
\end{figure}

The proof of Assumption \ref{SE-assumption} is beyond the scope of this paper. It would follow from a generalization to any memoryless channel of the analysis done in \cite{rush2015capacity}, that accounts for the $B$-dimensional prior of the SS codes, or more generally from the analysis of the non-separable priors as recently done in \cite{Berthier2017,Fletcher2018}.
Our assumption is, however, supported by numerical simulations \cite{BBD_ISIT2017} (see Fig. \ref{fig:SE_tracking}).
\subsection{Potential formulation}\label{subsec:potentials}
The fixed point solutions of SE can be reformulated as stationary points of a {\it potential function}. This potential function can be obtained from the replica method \cite{barbier2014replica} as shown
in Appendix \ref{sec:App_Bethe} or by directly integrating the SE fixed point equations with the correct ``integrating factor'' as done in \cite{6887298}. Our subsequent
analysis does not depend on the means of obtaining the potential function which is here a mere mathematical tool.  
\begin{definition}[Potential function of underlying ensemble]\label{def:pot_underlying}
The potential function of the underlying ensemble is given by
\begin{align*}
F_{\text{un}}( E) \defeq U_{\text{un}}(E) - S_{\text{un}}( \Sigma(E)),
\end{align*}
with
\begin{align*}
U_{\text{un}}(E) &\defeq -\frac{E}{2\ln(2)\Sigma^{2}(E)} - \frac{1}{R} \mathbb{E}_Z\Big[\int dy\, \phi(y|Z,E) \log_2 \phi(y|Z,E)\Big], \\
S_{\text{un}}( \Sigma( {E})) &\defeq \mathbb{E}_{\bS,\bZ}\Big[\log_B\int d^B{\bx} \,p_0(\bx) \theta(\bx,\bS,\mathbf{Z},\Sigma(E))\Big],
\end{align*}
where
\begin{align*} 
\phi(y|z,E) &\defeq \int dx P_{\text{out}}(y|x)\mathcal{N}(x|z \sqrt{1 - E}, E),\\
% = \psi(z\sqrt{1 - E},y,E).
\theta(\bx,\bs,\bz,\Sigma(E)) &\defeq \exp\bigg(-\frac{\|\bx - (\bs +\bz \Sigma(E)/\sqrt{\log_2 B})\|_2^2}{2\Sigma^{2}(E)/\log_2 B} \bigg).
\end{align*}
Replacing the prior distribution of SS codes \eqref{eq:sectionPrior} in the definition of $S_{\text{un}}( \Sigma( {E}))$, one gets
\begin{align*}
S_{\text{un}}( \Sigma( {E})) \defeq \mathbb{E}_{\textbf{Z}}\Big[\log_B\Big(1+\sum_{i=2}^B e_{i}\big(\textbf{Z},\frac{\Sigma(E)}{\sqrt{\log_2 B}}\big)\Big)\Big],
\end{align*}
where
\begin{align*}
e_{i}(\textbf{z},a) \defeq \exp\big(\frac{z_i - z_1}{a} - \frac{1}{a^2} \big).
\end{align*}
\end{definition}
\begin{definition}[Free energy gap] \label{def:freeEnergyGap}
For a fixed channel, the free energy gap is 
\begin{align*}
\Delta F_{\text{un}} \defeq {\rm inf} _{E \notin \mathcal{V}_0 } (F_{\text{un}}( E) - F_{\text{un}}(E_{\rm f})),
\end{align*}
with the convention that the infimum over the empty set is $\infty$ (i.e. when $R < R_{\text{un}}$).
Note that for a given channel, the free energy gap is a function of the rate as both $F_{\text{un}}$ and $\mathcal{V}_0$ vary with the rate.
\end{definition}
\begin{definition}[Potential threshold]
\label{def:potThresh}
The potential threshold is defined as
\begin{align*}
R_{\rm pot} \defeq {\rm sup}\{R>0\ \! |\ \! \Delta F_{\text{un}} > 0\}.
\end{align*} 
%The region $]R_{\text{BP}}^{\text{s}}, R_*]$ is refered as the \emph{hard phase}, where the free energy of the single %system posses a metastable local minimum at $E_{\text{meta}} > E_{\rm f}$.
\end{definition}
\begin{figure}[!t]
\centering
\includegraphics[draft=false,width=0.45\textwidth, height=152pt, trim={0pt 3 1 0},clip]{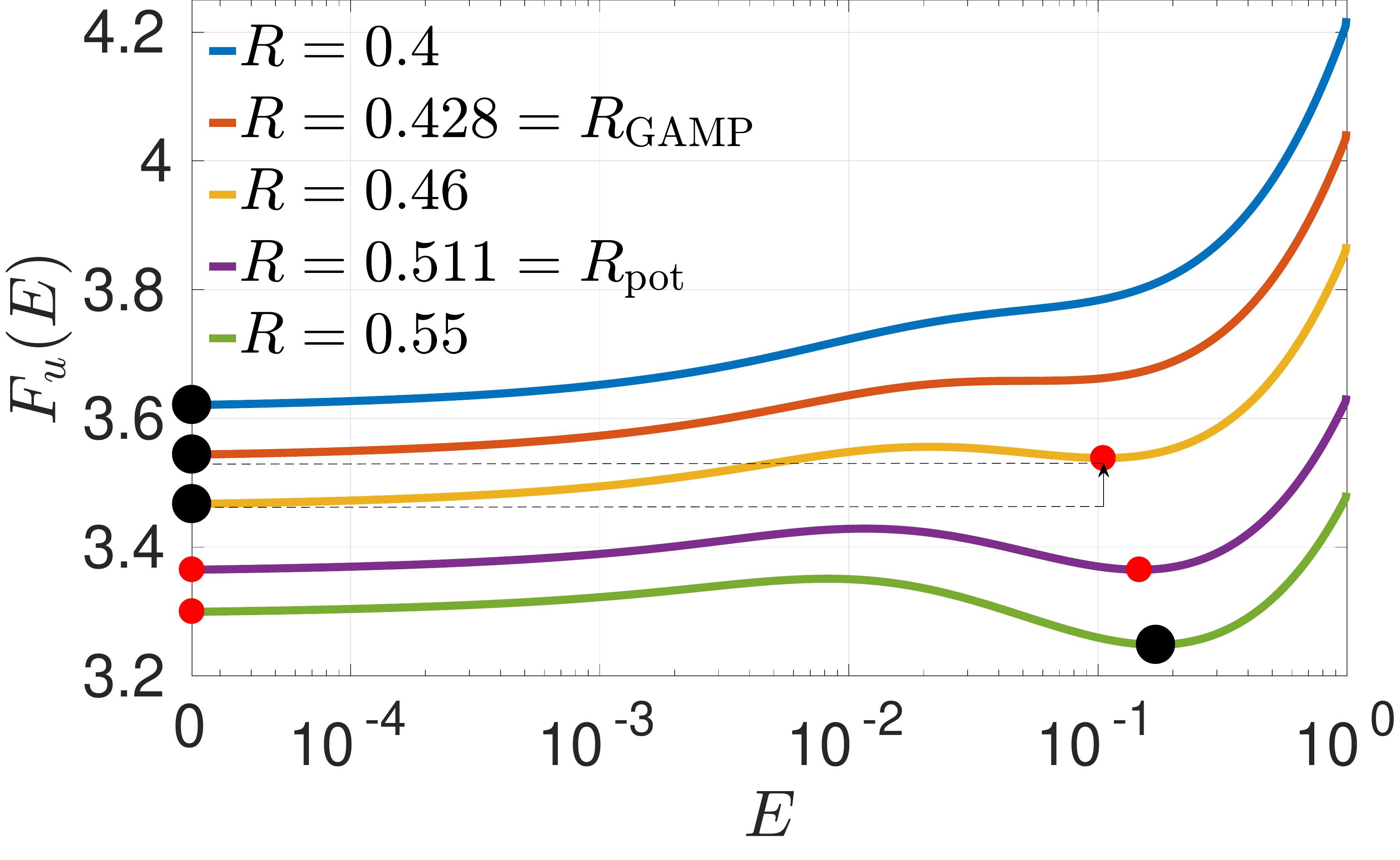}
\centering
\includegraphics[draft=false,width=0.45\textwidth, height=152pt, trim={0pt 3 1 0},clip]{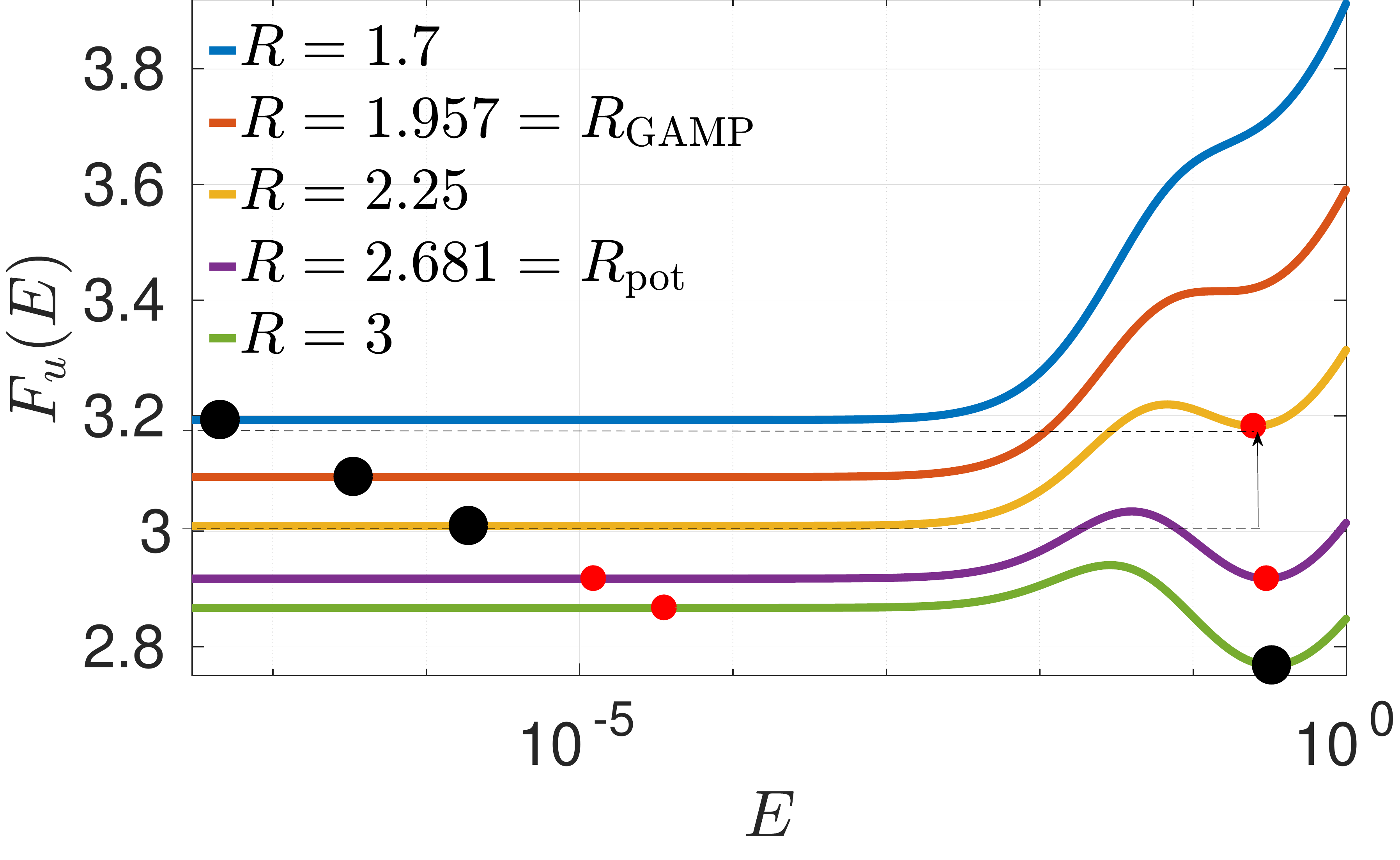}
\caption{The potential functions for the BEC with $\epsilon = 0.1$ (left) and the AWGN channel with $\rm{snr} = 100$ (right), in both cases with $B = 2$. The black dots correspond to the global minima while the red dots correspond to the local minima preventing GAMP to decode (e.g. yellow curves). For a given rate (yellow curves), the black arrows indicate the free energy gap $\Delta F_{\text{un}}$ for each channel. The x-axis is given in the log scale to differentiate between the BEC where there is no error floor and the AWGN channel with non-negligible error floor.}
\label{fig:potential}
\end{figure}

We give examples of potential functions for the BEC and the AWGN channel in Fig. \ref{fig:potential} for $B=2$. Because of Lemma \ref{lemma:fixedpointSE_extPot} below, the minimum that is 
in the basin of attraction of $E=0$ corresponds to the error floor $E_{\rm f}$. We observe that there is a non-vanishing error floor for the AWGN channel but a vanishing one for 
the BEC. The latter situation is also the case for the BSC and Z channel.

Similarly to the underlying ensemble, one can define the potential function of the spatially coupled ensemble that is applied on a vector indexed by the spatial dimension. 
\begin{definition}[Potential function of spatially coupled ensemble]\label{def:pot-coupled}
The potential function of the spatially coupled ensemble is given by
\begin{align*}
F_{\text{co}}({\tbf E}) &\defeq U_{\text{co}}({\tbf E}) - S_{\text{co}}({\tbf E}) = \sum_{r=1}^{\Gamma} U_{\text{un}}(E_r) - \sum_{c=1}^{\Gamma} S_{\text{un}}( \Sigma_c({\tbf E})).
\end{align*}
\end{definition}
The following lemma links the potential and SE formulations. 
\begin{lemma}\label{lemma:fixedpointSE_extPot}
If $T_{\text{un}}(\mathring E) = \mathring E$, then $\frac{\partial F_{\text{un}}}{\partial E}|_{\mathring E} =0$. Similarly for the spatially coupled system, if $[T_{\text{co}}(\mathring{{\tbf E}})]_r = \mathring{E}_r$ $\forall \ r\in \mathcal{R}^{\text{c}} =\{3w+1,\dots, \Gamma-3w\}$ then $\frac{\partial F_{\text{co}}}{\partial E_r}|_{\mathring{{\tbf E}}} = 0 \ \forall \ r\in \mathcal{R}^{\text{c}}$.
\end{lemma}
\begin{IEEEproof}
See Appendix \ref{sec:App_SE}.
\end{IEEEproof}

We end this section by pointing out that the terms composing the potentials have natural interpretations in terms of effective channels. 
The term $\mathbb{E}_Z[\int dy\, \phi \log_2(\phi)]$ in $U_{\text{un}}(E)$
is minus the conditional entropy $H(Y\vert Z)$ for the concatenation of the channels $\mathcal{N}(x\vert z\sqrt{1-E}, E)$
and $P_{\text{out}}(y|x)$ with a standardised input $Z \sim \mathcal{N}(0, 1)$. The term $S_{\text{un}}( \Sigma( {E}))$
is equal to minus the mutual information $I({\bf S};{\bf Y})/\log_{2}B$ for the Gaussian channel $\mathcal{N}(\mathbf{y}|\mathbf{s}, \tbf{I}_B\,\Sigma^2(E)/\log_2 B)$ 
and input distribution $p_0(\bf s)$, up to a constant factor $-(2\ln 2)^{-1}$.

\section{Threshold saturation}\label{sec:proofsketch}
We now prove threshold saturation for spatially coupled SS codes using methods from \cite{6887298}. The main strategy is to assume a ``bad'' fixed point solution of the spatially coupled SE and to 
calculate the change in potential due to a small \emph{shift} in two different ways: $i)$ by second order Taylor expansion (Lemma \ref{lemma:Fdiff_quadraticForm} 
and Lemma \ref{lemma:quadFormBounded}), $ii)$ by direct evaluation (Lemma \ref{lemma:diffShited_directEval}). We then show 
by contradiction that as long as $R < R_{\rm pot}$ the SE converges to the ``good'' fixed point (Theorem \ref{th:mainTheorem}).

Our threshold saturation proof follows the lines of \cite{6887298}. However, we consider a more general coupling construction. More specifically, we assume a general coupling strength which is not necessarily uniform or symmetric as in \cite{6887298}. This relaxation could significantly improve the performance in practice \cite{CaltagironeZ14}. Moreover, it is worth noting that carrying out step $i)$ presents some technical difficulties when bounding the second-order Taylor expansion of the coupled state evolution which do not appear in \cite{6887298}. This is due to the special form of the state evolution tracking the performance of the  GAMP algorithm for SS codes over general channels.

In Section \ref{sec:propCoupledSyst} we start by showing some essential properties of the spatially coupled SE operator. 
%A proper shift is applied and the necessary lemmas and theorem are then given details. 

\subsection{Properties of the coupled system}\label{sec:propCoupledSyst}
Monotonicity properties of the SE operators $T_{\text{un}}$ and $T_{\text{co}}$ are key elements in the analysis. 
%We start by proving some useful lemmas.
% \begin{lemma}
% The entropy $S^{\text{s}}( \Sigma)$ is a non-negative increasing function of the temperature $ \Sigma^2$. \label{lemmma:Sincreases}
% \end{lemma}
% %
% \begin{IEEEproof}
% \tbf{(TO DO)}.
% \end{IEEEproof}
% %
% \begin{lemma}
% \label{lemmma:Uincreases}
% The coupled energy $U_{\text{co}}({\tbf E})$ increases with $ E_r$.
% \end{lemma}
% %
% \begin{IEEEproof}
% %
% Knowing that $ E_r\ge 0$, we have $\partial_{ E_r} U_{\text{co}}({\tbf E})= E_r/\big[2R_r\ln(2)(1/{\rm{snr}} +  E_r)^2\big]\ge 0.$
% \end{IEEEproof}
%
\begin{lemma}
The SE operator of the coupled 
system maintains degradation in space, i.e.
%if $ {\tbf E} \preceq  {\tbf G}$, then $T_{\txt{co}}({\tbf E}) \preceq T_{\txt{co}}({\tbf{G}})$. 
 if  ${\tbf E} \succeq  {\tbf G}$, then $T_{\txt{co}}({\tbf E}) \succeq T_{\txt{co}}({\tbf{G}})$. 
This property is verified for $T_{\txt{un}}$ for a scalar error as well.
\label{lemma:spaceDegrad}
\end{lemma}
\begin{IEEEproof}
Combining Lemma~\ref{lemma:SigmaIncreases} with the first equality in Definition \ref{def:sigmac} implies that if ${\tbf E} \succeq  {\tbf G}$, then
$\Sigma_c({\tbf E}) \geq \Sigma_c({\tbf G})\ \forall \ c$. Now, the SE operator of Definition \ref{def:SEc} can be interpreted 
as an average over the spatial dimension of local MMSE's. The local MMSE's for each position $c=1,\cdots,\Gamma$ are the ones 
of $B$-dimensional equivalent AWGN channels with noise $\bxi \sim \mathcal{N}(\mathbf{0}, \tbf{I}_B\,\Sigma_c^2/\log_2B)$. These 
are non-decreasing functions of $\Sigma_c^2$: this is intuitively clear but we provide a justification based on an explicit formula for the derivative below.
Thus $[T_{\txt{co}}( {\tbf E})]_r \geq [T_{\txt{co}}( {\tbf G})]_r \ \forall \ r$, which means $T_{\txt{co}}({\tbf E}) \succeq T_{\txt{co}}({\tbf{G}})$.

The derivative of the MMSE of the Gaussian channel with i.i.d noise $\mathcal{N}(\mathbf{0}, \tbf{I}_B\,\Sigma^2)$ can be computed as 
\begin{align}
 \frac{d \ \! {\rm mmse}(\Sigma)}{d(\Sigma^{-2})} =  \frac{d}{d(\Sigma^{-2})} \mathbb{E}_{\bX, {\tbf Y}}\bigl[\|\bX - \mathbb{E}[\bX\vert {\tbf Y}]\|_2^2\bigl] = - 2 \mathbb{E}_{\bX, {\tbf Y}}\bigl[\|\bX - \mathbb{E}[\bX\vert {\tbf Y}]\|_2^2 {\rm Var}[\bX\vert {\tbf Y}]\bigl].
\end{align}
This formula is valid for vector distributions $p_0(\bx)$, and in particular, for our $B$-dimensional sections. 
It confirms that $T_{\text{un}}$ (resp. $[T_{\txt{co}}]_r$) is a non-decreasing function of $\Sigma$ (resp. $\Sigma_c$). In particular the local MMSE's for each
position $c = 1,\cdots, \Gamma$
in definition
\ref{def:SEc} are non-decreasing.
\end{IEEEproof}

\begin{figure}[!t]
\centering
\includegraphics[draft=false,width=0.7\textwidth, height=90pt]{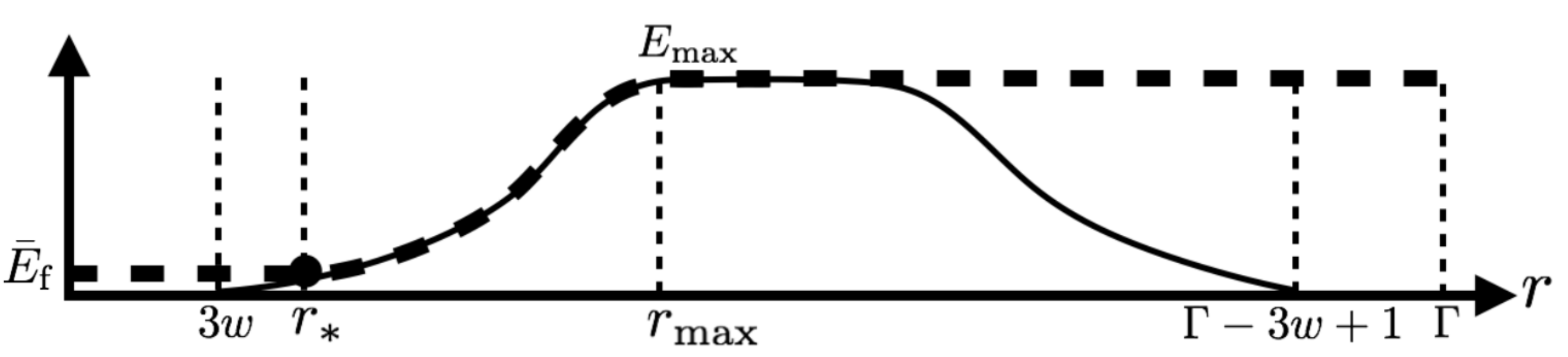}
\caption{A non-symmetric error profile in a typical SE iteration. The solid line corresponds to the original spatially coupled system and the dashed line to the {\it modified} system. 
The error profile of the original system has a $0$ plateau for all $r \le 3w$ and it increases until $r_{\rm max}$ where it 
reaches its maximum value $E_{\rm max} \in [0,1]$. It flattens after $r_{\rm max}$ then it decreases to reach $0$ at $\Gamma-3w+1$ and remains null after. 
The non-symmetric shape of the double-sided wave in Fig.~\ref{fig:errorProfile} emphasises that we are considering the 
generic case of non-symmetric coupling strength when designing spatially coupled matrices (see Section \ref{sec:codeens}).
The error profile of the modified system (dashed line) starts with a plateau at $\bar{E}_{\rm f}$ for all $r \le r_*$, where $r_*+1$ is the first position s.t the original profile is at least $\bar{E}_{\rm f}$, and then matches that of the original system for all $r\in\{r_*,\dots,r_{\rm max}\}$. It then saturates 
to $E_{{\rm max}}$ for all $r \ge r_{\rm max}$. Note that if ${\tbf E} \preceq \bar{{\tbf E}}_{\rm f}$ then $r_* = r_{\rm max}$. By construction, the error 
profile of the modified system is non-decreasing and degraded with respect to that of the original system.
}
\label{fig:errorProfile}
\end{figure}

\begin{corollary}
\label{cor:timeDegrad}
The SE operator of the coupled 
system maintains degradation in time, 
i.e. $T_{\txt{co}}({\tbf E}^{(t)}) \preceq  {\tbf E}^{(t)}$ 
implies $T_{\txt{co}}({\tbf E}^{(t+1)}) \preceq  {\tbf E}^{(t+1)}$. 
Similarly, $T_{\txt{co}}({\tbf E}^{(t)}) \succeq  {\tbf E}^{(t)}$ 
implies $T_{\txt{co}}({\tbf E}^{(t+1)}) \succeq  {\tbf E}^{(t+1)}$. 
Furthermore, if we take the initial conditions ${\bf E}^{(0)} = {\bf 1}$ (the all one-vector) or ${\bf E}^{(0)} = {\bf 0}$ (the all zero-vector)
the \emph{limiting profile}
\be
\lim_{t\rightarrow \infty} {\tbf E}^{(t)} \defeq T_{\txt{co}}^{(\infty)}({\tbf E}^{(0)}),
\label{26}
\ee
exists. Finally under Assumption \ref{continuity-assumption} the limiting profile is a fixed point of $T_{\txt{co}}$, i.e., 
\be
T_{\txt{co}} (T_{\txt{co}}^{(\infty)}({\tbf E}^{(0)})) = T_{\txt{co}}^{(\infty)}({\tbf E}^{(0)}).
\label{27}
\ee
These properties are verified by $T_{\txt{un}}$ for the underlying system as well.
\end{corollary}

\begin{IEEEproof}
First we note $T_{\txt{co}}({\tbf E}^{(t)}) \preceq  {\tbf E}^{(t)}$ 
means ${\tbf E}^{(t+1)} \preceq {\tbf E}^{(t)}$ and thus by Lemma~\ref{lemma:spaceDegrad} $T_{\txt{co}}({\tbf E}^{(t+1)}) \preceq T_{\txt{co}}({\tbf E}^{(t)})$ which means
$T_{\txt{co}}({\tbf E}^{(t+1)}) \preceq  {\tbf E}^{(t+1)}$. The same argument shows that $T_{\txt{co}}({\tbf E}^{(t)}) \succeq  {\tbf E}^{(t)}$ 
implies $T_{\txt{co}}({\tbf E}^{(t+1)}) \succeq  {\tbf E}^{(t+1)}$. Let us show the existence of the limit \eqref{26} when we start with the initial condition 
${\tbf E}^{(0)}={\bf 1}$. This flat profile is maximal at every position thus after one iteration we necessarily have ${\tbf E}^{(1)}\preceq {\tbf E}^{(0)}$.
Applying $t$ times the operator $T_{\txt{co}}$ we get ${\tbf E}^{(t+1)}\preceq {\tbf E}^{(t)}$ which means $E_r^{(t+1)}\leq   E_r^{(t)}$. Thus for every position we have 
a non-increasing sequence which is non-negative. Thus the sequence converges and $\lim_{t\to \infty} {\tbf E}^{(t)} = T_{\txt{co}}^{(\infty)}({\bf 1})$
exists. The same argument applies if we start from the initial condition ${\tbf E}^{(0)}={\bf 0}$ (the limit may be different of course). 
To show the last statement \eqref{27} we argue that $T_{\txt{co}}$ is continuous with respect to ${\tbf E}$.  
We already noted after Definition \ref{def:SE} that the denoiser $[g_{{\rm in}}]_i$ is a continuous function of $\Sigma\geq 0$. Clearly, the denoiser satisfies $0 \le [g_{{\rm in}}]_i \le 1$ also,  and so does the expression $([g_{{\rm in}}]_i - s_i)^2$. 
A look at the Definition \ref{def:SEc} of $[T_{\txt{co}}({\bf E})]_r$ thus shows, by Lebesgue's dominated convergence theorem, that
$[T_{\txt{co}}({\bf E})]_r$ is jointly continuous in $\Sigma_c({\bf E})$, $c=1, \cdots, \Gamma$. Thanks to Definition \ref{def:sigmac} and the Assumption \ref{continuity-assumption} 
of continuity of $\Sigma(E)$, we conclude that $T_{\txt{co}}$ is a continuous function of ${\bf E}$.
\end{IEEEproof}
%

%We assume that the initial profile is $E_r^{* (0)} = 1 \ \forall \ r$.
\begin{corollary}\label{rk:shape}
%\begin{proposition}\label{prop:shape}
Starting from the error profile $\tbf{E}^{(0)} = \tbf{1}$ and due to the pinning condition, as the SE progresses the perfect side information propagates inwards and the error profile adopts the shape of the {\it solid line} 
shown on Fig.~\ref{fig:errorProfile} for every iteration $t>1$:
%it is $0$ for $r \le 3w$, non-decreasing for $3w \le r \le r_{\rm max}$, 
%non-increasing for $r_{\rm max} \le r \le \Gamma -3w +1$ and $0$ for $\Gamma -3w +1 \le r \le \Gamma$; for $r_{\rm max}\in\{3w,\dots,\Gamma-3w+1\}$.
it is non-decreasing for $r \le r_{\rm max}$ and non-increasing for $r \ge r_{\rm max}$ for some value of $r_{\rm max}\in\{3w,\dots,\Gamma-3w+1\}$.   
%\end{proposition}
\end{corollary}

\begin{IEEEproof}
For a large enough $\Gamma$, the pinning condition \eqref{eq:pinningConcition1} and the variance symmetry \eqref{eq:varianceSymmetry} ensure that in 
the first SE iteration $\Sigma^2_c(\tbf{E}^{(0)}={\bf 1})$ satisfies the following ordering along the positions: $i)$ it is non-decreasing for all $c\in\{1,\dots,4w+1\}$, $ii)$ it is non-increasing for all $c\in\{\Gamma- 4w,\dots,\Gamma\}$, $iii)$ it is constant elsewhere.
Using the pinning condition again and the fact that the componentwise SE operator is non-decreasing in $\Sigma^2_c$ (see the proof of Lemma \ref{lemma:spaceDegrad}), one 
can show that after the first SE iteration the error profile $\tbf{E}^{(1)}$  must adopt the following ordering: $i)$ it is non-decreasing 
for all $r\in\{1,\dots,5w+1\}$, $ii)$ it is non-increasing for all $r\in\{\Gamma- 5w,\dots,\Gamma\}$, $iii)$ it is constant elsewhere. Repeating 
the same argument by recursion one deduces that a double-sided 
wave (solid line shown in Fig. \ref{fig:errorProfile}) propagates inwards as the SE progresses. 
\end{IEEEproof}

Recall that state evolution is initialized with ${\tbf E}^{(0)}={\tbf 1}$. The iterations will eventually converge to a {\it fixed point profile}
\be
{\tbf E}^{(\infty)} \defeq T_{\txt{co}}^{(\infty)}({\tbf 1}).
\ee
The fixed point reached by SE may be the ``good''  MSE floor profile $\tbf{E}_{\rm f}$ or may be a ``bad'' profile which is strictly degraded 
with respect to $\tbf{E}_{\rm f}$.

\subsection{Proof of threshold saturation}

The goal of this section is to arrive at a proof of the two main results, namely Theorem \ref{th:mainTheorem} and Corollary \ref{cor:maincorollary},
both formulated at the end of the section. In this section we consider rates in the range $0< R < R_{\rm pot}$. 
Thus the gap given in Definition \ref{def:freeEnergyGap} is strictly positive and finite, i.e., $0< \Delta F_{\text{un}} < +\infty$. 

\begin{definition}[The pseudo error floor]\label{def:pseudo}
We fix $0<\eta <1$ (the reader may as well think of $\eta =1/2$ in all subsequent arguments of this section). It can be shown that continuity of $\Sigma(E)$ (Assumption \ref{continuity-assumption}) implies that the potential function $F_{\text{un}}(E)$ is continuous 
for $E\in [0, 1]$. In particular it is continuous at the error floor $E_{\rm f}$. Therefore we can find $\delta(\eta, B, R) > 0$ such that 
$\vert F_{\text{un}}(E) - F_{\text{un}}(E_{\rm f})\vert \leq \eta \Delta F_{\text{un}}$ whenever $\vert E - E_{\rm f}\vert \leq \delta(\eta, B, R)$. Now 
we take {\it any} $0 <\epsilon <\delta(\eta, B, R)$ and set $\bar E_{\rm f} = E_{\rm f} +\epsilon$. We have in particular
$\vert F_{\text{un}}(\bar E_{\rm f}) - F_{\text{un}}(E_{\rm f})\vert \leq \eta \Delta F_{\text{un}}$.
This number $\bar E_{\rm f}$, will serve as a "pseudo error floor" in the analysis. 
\end{definition}

\begin{definition}[The modified system]\label{def:modSystem}
The modified system is a modification of the SE iterations defined by applying {\it two saturation constraints} to the error profile of the original system {\it at every iteration}. 
First recall that the error profile of the original system has a $0$ plateau for all $r \le 3w$ and increases until $r_{\rm max}$ where it 
reaches its maximum value $E_{\rm max} \in [0,1]$. It flattens after $r_{\rm max}$ then it decreases to reach $0$ at $\Gamma-3w+1$ and remains null after.  Now take {\it any} $0< \epsilon < \delta(\eta, B, R)$ and set $\bar E_{\rm f} = E_{\rm f} + \epsilon$
where $E_{\rm f}$ is the true error floor. At each iteration the profile of the {\it modified system} is defined by applying the following two saturation constraints: (i) the profile is set to the pseudo error floor $\bar E_{\rm f}$ for all $r \le r_*$, where $r_*+1$ is the first position s.t the original profile is at least $\bar{E}_{\rm f}$;  (ii) the profile is set to $E_{{\rm max}}$ for all $r \ge r_{\rm max}$. For $r\in\{r_*,\dots,r_{\rm max}\}$ the profiles of the modified and original systems are equal.
\end{definition}

Figure~\ref{fig:errorProfile} gives an illustration of this definition: the full line corresponds to the original system and the  dashed one to the modified system. 
By construction, the error profile of the modified system is non-decreasing and degraded with respect to that of the original system. 
We note that when the error floor is non-vanishing (e.g. on the AWGN channel) we could take in the analysis $\bar E_{\rm f} = E_{\rm f} +\epsilon \to E_{\rm f}$ for fixed 
code parameters. However for zero error floor we need to have $\epsilon > 0$ in the analysis. For code parameters $w$ and $\Gamma$ large enough we can make $\epsilon$ arbitrarily small. 

The fixed point profile of the modified system is degraded with respect to ${\tbf E}^{(\infty)}$, thus the modified system serves as an upper bound in our proof. Note that the SE iterations of 
the modified system also satisfy the monotonicity properties of $T_{\text{co}}$ (see Section \ref{sec:propCoupledSyst}). Moreover, the modified system preserves the shape of the single-sided wave at all times. In the rest of this section we shall work with the modified system.

We now choose a proper shift of the saturated profile 
in Definition \ref{def:shift}, and then evaluate the change in potential due to this shift in two 
different ways in Lemma \ref{lemma:quadFormBounded} and Lemma \ref{lemma:diffShited_directEval}. 
Theorem \ref{th:mainTheorem} and Corollary \ref{cor:maincorollary} will then be easy consequences.  
\begin{definition}[Shift operator]\label{def:shift}
The \emph{shift operator} is defined pointwise as
$[\text{S}({\tbf E})]_1 \defeq \bar E_{\rm f}, \ [\text{S}({\tbf E})]_r \defeq  E_{r-1}$.
\end{definition}

\begin{lemma}\label{leminterp}
Let ${\tbf E}$ be a fixed point profile of the modified system initialized with ${\tbf E}^{(0)}=\tbf{1}$. Then there exist $\hat{t}\in[0,1]$ such that
\begin{align*}
F_{\text{co}}(\text{S}({\tbf E})) &-F_{\text{co}}({\tbf E}) = \frac{1}{2} \sum_{r,r'=1}^{\Gamma} \Delta {E}_r \Delta {E}_{r'} \left[\frac{\partial^2 F_{\text{co}}}{\partial  E_{r}\partial  E_{r'}}\right]_{\hat {\tbf E}}.
\end{align*}
where $\Delta {E}_{r} \defeq {E}_{r} -{E}_{r-1}$ and $\hat {\tbf E} \defeq (1-\hat{t}) {\tbf E} + \hat{t}\text{S}({\tbf E})$.
Note that $\hat{t}$ depends in a non-trivial fashion on ${\tbf E}$. 
\label{lemma:Fdiff_quadraticForm}
\end{lemma}
\begin{IEEEproof}
Consider $F_{\text{co}}(t)\defeq F_{\text{co}}({\tbf E} + t(\text{S}({\tbf E}) - {\tbf E}))$ and note that 
$F_{\text{co}}(0) = F_{\text{co}}({\tbf E})$, $F_{\text{co}}(1) = F_{\text{co}}(\text{S}({\tbf E}))$. 
Since $[\text{S}({\tbf E})]_r = {E}_{r} + \Delta {\tbf E}_r$ the mean value theorem yields 
\begin{align}
F_{\text{co}}(\text{S}({\tbf E})) - F_{\text{co}}({\tbf E}) =
 - \sum_{r=1}^{\Gamma} \Delta {E}_{r}  \left[\frac{\partial F_{\text{co}}}{\partial  E_r}\right]_{{\tbf E}}
+\frac{1}{2}\sum_{r,r'=1}^{\Gamma} \Delta {E}_{r} \Delta {E}_{r'}  \left[\frac{\partial^2 F_{\text{co}}}{\partial  E_{r}\partial  E_{r'}}\right]_{\hat {\tbf E}}, \label{eq:expFbsminusFb}
\end{align}
for some suitable $\hat t\in [0,1]$. 
By saturation of $\tbf E$, $\Delta E_r=0 \ \forall \ r \in\mathcal{B}\defeq\{1,\dots,r_*\}\cup \{r_{\rm max}+1,\dots,\Gamma\}$. 
Moreover for $r\notin\mathcal{B}$, $E_r=[T_{\text{co}}(\tbf E)]_r$, and thus by Lemma~\ref{lemma:fixedpointSE_extPot} the potential 
derivative cancels at these positions. 
Hence the first sum in the right hand side of (\ref{eq:expFbsminusFb}) cancels.
\end{IEEEproof}
\begin{lemma}\label{lemma:Evariesslowly}
The fixed point profile of the modified system initialized with $\tbf{E}^{(0)}=\tbf{1}$ is \emph{smooth}, meaning that  $\Delta {E}_{r} $ satisfies the following
\begin{align*}
|\Delta {E}_{r} | & \leq  \frac{g_* + \bar g}{w\underline g}   \exp({-c(B) \Sigma^{-2}(E_{r+w})}) 
\nonumber \\ &
\leq 
\frac{g_* + \bar g}{w\underline g} ,
\end{align*}
where $w$ is the coupling window and $c(B)>0$ is a constant depending only on $B$; whereas $g_*$, $\bar g$ and $\underline g$ correspond to the design function defined in Section \ref{subsec:SC_SS}. 
\end{lemma}
\begin{IEEEproof}
$\Delta E_r=0$ for all $r \in\mathcal{B}$. By construction of $\{J_{r,c}\}$ we have 
$$
J_{r,c} \leq \frac{g_w((r-c)/w)}{\underline g (2w+1)}.
$$
Moreover from Definitions \ref{def:sigmac} and \ref{def:SEc} of the coupled state evolution operator, 
the fact that ${\rm mmse}$ is an increasing function of the noise and Lemma \ref{lemma:SigmaIncreases}, we have
${\rm mmse}(\Sigma_c({\tbf E})) \leq {\rm mmse}(\Sigma(E_{r+w}))$ for $c= r-w, \cdots, r+w$.
Thus using Lipschitz continuity of $g_w$, 
we have for all $r \notin\mathcal{B}$ that
\begin{align}\label{bad-bound}
|\Delta {E}_{r}| = \Big|[T_{\text{co}}({\tbf E})]_r - [T_{\text{co}}({\tbf E})]_{r-1}\Big| & =\Big|\sum_{c=1}^{\Gamma} (J_{r,c} - J_{r-1,c}) {\rm mmse}(\Sigma_c({\tbf E})) \Big|\nonumber \\
& \le \frac{{\rm mmse}(\Sigma(E_{r+w}))}{(2w+1)\underline g}\sum_{c=1}^{\Gamma} \Big| g_{w}\big(\frac{r-c}{w}\big) - g_{w}\big(\frac{r-1-c}{w}\big)\Big|\nonumber \\ 
&\le \frac{{\rm mmse}(\Sigma(E_{r+w}))}{(2w+1)\underline g} \Big(\, 2w \frac{g_*}{w} + |g_{w}(1)| + |g_{w}(-1)|\, \Big)\nonumber\\
&< \frac{{\rm mmse}(\Sigma(E_{r+w}))}{2w\underline g} \Big(\, 2g_* + 2\bar g\, \Big)\nonumber\\
&\le \frac{g_* + \bar g}{w\underline g}  \exp({-c(B) \Sigma^{-2}(E_{r+w})}).
\end{align}
The last inequality is obtained by knowing that for an equivalent AWGN channel of variance $\Sigma^2$ and under {\it discrete prior}, ${\rm mmse}(\Sigma) \leq \exp({-c\Sigma^{-2}})$ where $c$ is some positive number that depends on the prior (see e.g. Appendix D of \cite{BMDK_2017} for an explicit proof). Here the prior is uniform over sections so this number depends only on $B$. 
\end{IEEEproof}
\begin{lemma}
\label{lemma:quadFormBounded}
Let ${\tbf E}$ be a fixed point profile of the modified system initialized with $\tbf{E}^{(0)}=\tbf{1}$. Then the coupled potential verifies
\begin{align*}  
\frac{1}{2}\Big|\sum_{r,r'=1}^{\Gamma} \Delta {E}_{r} \Delta {E}_{r'}\left[\frac{\partial^2 F_{\text{co}}}{\partial  E_{r}\partial  E_{r'}}\right]_{\hat{\tbf E}}\Big| < \frac{K(B, \bar g, \underline g, g_*)}{(E_{\rm f} +\epsilon)^{2\beta} R w}.
\end{align*}
where $K(B, \bar g, \underline g, g_*)>0$. In particular, the RHS is $\mathcal{O}(w^{-1})$.
\end{lemma}

\begin{IEEEproof}
First remark that a fixed point of the modified system satisfies ${\tbf E} \succeq {\tbf E}_{\rm f}$. For ${\tbf E} = {\tbf E}_{\rm f}$ the result is immediate since $\Delta {E}_{r}=0$. It remains 
to prove this lemma for ${\tbf E}$ a fixed point of the modified system such that ${\tbf E} \succ {\tbf E}_{\rm f}$. In Appendix \ref{sec:App_Bound} we prove that  
\begin{align}
\Big[\frac{\partial^2 F_{\text{co}}}{\partial  E_{r}\partial  E_{r'}}\Big]_{\hat{\tbf E}}
\leq
\delta_{r, r'} \frac{K_1(B, \bar g, \underline g)}{(E_{\rm f} +\epsilon)R} + 1_{\vert r - r' \vert \leq 2w+1} 
\frac{K_2(B, \bar g, \underline g)}{(E_{\rm f} + \epsilon)^{2\beta}R(2w+1)}
\label{proved-in-appendix}
\end{align}
for some finite positive $K_1(B, \bar g, \underline g)$ and $K_2(B, \bar g, \underline g)$ independent of $w$ and $\Gamma$.
Since $\Delta E_r \geq 0$, using the triangle inequality we get
\begin{align*}
\frac{1}{2}\Big|\sum_{r,r'=1}^{\Gamma} \Delta {E}_{r} \Delta {E}_{r'}\left[\frac{\partial^2 F_{\text{co}}}{\partial  E_{r}\partial  E_{r'}}\right]_{\hat{\tbf E}}\Big|
& 
\leq 
\frac{K_1(B, \bar g, \underline g)}{2(E_{\rm f} +\epsilon)R}\sum_{r=1}^{\Gamma} \Delta {E}_{r}^2
+
\frac{K_2(B, \bar g, \underline g)}{2(E_{\rm f} + \epsilon)^{2\beta}R(2w+1)}
\sum_{r=1}^\Gamma \Delta {E}_{r} \sum_{r'= r - w}^{r+w} \Delta {E}_{r'}
\nonumber \\ &
\leq 
\frac{K_1(B, \bar g, \underline g)}{2(E_{\rm f} +\epsilon)R}\max_{r'} \Delta E_{r'} \sum_{r=r_*+1}^{r_{\rm max}} \Delta {E}_{r}
+
\frac{K_2(B, \bar g, \underline g)}{2(E_{\rm f} + \epsilon)^{2\beta}R}\max_{r'} \Delta E_{r'}
\sum_{r=r_*+1}^{r_{\rm max}}  \Delta {E}_{r}
\nonumber \\ &
\leq 
\frac{K_1^{\prime}(B, \bar g, \underline g, g_*)}{2(E_{\rm f} +\epsilon)R w}
+
\frac{K_2^{\prime}(B, \bar g, \underline g, g_*)}{2(E_{\rm f} + \epsilon)^{2\beta}R w}.
\end{align*}
To get the last inequality we used Lemma \ref{lemma:Evariesslowly} and 
$\sum_{r=r_*+1}^{r_{\rm max}}  \Delta {E}_{r} = E_{\rm max} - E_{r_*+1} < 1$. Finally, one can find $K(B, \bar g, \underline g, g_*) > 0$ such that the last estimate is smaller than 
$$
\frac{K(B, \bar g, \underline g, g_*)}{(E_{\rm f} +\epsilon)^{2\beta} R w}
$$. 
\end{IEEEproof}

The change in potential due to the shift can be also computed by direct evaluation as shown in the following lemmas.   
\begin{lemma}
Let ${\tbf E}$ be a fixed point profile of the modified system initialized with $\tbf{E}^{(0)}=\tbf{1}$. 
If ${\tbf E}\succ \bar{{\tbf E}}_{\rm f}$, then $E_{\rm max}$ cannot be in the basin of attraction to the MSE floor, i.e., $E_{\rm max} \notin \mathcal{V}_0$.
\label{lemma:outside_basin}
\end{lemma}
\begin{IEEEproof}
Knowing that ${\tbf E} \succ \bar{{\tbf E}}_{\rm f}$ and also that ${\tbf E}$ is non-decreasing implies $\bar E_{\rm f} < E_{\rm max}$. Moreover, we have that
\be \label{eq:outside_basin}
E_{\rm max} = [T_\text{co}({\tbf E})]_{r_{\rm max}} =  \sum_{c=1}^{\Gamma} J_{r_{\rm max},c} \, {\rm mmse}(\Sigma_c({\tbf E})) \le  \sum_{c=1}^{\Gamma} J_{r_{\rm max},c} \, {\rm mmse}(\Sigma(E_{\rm max})) \le T_\text{un}(E_{\rm max}),
\ee
where the first inequality follows from the fact that $\Sigma_c({\tbf E}) \le \Sigma(E_{\rm max})$ due to the variance symmetry \eqref{eq:varianceSymmetry} at $r_{\rm max}$ and the fact that $\tbf E$
is non-decreasing. The second inequality follows from the variance normalization \eqref{eq:varianceNormalization}. Applying the monotonicity of $T_\text{un}$ on \eqref{eq:outside_basin} yields
\begin{align}
E_{\rm f} < \bar E_{\rm f} < E_{\rm max} \le T_{\text{un}}^{(\infty)}( E_{\rm max}),
\end{align}
which implies that $ E_{\rm max} \notin \mathcal {V}_0$.
\end{IEEEproof}
\begin{lemma}
Let $0< \eta <1$ fixed and $\bar E_{\rm f} = E_{\rm f}+\epsilon$ with any $0<\epsilon<\delta(\eta, B, R)$ where $\delta(\eta, B, R)$ has been constructed in Definition \ref{def:pseudo}. 
Let ${\tbf E} \succ \bar{{\tbf E}}_{\rm f}$ be a fixed point profile of the modified system initialized with $\tbf{E}^{(0)}=\tbf{1}$. Then ${\tbf E}$ satisfies
\begin{align*}
F_{\text{co}}(\text{S}({\tbf E})) - F_{\text{co}}({\tbf E}) \leq - (1-\eta)\Delta F_{\text{un}},
\end{align*}
where $\Delta F_{\text{un}}$ is the free energy gap of the underlying system given in Definition \ref{def:freeEnergyGap}.
\label{lemma:diffShited_directEval}
\end{lemma}
\begin{IEEEproof}
The contribution of the change in the ``energy'' term is a perfect telescoping sum:
\begin{align}  
U_{\text{co}}(\text{S}({\tbf E})) - U_{\text{co}}({\tbf E}) = U_{\text{un}}(\bar E_{\rm f}) - U_{\text{un}}(E_{\rm max}). \label{eq:DeltaU}
\end{align}
We now deal with the contribution of the change in the ``entropy'' term. Using the properties of the construction of $J_{r, c}$ we notice that
for all $c \in \{2w + 1,\dots,\Gamma-2w-1\}$
\be
\Sigma_{c+1}^{-2}({\text{S}(\tbf E})) =  \sum_{r=c+1-w}^{c+1+w} \frac{J_{r,c+1}}{\Sigma^{2}(E_{r-1})} =  \sum_{r=c-w}^{c+w} \frac{J_{r+1,c+1}}{\Sigma^{2}(E_{r})} =  \sum_{r=c-w}^{c+w} \frac{J_{r,c}}{\Sigma^{2}(E_{r})} =  \Sigma_{c}^{-2}({\tbf E})
\ee
which yields
\begin{align}  
&S_{\text{co}}({\tbf E})-S_{\text{co}}(\text{S}({\tbf E}))= S_{\text{un}}( \Sigma_{\Gamma-2w}({\tbf E}))-S_{\text{un}}( \Sigma_{2w+1}(\text{S}{(\tbf E)}))  - \sum_{c\in \mathcal{S}} [S_{\text{un}}( \Sigma_{c}(\text{S}{(\tbf E)})) - S_{\text{un}}( \Sigma_{c}({\tbf E}))], \label{eq:diffSB}
\end{align}
where $\mathcal{S}\defeq\{1,\dots,2w\}\cup\{\Gamma-2w+1,\dots,\Gamma\}$. By the saturation of the modified system, $\tbf E$ possesses the following property
\begin{align} 
&[\text{S}( {\tbf E})]_r = [ {\tbf E}]_r \quad {\rm for~all} \quad  r \in \{1,\dots,r_*\}\cup\{r_{\rm max}+1,\dots,\Gamma\}.
\end{align}
Hence, $\Sigma_c(\text{S}( {\tbf E})) =  \Sigma_c( {\tbf E})$ for all $c \in \mathcal{S}$ and thus the sum in (\ref{eq:diffSB}) cancels. Furthermore, one can show, using the saturation of $\tbf E$ and the variance symmetry \eqref{eq:varianceSymmetry}, that $\Sigma_{2w+1}(\text{S}({\tbf E})) = \Sigma( \bar E_{\rm f})$. The same arguments and the fact that $r_{\rm max} \le \Gamma-3w$ for $\tbf E\succ \bar{\tbf{E}}_{\rm f}$ lead to $\Sigma_{\Gamma-2w}({\tbf E}) = \Sigma( E_{\rm max})$. Hence, (\ref{eq:diffSB}) yields
\begin{align}  
S_{\text{co}}({\tbf E}) - S_{\text{co}}(\text{S}({\tbf E})) = S_{\text{un}}(\Sigma( E_{\rm max})) - S_{\text{un}}(\Sigma(\bar E_{\rm f})). \label{eq:DeltaS}
\end{align}
Combining (\ref{eq:DeltaU}) with (\ref{eq:DeltaS}) gives
\begin{align*}  
F_{\text{co}}(\text{S}({\tbf E})) - F_{\text{co}}({\tbf E}) & = - (F_{\text{un}}( E_{\rm max}) -  F_{\text{un}}( \bar E_{\rm f}))
\nonumber \\ &
= 
- (F_{\text{un}}( E_{\rm max}) -  F_{\text{un}}(E_{\rm f}))
+ (F_{\text{un}}({\bar E_{\rm f}}) - F_{\text{un}}(E_{\rm f})).
\end{align*}
Using the definition of the free energy gap (Definition \ref{def:freeEnergyGap}), the fact that $E_{\rm max} \notin \mathcal{V}_0$ (Lemma~\ref{lemma:outside_basin}), and    $F_{\text{un}}({\bar E_{\rm f}}) - F_{\text{un}}(E_{\rm f})\leq \eta \Delta F_{\text{un}}$ we find
\begin{align*}  
F_{\text{co}}(\text{S}({\tbf E})) - F_{\text{co}}({\tbf E}) \leq - (1-\eta)\Delta F_{\text{un}}.
\end{align*}
\end{IEEEproof}

Using Lemmas \ref{lemma:Fdiff_quadraticForm}, \ref{lemma:quadFormBounded}, \ref{lemma:diffShited_directEval} 
we now prove threshold saturation. 

\begin{thm}\label{th:mainTheorem}
Let $0< \eta <1$ fixed and $\bar E_{\rm f} = E_{\rm f}+\epsilon$ with any $0<\epsilon<\delta(\eta, B, R)$ where $\delta(\eta, B, R)$ has been constructed in Definition \ref{def:pseudo}. Fix
\begin{align}\label{condition-on-w}
R < R_{\rm pot}\quad {\rm and} \quad w > \frac{K(B, \bar g, \underline g, g_*)}{(E_{\rm f} + \epsilon)^{2\beta} R (1-\eta)\Delta F_{\text{un}}}
\end{align}
Then the fixed point profile ${\tbf{E}^{(\infty)}}$ of the coupled SE must satisfy ${\tbf{E}}^{(\infty)} \preceq \bar{{\tbf{E}}}_{\rm f}$.
%, and the MSE profile $\tilde {\tbf E}^* \prec {\tbf{E}}_{\rm f}$ as well.
\end{thm}

\begin{IEEEproof}
Assume that, under these hypotheses, the fixed point profile of the modified system initialized with  ${\tbf E}^{(0)}={\tbf 1}$ is such that $\tbf E\succ \bar{{\tbf{E}}}_{\rm f}$. On one 
hand by Lemma~\ref{lemma:diffShited_directEval} 
we have for $R < R_{\rm pot}$ a positive $\Delta F_{\text{un}}$ and 
$$
|F_{\text{co}}({\tbf E}) - F_{\text{co}}(\text{S}({\tbf E}))| \ge (1-\eta)\Delta F_{\text{un}}.
$$
On the other hand by Lemmas~\ref{lemma:Fdiff_quadraticForm} and \ref{lemma:quadFormBounded}  
$$
|F_{\text{co}}({\tbf E}) - F_{\text{co}}(\text{S}({\tbf E}))| \leq \frac{K(B, \bar g, \underline g, g_*)}{(E_{\rm f} + \epsilon)^{2\beta} R w}.
$$
Thus we get 
$$
w \leq \frac{K(B, \bar g, \underline g, g_*)}{(E_{\rm f} + \epsilon)^{2\beta} R (1-\eta)\Delta F_{\text{un}}}
$$
which is a contradiction.
Hence, ${\tbf{E}} \preceq \bar{{\tbf{E}}}_{\rm f}$. Since $\tbf E \succeq {\tbf E}^{(\infty)}$ we have ${\tbf E}^{(\infty)} \preceq \bar{{\tbf{E}}}_{\rm f}$.
\end{IEEEproof}

The most important consequence of this theorem is a statement on the GAMP threshold,

\begin{corollary}\label{cor:maincorollary}
By first 
taking $\Gamma \to \infty$ and then $w\to\infty$, the GAMP threshold of the coupled 
ensemble  satisfies $R_{\text{co}}\geq R_{\rm pot}$.
\end{corollary}
This result follows from Theorem~\ref{th:mainTheorem} and Definition~\ref{def:AMPcoupled}. Once the limit $w\to +\infty$ is taken 
we can send $\epsilon\to 0$ and the pseudo error floor tends to the true error floor $\bar E_{\rm f} \to E_{\rm f}$. 

\subsection{Discussion}
Corollary \ref{cor:maincorollary} says that the GAMP threshold for the coupled codes saturates the potential threshold in the limit $w\to +\infty$. It is in fact not possible to have the strict inequality
$R_{\text{co}} > R_{\rm pot}$, so in fact equality holds, but the proof would require a separate argument that we omit here because it is not so informative. 
Besides, this equality is not needed in order to argue that sparse superposition codes universally achieve capacity under GAMP decoding when $B\to +\infty$.
Indeed 
we have necessarily $R_{\text{co}} < C$ and we show in Section~\ref{sec:larg_B}, using non-rigorous asymptotic computations, that $\lim_{B\to \infty}R_{\rm pot}=C$. Thus $\lim_{B\to \infty} R_{\text{co}} = C$.

We emphasize that Theorem~\ref{th:mainTheorem} and 
Corollary~\ref{cor:maincorollary} hold for a large class of estimation 
problems with random linear mixing \cite{rangan2011generalized}. Both the SE and potential formulations 
of Section~\ref{sec:stateandpot} as well as the proof given in the present section are not restricted to SS codes. Indeed all the definitions and 
results are obtained for any memoryless channel $P_{\text{out}}$ and can be generalized for any 
factorizable (over $B$-dimensional sections) prior of the message (or signal) $\bs$.

Theorem \ref{th:mainTheorem} states that for $w$ large enough the state evolution iterations will drive the MSE profile below some pseudo error floor $\bar E_{\rm f}=E_{\rm f}+\epsilon$. This is then enough
information to deduce that the threshold saturation phenomenon happens in the limit where $w\to +\infty$ (and note we do not expect full threshold saturation, i.e., 
$R_{\rm co}\to R_{\rm pot}$ for finite $w$). 
However, it is worth pointing out that the condition \eqref{condition-on-w} in Theorem \ref{th:mainTheorem} on the size of the coupling window is most probably {\it not} optimal. We conjecture that a better bound 
should hold where $w > C/ \Delta F_{\rm un}$ for some $C>0$ which does {\it not} diverge when $E_{\rm f} +\epsilon \to 0$.
The appearance of the error floor
in the denominator can be traced back to inequality \eqref{proved-in-appendix} whose derivation is detailed in Appendix \ref{sec:App_Bound}. One possible way to cancel this divergence would be 
to obtain a better bound on $\Delta E_r$ than the one given by \eqref{bad-bound}. More precisely if $E_{r+w}$ can be replaced by $E_r$ then the proof of 
Lemma \ref{lemma:quadFormBounded} and Theorem \ref{th:mainTheorem} would give a more resonable 
lower bound for $w$. Carrying out this program presents technical difficulties in the analysis of coupled state evolution which we have not overcome in this work. 
The present difficulties do not appear in the analysis of spatially coupled LDPC codes \cite{PfisterMacrisBMS}.

\section{Large alphabet size analysis and connection with Shannon's capacity}
\label{sec:larg_B}
\begin{figure}[!t]
\centering
\includegraphics[draft=false,width=0.38\textwidth, height=162pt, trim={0pt 2 3 0},clip]{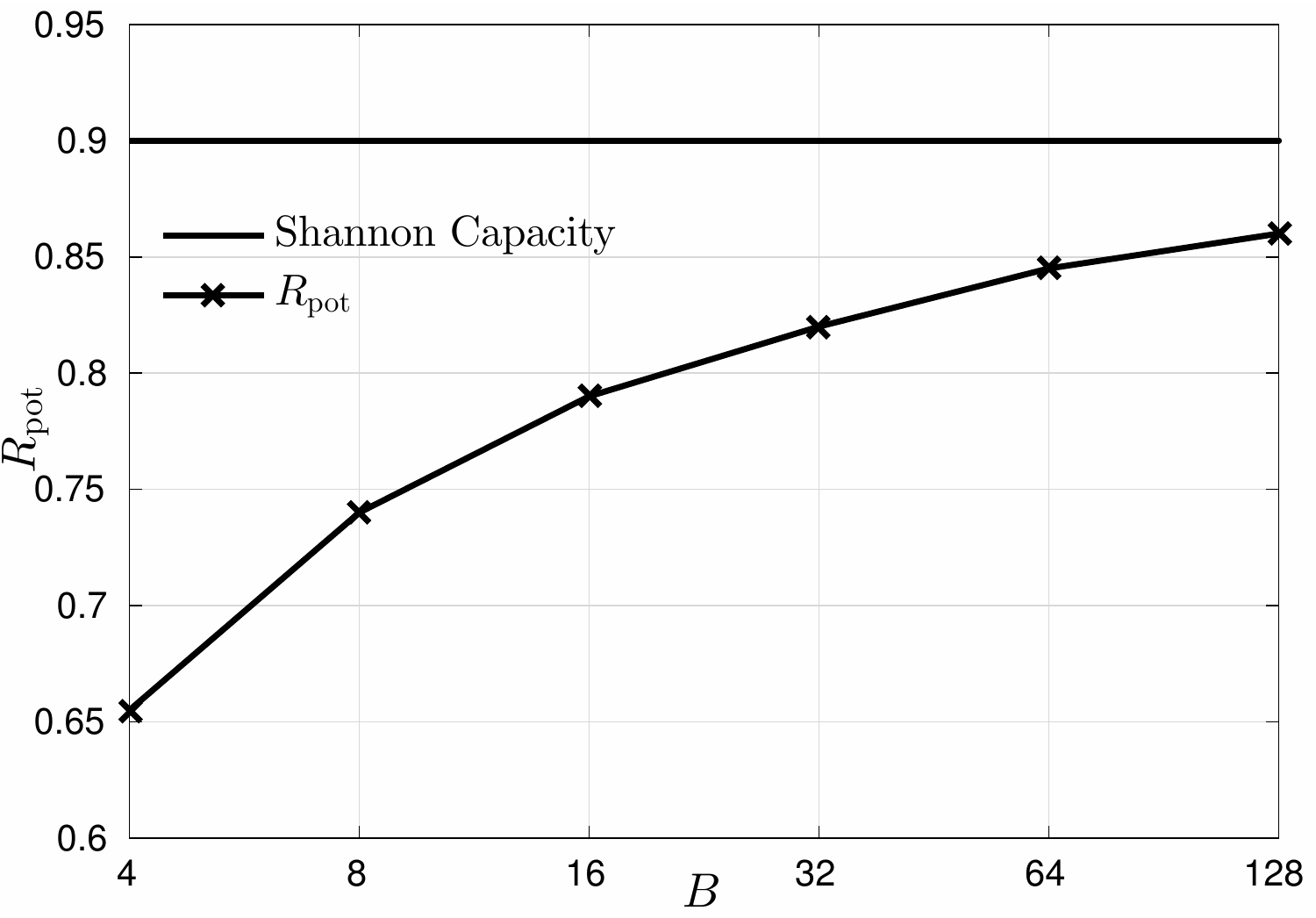}
\caption{The potential threshold v.s the alphabet size $B$ for the BEC with erasure probability $\epsilon = 0.1$.}
\label{fig:potential_BEC}
\end{figure}

We now show, through non-rigorous analytical computations, that as the alphabet size $B$ increases, the potential threshold of SS codes approaches Shannon's 
capacity $R_{\rm pot}^{\infty}\defeq\lim_{B\to \infty} R_{\rm pot} = C$ (Fig. \ref{fig:potential_BEC}), and also that $\lim_{B\to \infty}E_{\rm f} = 0$. These 
are ``static'' or ``information theoretic'' properties of the code independent of the decoding algorithm. Nevertheless this result has
an algorithmic consequence. The threshold 
saturation established in Corollary~\ref{cor:maincorollary} for spatially coupled SS codes suggests that optimal decoding can actually be performed using the GAMP decoder, i.e. $\lim_{B\to \infty} R_{\text{co}} = C$, because $R_{\rm pot} \le R_{\text{co}} \leq C$.

The potential of the underlying system contains all the information about $R_{\rm pot}$ and $R_{\text{un}}$. Hence, we proceed by computing the potential in the large $B$ regime,
\begin{align} \label{eq:largeBpot_1}
\varphi_{\text{un}}(E) \defeq \lim_{B\to\infty} F_{\text{un}}(E). 
\end{align}
The limit (\ref{eq:largeBpot_1}) 
was heuristically
computed in \cite{BarbierK15,phdBarbier} for the AWGN channel. Extending this computation to the present setting, one obtains
\begin{align} \label{eq:largeBpot}
\varphi_{\text{un}}(E) = U_{\text{un}}(E) - {\rm max}\Big(0,1 - \frac{1}{2\ln(2)\Sigma^{2}(E)}\Big). 
\end{align}
The extension from the AWGN case is straightforward, the $U_{\text{un}}(E)$ term in $F_{\text{un}}(E)$ is independent of $B$ while the $S_{\text{un}}(\Sigma(E))$ term remains the same. The difference is only in the computation of the effective noise $\Sigma(E)$, which is independent of $B$. We note that \eqref{eq:largeBpot} is not a trivial asymptotic calculation because the ``entropy'' term $S_{\text{un}}(\Sigma(E))$ involves a $B$-dimensional integral (see Definition \ref{def:pot_underlying}). Since $B \rightarrow \infty$, this amounts to compute a ``partition function'' (or equivalently solve a non-linear estimation problem where the signal has one non-zero component). We have not 
attempted to make this asymptotic computation rigorous but we expect that
%at least for the AWGN channel the results of \cite{BarbierK15,phdBarbier} could be made rigorous using the recent work \cite{BDMK_alerton2016,BMDK_2017,ReevesPfister_isit16}.
such computation
could be made rigorous using the recent work \cite{BDMK_alerton2016,BMDK_2017,ReevesPfister_isit16,ReevesPfister_trans,pmlr-v75-barbier18a}.% Note that the results of \cite{ReevesPfister_isit16,BMDK_trans2017,BDMK_alerton2016,pmlr-v75-barbier18a} ensure that the potential threshold $R_{\rm pot}$ is equal to the Bayes optimal threshold.

The analysis of (\ref{eq:largeBpot}) for $E\in[0,1]$ leads to the following 

\begin{claim}\label{claim:potential}
For a fixed rate $R$ and $E\in[0,1]$, the only possible local minima of $\varphi_{\rm {un}}(E)$ are at $E=0$ and $E=1$. Furthermore, for $E^{\prime} \in \big\{ E\in[0,1] \mid 2\ln(2)\Sigma^{2}(E) <1 \big\}$ the minimum is at $E^{\prime} =0$ and for $E^{\prime} \in \big\{ E\in[0,1] \mid 2\ln(2)\Sigma^{2}(E) > 1 \big\}$ the minimum is at $E^{\prime} =1$. 
\end{claim}

Note that this result was rigorously proven for the AWGN channel in \cite{BarbierK15} and then verified for several memoryless channels in \cite{barbierDiaMacris_itw2016}. A fully rigorous analysis of the function 
$\varphi_{\text{un}}(E)$ would be lengthy; we thus only claim the result here, which is confirmed by numerical analysis.

\begin{figure}[!t]
\centering
\includegraphics[draft=false,width=0.38\textwidth, height=165pt, trim={0pt 3 3 0},clip]{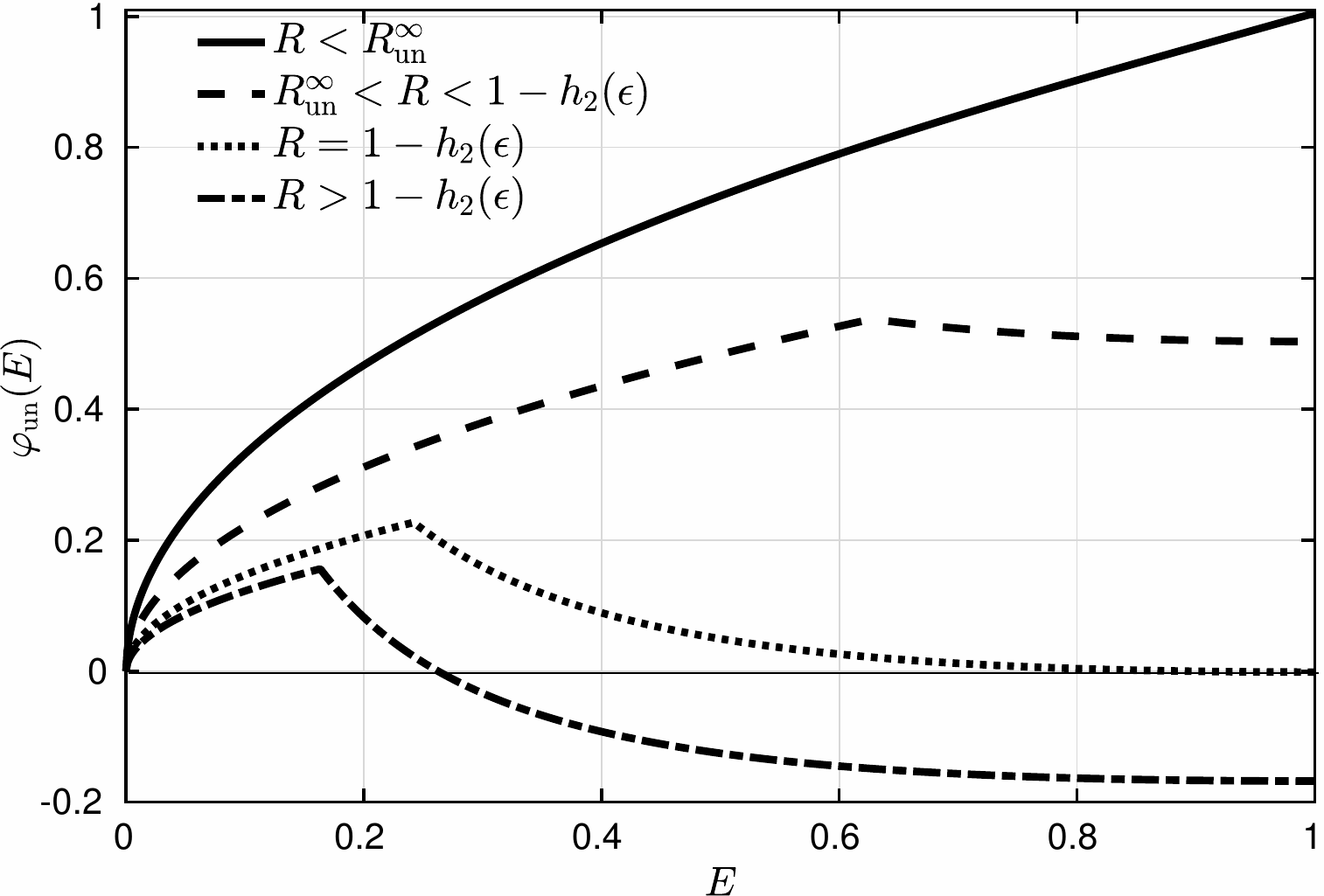}
\centering
\includegraphics[draft=false,width=0.38\textwidth, height=163pt, trim={0pt 3 3 0},clip]{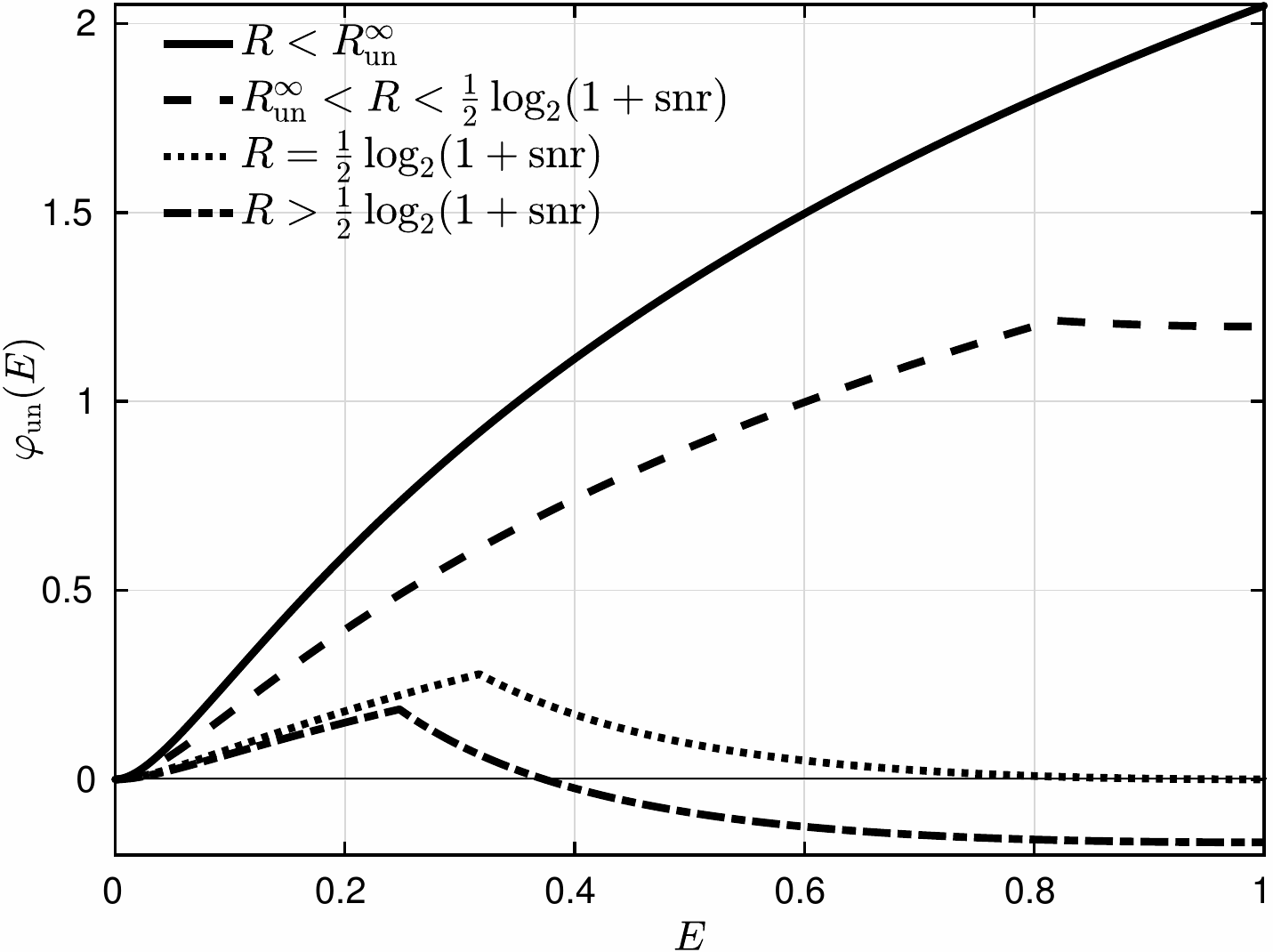}
\caption{The large alphabet potential $\varphi_{\text{un}}(E)$ (\ref{eq:largeBpot}) as a function of the error parameter $E$ for the BSC (left) and AWGN (right) channels with $\epsilon=0.1$ and ${\rm snr}=10$ respectively. $\varphi_{\text{un}}(E)$ is scaled such that $\varphi_{\text{un}}(0)=0$. For $R$ below the ``asymptotic'' GAMP threshold $R_{\text{un}}^{\infty}$, there is a unique minimum at $E=0$ while just above $R_{\text{un}}^{\infty}$, this minimum coexists with a local one at $E=1$. At the optimal threshold of the code, that coincides with the Shannon capacity, the two minima are equal. Then, for $R>C$ the minimum at $E=1$ becomes the global one, and thus decoding is impossible.}
\label{fig:largeB}
\end{figure}

The existence of a minimum at $E=0$ means that the error floor $E_{\rm f}$, if it exists, vanishes as $B$ increases (Fig. \ref{fig:largeB}). Moreover, if $\Sigma^{2}(E) < (2\ln(2))^{-1}\ \forall\ E\in[0,1]$, 
which corresponds to the region $R<(2\ln(2))^{-1}\mathbb{E}_{p|1} [\mathcal{F}(p|1)]$, then $\varphi_{\text{un}}(E)$ has a unique minimum at $E = 0$. 
Similarly if $\Sigma^2(E) > (2\ln 2)^{-1}\ \forall\ E\in[0,1]$, corresponding to $R>(2\ln(2))^{-1}\mathbb{E}_{p|0} [\mathcal{F}(p|0)]$, then 
$\varphi_{\text{un}}(E)$ has a unique minimum at $E=1$. For \emph{intermediate rates} both minima exist. 

Therefore, we identify the algorithmic GAMP threshold, when $B\to +\infty$, as the smallest rate such that a second minimum appears,
\begin{align}
R_{\text{un}}^{\infty}\defeq \lim_{B\to \infty} R_{\text{un}} = \frac{\mathbb{E}_{p|1} [\mathcal{F}(p|1)]}{2\ln(2)} = \frac{\mathcal{F}(0|1)}{2\ln(2)}.
\label{Ru}
\end{align}
Recall $R_{\rm pot}$ is defined by the point where $\Delta F_{\text{un}}$ switches sign (Definition \ref{def:potThresh}). Thus $R_{\rm pot}^{\infty}$ can be obtained by equating the two minima of $\varphi_{\text{un}}(E)$. The potential (\ref{eq:largeBpot}) takes the following values at the two minimizers 
\begin{align*}
\varphi_{\text{un}}(0) &= - \frac{1}{R} \mathbb{E}_z\bigg[\int dy\, \phi(y|z,0) \log_2\big(\phi(y|z,0)\big)\bigg], \\
\varphi_{\text{un}}(1) &= - \frac{1}{R} \mathbb{E}_z\bigg[\int dy\, \phi(y|z,1) \log_2\big(\phi(y|z,1)\big)\bigg] -1,
\end{align*}
where $\phi(y|z,E)$ is given in Definition \ref{def:pot_underlying}.
Then, setting $\varphi_{\text{un}}(1) = \varphi_{\text{un}}(0)$ yields
\begin{align}\label{eq:large_B}
R_{\rm pot}^{\infty}= &- \int \int dz dy \mathcal{N}(z\vert 0, 1) P_{\text{out}}(y|z) \log_2\Big( \int d\tilde{z} \mathcal{N}(\tilde{z}\vert 0, 1) P_{\text{out}}(y|\tilde z)\Big) \nonumber \\
&+ \int \int dz dy \mathcal{N}(z\vert 0, 1) P_{\text{out}}(y|z) \log_2\Big(P_{\text{out}}(y|z)\Big).
\end{align}
We will now recognize that this expression is the Shannon capacity of $W$ for a proper choice of the map $\pi$.

Let $\mathcal{A}$ and $\mathcal{B}$ be the input and output alphabet of $W$ respectively, where $\mathcal{A}, \mathcal{B} \subseteq \mathbb{R}$ have discrete or continuous supports. Call $\mathcal{P}$ the capacity-achieving input distribution associated with $W$.
Choose $\pi:\mathbb{R}\to\mathcal{A}$ such that $i)$ $P_{\text{out}}(y|z) = W(y|\pi(z))$ and $ii)$ if $Z\sim\mathcal{N}(0,1)$, then $\pi(Z) \sim \mathcal{P}$. This map converts a standard Gaussian random variable $Z$ onto a channel-input random variable $\pi(Z)=A$ with capacity-achieving distribution $\mathcal{P}(a)$. Recall that $\pi$ can be viewed equivalently as part of the code or of the channel. 

Now using the relation
\begin{align*}
\int dz \mathcal{N}(z\vert 0, 1) P_{\text{out}}(y|z) = \int dz \mathcal{N}(z\vert 0, 1) W(y|\pi(z)) =  \int da \mathcal{P}(a) W(y|a),
\end{align*}
(\ref{eq:large_B}) can be expressed equivalently as
\begin{align}\label{eq:large_B_symm}
R_{\rm pot}^{\infty}= &- \int \int dy da \mathcal{P}(a) W(y|a) \log_2\Big( \int d\tilde a \mathcal{P}(\tilde a) W(y|\tilde a)\Big) \nonumber \\
&+ \int \int dy da \mathcal{P}(a) W(y|a) \log_2\Big(W(y|a)\Big).
\end{align}
The first term in (\ref{eq:large_B_symm}) is nothing but the Shannon entropy $H(Y)$ of the channel output-distribution. The second term eaquals minus
 the conditional entropy $H(Y|A)$ of the channel-output distribution given the input $A=\pi(Z)$ with capacity-achieving distribution. Thus, $R_{\rm pot}^{\infty}$ is the Shannon capacity of $W$.
Combining this result with Corollary~\ref{cor:maincorollary}, we can argue that spatially coupled SS codes allow to communicate reliably up to Shannon's capacity over any memoryless channel under low complexity GAMP decoding.
%Combining this result with Theorem~\ref{th:mainTheorem}, we can assert that spatially coupled SS codes are not only capacity-achieving over any memoryless channels but can also be decoded at low computational cost using GAMP.

\begin{figure}[!t]
\centering
\includegraphics[draft=false,width=0.38\textwidth, height=160pt, trim={0pt -1.3 3 0},clip]{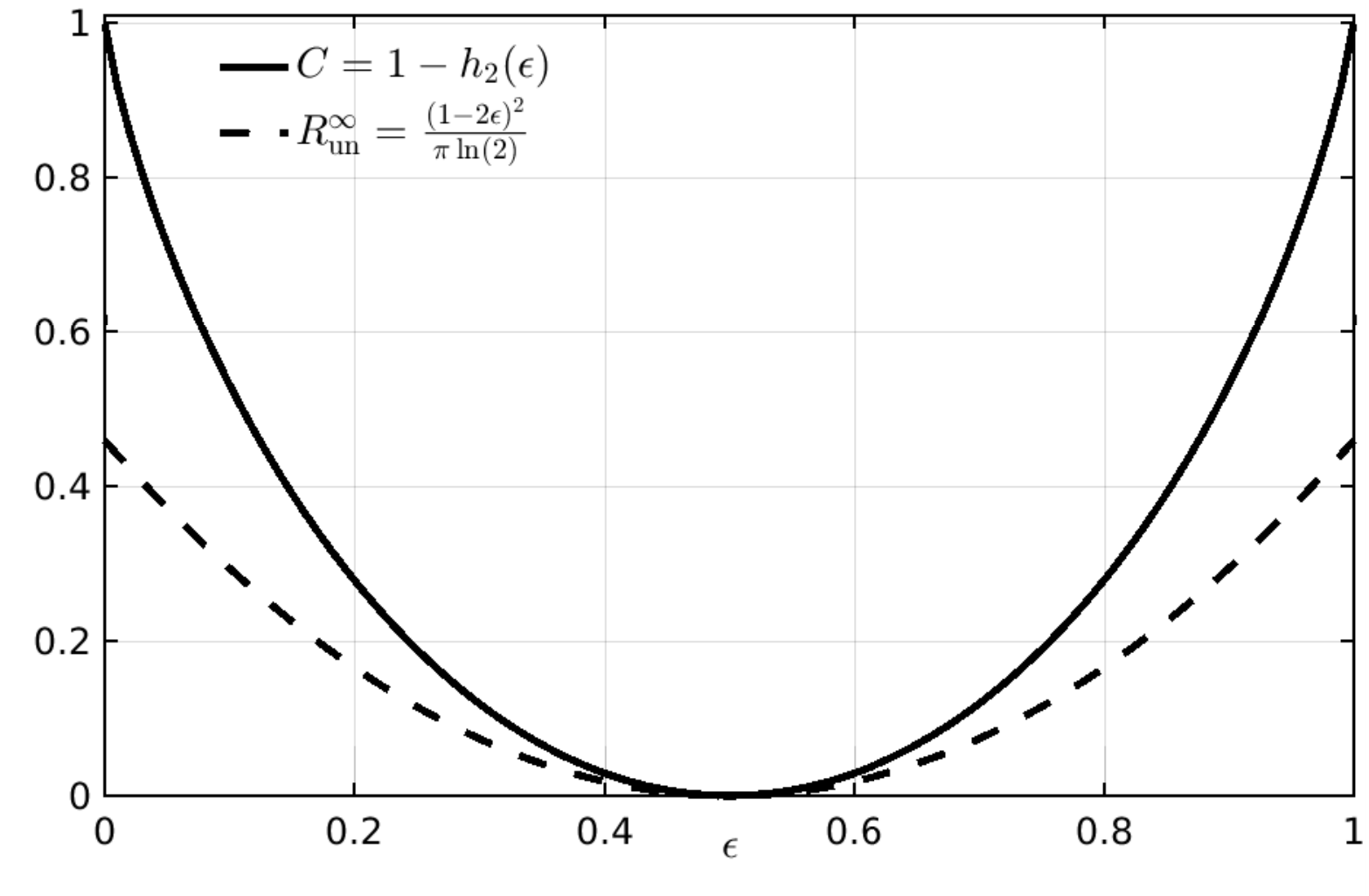}
\includegraphics[draft=false,width=0.38\textwidth, height=160pt, trim={0pt 0 1 0},clip]{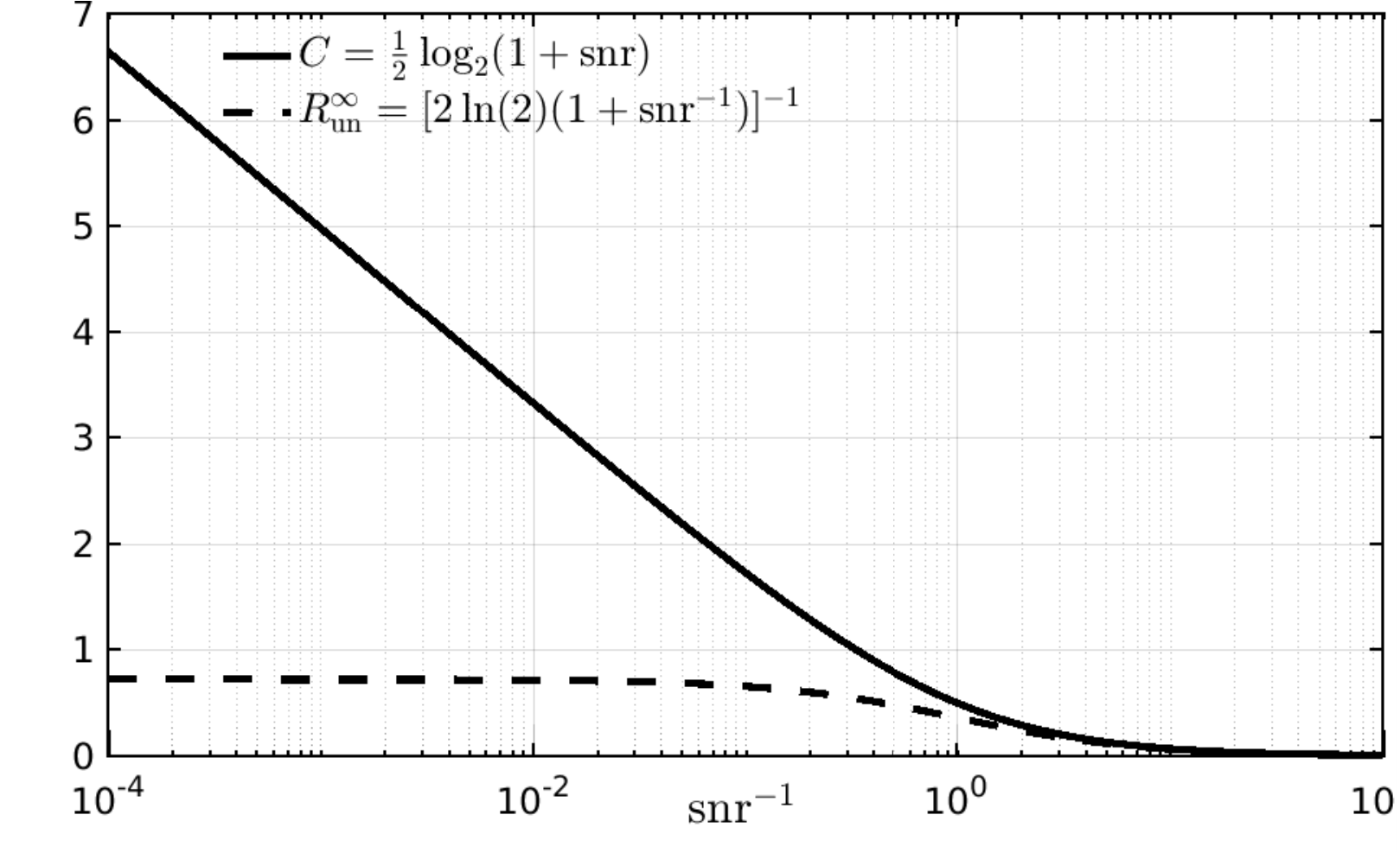}
\caption{The capacities and GAMP thresholds in the infinite alphabet limits for the BSC (left) and AWGN (right) channels.}
\label{fig:capacities}
\end{figure}

An essential question remains on how to find the proper map $\pi$ for a given memoryless channel. In the case of discrete input memoryless symmetric channels, 
Shannon's capacity can be attained by inducing a uniform input distribution $\mathcal{P} = \mathcal{U}_\mathcal{A}$. Let us call $q$ the 
cardinality of $\mathcal{A}=\{a_1,\dots,a_q\}$. In this case the mapping $\pi$ is simply $\pi(z) = a_i$ if $z\in \, ]z_{(i-1)/q}, z_{i/q}]$, 
where $z_{i/q}$ is the $i^{th}$ $q$-quantile\footnote{With $z_{i/q} = Q^{-1}(1- {i}/{q})$, where $Q^{-1}(\cdot)$ is the inverse of the Gaussian $Q$-function defined by $Q(x)=\int_{x}^{+\infty} dt\, \frac{e^{-\frac{t^2}{2}}}{\sqrt{2\pi}}$.} of the Gaussian distribution, with $z_{0} = -\infty, z_{1}=\infty$. For asymmetric 
channels, one can use some standard methods such as Gallager's mapping or more advanced ones \cite{MondelliUrbankeHassani_assymetricChannels} that introduce bias in the channel-input distribution in order to match the capacity-achieving one.
%
%As the SS codes are using a uniform prior over the messages, or equivalently over the codewords, it implies that $R_{\rm pot}^{\infty} = {\rm max}_{P(z) \in \mathcal{S}}\, I(Y,Z)$ is the \emph{symmetric capacity} of the channel, that is the maximum of the input-output mutual information over the set $\mathcal{S}$ of symmetric input probability densities. The maximum has already been taken in (\ref{eq:large_B}), and is obtained for $z\sim \mathcal{N}(z|0,1)$ due to the dense random coding matrix as explained. 
%
We now illustrate these findings for various channels as depicted in Fig.~\ref{fig:capacities} and Fig.~\ref{fig:capacity_Z}.
\subsection{AWGN channel} \label{sec:AWGN}
We start showing that our results for the AWGN channel \cite{barbierDiaMacris_isit2016} are a special case of the present general framework. No map $\pi$ is required and the Shannon 
capacity is directly obtained from (\ref{eq:large_B}) because the capacity-achieving input distribution for the AWGN channel is Gaussian. Thus, by replacing
$P_{\text{out}}(y|z)=\mathcal{N}(y|z,1/{\rm snr})$ in (\ref{eq:large_B}), one recovers the Shannon capacity $R_{\rm pot}^{\infty} = \frac{1}{2}\log_2(1+{\rm snr})$. 
Furthermore, from \eqref{Ru} one obtains the following algorithmic threshold as $B \rightarrow \infty$
\be\label{eq:Ru_AWGN}
R_{\text{un}}^{\infty}=\frac{1}{2\ln(2)(1+\rm {snr}^{-1})}.
\ee
%
%This is the case whatever the factorizable (over the sections) prior $p_0(\bs)$ (as long as it does not give all weight to the all zeros vector) and $L\gg B$. Recall this limit is assumed when deriving all the present analysis.
%
\subsection{Binary symmetric channel} \label{sec:BSC}
The BSC with flip probability\footnote{With a slight abuse of notation, we use $\epsilon$ here as a channel parameter. Not to confuse with $\epsilon$ of Section \ref{sec:proofsketch} (Definition \ref{def:pseudo}).} $\epsilon$ has transition probability $W(y|a) = (1-\epsilon) \delta(y-a) + \epsilon \delta(y+a)$, where  $\mathcal{A} = \mathcal{B} = \{-1,1\}$. The 
proper map is $\pi(z) = {\rm sign}(z)$. For $Z \sim \mathcal{N}(0,1)$, this map induces uniform input distribution $\mathcal{U}_{\mathcal{A}}=1/2$. So by replacing
$W$ and $\mathcal{U}_{\mathcal{A}}$ in (\ref{eq:large_B_symm}), or equivalently $P_{\text{out}}(y|z) = (1-\epsilon) \delta(y-\pi(z)) + \epsilon \delta(y+\pi(z))$ into (\ref{eq:large_B}), one obtains 
the Shannon capacity of the BSC channel $R_{\rm pot}^{\infty} = 1-h_2(\epsilon)$ where $h_2$ is 
the binary entropy function. Using \eqref{Ru} this map also gives the algorithmic threshold
\be\label{eq:Ru_BSC}
R_{\text{un}}^{\infty}=\frac{(1-2\epsilon)^{2}}{\pi\ln(2)}.
\ee
%
%%  %Figure for two columns conference  
%%\begin{figure}[!t]
%%\centering
%%\includegraphics[draft=false,width=0.24\textwidth, height=110pt, trim={0pt -1.3 3 0},clip]{./figures/BSC_capacity2.eps}
%%\includegraphics[draft=false,width=0.24\textwidth, height=110pt, trim={0pt 0 1 0},clip]{./figures/AWGN_capacity2.eps}
%%\vspace*{-16pt}
%%\caption{Large alphabet limits of the capacities and GAMP thresholds for the BSC (left) and AWGN (right) channels.}
%%\label{fig:capacities}
%%\end{figure}
%
\subsection{Binary erasure channel} \label{sec:BEC}

Note that the BEC is also symmetric. Therefore, the same mapping $\pi(z) = {\rm sign}(z)$ is used and leads to 
the Shannon capacity $R_{\rm pot}^{\infty} =1-\epsilon$, where $\epsilon$ is the erasure probability. Moreover, from \eqref{Ru} the algorithmic threshold for the BEC when $B \rightarrow \infty$ is
\be\label{eq:Ru_BEC}
R_{\text{un}}^{\infty}=\frac{1-\epsilon}{\pi\ln(2)}.
\ee
\subsection{Z channel} \label{sec:Z}
The Z channel is the ``most asymmetric'' discrete channel. It has binary input and output $\mathcal{A} = \mathcal{B} = \{-1,1\}$ with transition probability $W(y|a) = \delta(a-1)\delta(y-a) + \delta(a+1)[(1-\epsilon) \delta(y-a) + \epsilon \delta(y+a)]$, where $\epsilon$ is the flip probability of the $-1$ input. The map $\pi(z)={\rm sign}(z)$ leads to the \emph{symmetric capacity} of the Z channel
\be
R_{\rm pot}^{\infty} \big(\frac{1}{2}\big) = C\big(\frac{1}{2}\big) = h_2((1-\epsilon)/2) - h_2(\epsilon)/2,
\ee
where $C(\frac{1}{2})$ denotes the symmetric capacity, in other words the input-output mutual information when the input is 
uniformly distributed with $\mathcal{U}_{\mathcal{A}}=1/2$. Under the same map $\pi(z)$, one obtains the following algorithmic threshold in the limit $B\to +\infty$
\be\label{eq:Ru_Z_1}
R_{\text{un}}^{\infty}\big(\frac{1}{2}\big)=\frac{1-\epsilon}{\pi\ln(2)(1+\epsilon)}.
\ee
Note that the expression of $R_{\rm pot}^{\infty}(\frac{1}{2})$ differs from Shannon's capacity. However, one can introduce bias in the input distribution and hence match the capacity-achieving one. To do so, 
the proper map defined in terms of the $Q$-function\footnote{Here $Q(x)=\int_{x}^{+\infty} dt\, \frac{e^{-\frac{t^2}{2}}}{\sqrt{2\pi}}$.} is 
$\pi(z) = {\rm sign}(z - Q^{-1}(p_1))$, where $p_1$ is the induced input probability of the bit $1$. 

\begin{figure}[!t]
\centering
\includegraphics[draft=false,width=0.38\textwidth, height=160pt, trim={0pt 2 3 0},clip]{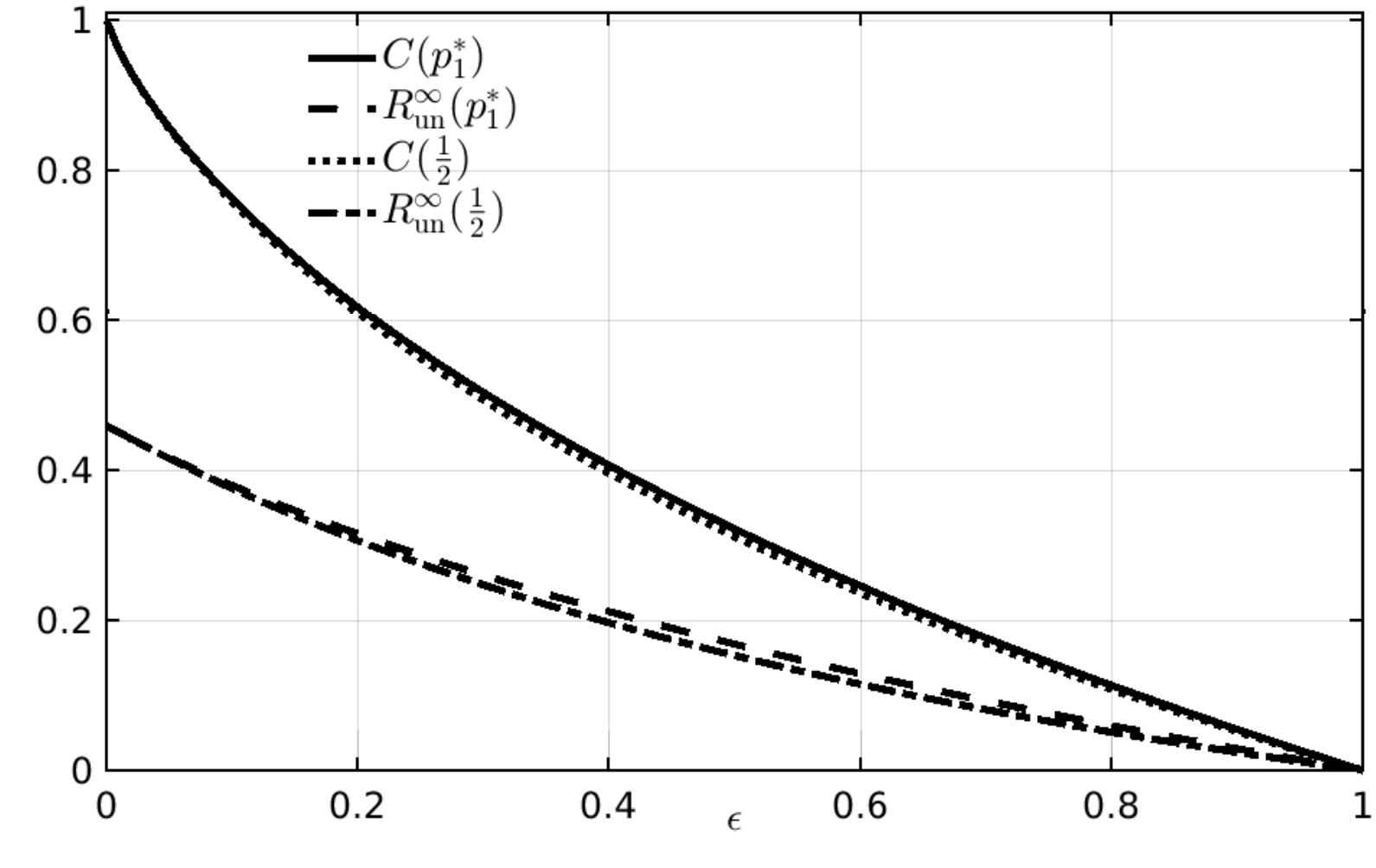}
\caption{Capacity and GAMP threshold of the Z channel in the infinite alphabet limits. $C(p_1^{*})$ and $R_{\text{un}}^{\infty}(p_1^{*})$ are the values under capacity-achieving input distribution, whereas $C(\frac{1}{2})$ and $R_{\text{un}}^{\infty}(\frac{1}{2})$ are the values under uniform distribution.}
\label{fig:capacity_Z}
\end{figure}

By optimizing over $p_1$, one can obtain Shannon's capacity of the Z channel
\be
R_{\rm pot}^{\infty} (p_1^{*}) = C(p_1^{*})= h_2((1-p_1^{*})(1-\epsilon)) - (1-p_1^{*})h_2(\epsilon),
\ee
with 
\be
p_1^{*} = 1 - [(1-\epsilon)(1+2^{h_2(\epsilon)/(1-\epsilon)})]^{-1}.
\ee
Using this optimal map, one obtains the following algorithmic threshold as depicted in Fig. \ref{fig:capacity_Z}
\be\label{eq:Ru_Z_2}
R_{\text{un}}^{\infty} (p_1^{*}) = \frac{(1-\epsilon)\big(e^{-[Q^{-1}(p_1^{*})]^2/2}\big)^{2}}{4 \pi \ln (2)(1-p_1^{*})((1-p_1^{*})\epsilon+p_1^{*})}.
\ee
%
%
%%  %Figure for two columns conference
%%\begin{figure}[!t]
%%\centering
%%\includegraphics[draft=false,width=0.28\textwidth, height=110pt, trim={0pt 2 3 0},clip]{./figures/Z_capacity2.eps}
%%\vspace*{-5pt}
%%\caption{Capacity and GAMP threshold of the Z channel in the infinite alphabet limits. $C(p_1^{*})$ and $R_{\rm un}^{\infty}(p_1^{*})$ are the values under capacity-achieving input distribution, whereas $C(\frac{1}{2})$ and $R_{\rm un}^{\infty}(\frac{1}{2})$ are the values under uniform distribution.}
%%\label{fig:capacity_Z}
%%\end{figure}
%

\section{Conclusion and open challenges} \label{sec:openChallenges}
In this work, we argue that spatially coupled SS codes universally achieve capacity over any memoryless channel under GAMP decoding. In particular, 
we prove that spatial coupling allows the algorithmic GAMP performance to saturate the potential threshold of the underlying code ensemble. Moreover, we 
show by analytical calculation that the potential threshold tends to capacity and the error floor vanishes in the proper limit. The approach taken in 
this work relies on the SE analysis and the application of the potential method.

We end up pointing out some open problems. In order to have a fully rigorous capacity achieving scheme over any memoryless channel, 
using spatially coupled SS codes and GAMP decoding, it must be shown that SE tracks the asymptotic performance of GAMP for the $B$-dimensional prior. 
We conjecture that this is indeed the case. The proof is beyond the scope of this work and would follow by extending the 
analysis of \cite{BayatiMontanari10,Montanari-Javanmard} to the SS codes setting as done in \cite{rush2015capacity} for AMP. Moreover, a rigorous proof of the asymptotic alphabet size analysis of Section \ref{sec:larg_B} is also needed.

It is also desirable to consider 
practical coding schemes using Hadamard-based operators or, more generally, row-orthogonal matrices. 
Another important point is to estimate at what rate the error floor vanishes as $B$ increases (when it exists e.g., in the AWGN channel). 
Finally, finite size effects should be considered in order to assess the practical performance of these codes, a direction which was recently pursued for power allocated codes in \cite{Greig18}, \cite{Rush_ISIT16} and \cite{Rush_ISIT17}.

\section*{Acknowledgment}
The authors would like to thank Florent Krzakala, Christophe Schülke and Rüdiger Urbanke for helpful discussions and Erdem Bıyık for numerical implementation of the GAMP algorithm.
The work of Jean Barbier and Mohamad Dia is supported by the Swiss national Foundation Grant no 200021-156672.

\appendices
\section{The output function in the GAMP Algorithm}\label{sec:App_GAMP}
The GAMP algorithm was introduced for general estimation with random linear mixing in \cite{rangan2011generalized}. The extension 
to the present context of SS codes with $B$-dimensional prior was given in Section \ref{sec:GAMP} of this paper. On a dense graphical model, an important notion of equivalent AWGN channel is used to simplify the BP messages. This notion is due to the linear mixing and it is independent of the physical channel. The physical channel $P_{\rm out}$ is reflected in the computation of the equivalent AWGN channel's parameter through the function $g_{\rm out}(\textbf{{p}}, \textbf{y}, \boldsymbol{\tau})$. 
This function is acting componentwise and can be 
interpreted as a \emph{score function} of the parameter ${p}_i$ associated with the distribution of $Y_i$. The general expression is
\begin{align}
 [g_{\rm out}(\textbf{{p}}, \textbf{y}, \boldsymbol{\tau})]_i & = (\mathbb{E}[{z}_i | {p}_i,{y}_i, {\tau}_i] - {p}_i)/{\tau}_i
 \nonumber \\ &
 =
 \frac{\int dz_i P_{\rm out}(y_i|z_i) \mathcal{N}(z_i| p_i, \tau_i) (z_i-  p_i)/\tau_i}{\int dz_i P_{\rm out}(y_i|z_i) \mathcal{N}(z_i| p_i, \tau_i)},
 \label{explicitgout}
\end{align}
where $Z_i\sim\mathcal{N}(p_i,\tau_i)$.
This expression is also equal to $\partial_{ p_i} \ln f(y_i| p_i, \tau_i)$ where $f$ is the function occurring in Definition \ref{def:effNoise} of 
the Fisher information.

In Table \ref{table:gout}\footnote{Based on a joint work with Erdem Bıyık \cite{BBD_ISIT2017}.} we give the explicit expressions for various channels as well as their derivatives used in the
GAMP algorithm of Section \ref{sec:GAMP} (where $\textrm{snr}$ is the signal-to-noise ratio of the AWGN channel, $\epsilon$ the erasure or flip probability of the BSC, BEC and ZC). The expressions of the Fisher information used in SE of Section \ref{sec:stateandpot} are given as well. These involve the Gaussian error function
${\rm erf}(x) = \frac{\sqrt 2}{\pi}\int_0^x dt\,e^{-t^2}$ and its complement ${\rm erfc}(x)= 1- {\rm erf}(x)$. Note that, for the sake of simplicity, all the expressions 
for the binary input channels of Table \ref{table:gout} (BSC, BEC and ZC) are given using the map $\pi(z)={\rm sign}(z)$. This map leads to a sub-optimal performance for the asymetric Z channel. The optimal map would require a bias in the input distribution as explained in Section \ref{sec:Z}.
\begin{table*}[t]
	\caption{The expressions for $g_{\rm out}$, $-\frac{\partial}{\partial \textbf{{p}}}g_{\rm out}$ and $\mathcal{F}$.}\label{table:gout}
	\centering
	{
	\vspace*{-5pt}
	\begin{tabular}{|M{0.95cm}|M{4.2cm}|M{7.95cm}|M{2.60cm}|N}
	\hline
	&
	$[g_{\rm out}(\textbf{{p}},\textbf{y},\boldsymbol{\tau})]_i$ & $[\!-\frac{\partial}{\partial \textbf{{p}}}\!g_{\rm out}(\textbf{{p}}, \textbf{y}, \boldsymbol{\tau})]_i$
	& $\mathcal{F}(p|E)$
	&\\[5pt]
	\hline
	\textbf{General} & 
	$(\mathbb{E}[Z_i | p_i,y_i, \tau_i] \!-\! {p}_i)\!/\!{\tau}_i$
	$Y_i\sim P_{\rm out}(\cdot|z_i), Z_i\sim \mathcal{N}(p_i,\tau_i)$
	& $({{\tau}_i\!-\! \textrm{Var}[Z_i|p_i,y_i,\tau_i]})\!/\!{\tau}_i^2$

	$Y_i\sim P_{\rm out}(\cdot|z_i), Z_i\sim \mathcal{N}(p_i,\tau_i)$
	& See Definition \ref{def:effNoise}
	& \\[8pt]
	\hline
	\textbf{AWGNC} & 
	$\frac{{y}_i\!-\!p_i}{{\tau}_i\!+\!1\!/\!\textrm{snr}}$
	& $\frac1{{\tau}_i\!+\!1\!/\!\textrm{snr}}$
	&
	$\frac1{1\!/\!\textrm{snr}\!+\!E}$
	&\\[8pt]
	\hline
	\textbf{BSC} & 
	$\frac{(p_i\!-\!k_i)v^+_i\!+\!(p_i\!+\!k_i)v^-_i}{{\cal Z}_{\textrm{BSC}}\tau_i}\!-\!\frac{p_i}{\tau_i}$
	& $\frac1{{\tau}_i}\!-\!\frac{(p_i^2\!+\!{\tau}_i\!-\!k'_i)v^+_i\!+\!(p_i^2\!+\!{\tau}_i\!+\!k'_i)v^-_i}{{\cal Z}_{\textrm{BSC}}{{\tau}_i}^2}\!+\!\big(\![g_{\rm out}\!(\!\textbf{{p}},\!\textbf{y},\!\boldsymbol{\tau}\!)]_i\!+\!\frac{p_i}{{\tau}_i}\!\big)\!^2$
	& 
	$\frac{Q'^2(1\!-\!2\epsilon)^2}{(Q\!+\!\epsilon\!-\!2\epsilon Q)(1\!-\!Q\!-\!\epsilon\!+\!2\epsilon Q)}$
	&\\[8pt]
	\hline
	\textbf{BEC} & 
	$\frac{(p_i\!-\!k_i)h^+_i\!+\!(p_i\!+\!k_i)h^-_i\!+\!2\epsilon\delta({y}_i)p_i}{{\cal Z}_{\textrm{BEC}}\tau_i}\!-\!\frac{p_i}{\tau_i}$
	& $\frac1{{\tau}_i}\!-\!\frac{(p_i^2\!+\!{\tau}_i\!-\!k'_i)h^+_i\!+\!(p_i^2\!+\!{\tau}_i\!+\!k'_i)h^-_i\!+\!2\epsilon\delta({y}_i)({\tau}_i\!+\!p_i^2)}{{\cal Z}_{\textrm{BEC}}{{\tau}_i}^2}\!+\!\big(\![g_{\rm out}\!(\!\textbf{{p}},\!\textbf{y},\!\boldsymbol{\tau}\!)]_i\!+\!\frac{p_i}{{\tau}_i}\!\big)\!^2$
	&
	$\frac{Q'^2(1\!-\!\epsilon)}{Q(1\!-\!Q)}$
	&\\[8pt]
	\hline
	\textbf{ZC} & 
	$\frac{(p_i\!-\!k_i)v^+_i\!+\!(p_i\!+\!k_i)\delta({y}_i\!-\!1)}{{\cal Z}_{\textrm{ZC}}\tau_i}\!-\!\frac{p_i}{\tau_i}$
	& $\frac1{{\tau}_i}\!-\!\frac{(p_i^2\!+\!{\tau}_i\!-\!k'_i)v^+_i\!+\!(p_i^2\!+\!{\tau}_i\!+\!k'_i)\delta({y}_i\!-\!1)}{{\cal Z}_{\textrm{ZC}}{{\tau}_i}^2}\!+\!\big(\![g_{\rm out}\!(\!\textbf{{p}},\!\textbf{y},\!\boldsymbol{\tau}\!)]_i\!+\!\frac{p_i}{{\tau}_i}\!\big)\!^2$
	& 
	$\frac{Q'^2(1\!-\!\epsilon)^2}{Q\!+\!\epsilon(1\!-\!Q)}\!+\!\frac{Q'^2(1\!-\!\epsilon)}{1\!-\!Q}$
	&\\[8pt]
	\hline
	\multicolumn{4}{|c|}{$h^+_i\!=\!(1\!\!-\!\!\epsilon)\delta({y}_i\!\!+\!\!1)$,\quad $h^-_i\!=\!(1\!\!-\!\!\epsilon)\delta({y}_i\!\!-\!\!1)$,\quad $v^+_i\!=\!(1\!\!-\!\!\epsilon)\delta({y}_i\!\!+\!\!1)\!\!+\!\!\epsilon\delta({y}_i\!\!-\!\!1)$,\quad $v^-_i\!=\!(1\!\!-\!\!\epsilon)\delta({y}_i\!\!-\!\!1)\!\!+\!\!\epsilon\delta({y}_i\!\!+\!\!1)$,}
	&\\[8pt]
	\multicolumn{4}{|c|}{ ${k}_i\!=\!\textrm{exp}\big(\frac{-p_i^2}{2\tau_i}\big)\sqrt{2{{\tau}_i}/\pi}\!\!+\!\!\textrm{erf}\big(\frac{p_i}{\sqrt{2{{\tau}_i}}}\big)p_i$,\quad $k'_i\!=\!{k}_i p_i\!\!+\!\!\textrm{erf}\big(\frac{p_i}{\sqrt{2{{\tau}_i}}}\big){\tau}_i$,\quad $Q \!=\! \frac12 \textrm{erfc}(\frac{\!-\!p}{\sqrt{2E}})$, \quad $Q'\!=\!\textrm{exp}\big(\frac{-p^2}{2E}\big)\big/{\sqrt{2\pi E}}$}
	&\\[8pt]
	\multicolumn{4}{|c|}{${\cal Z}_{\textrm{BEC}}\!=\! {\textrm{erfc}\big(\frac{p_i}{\sqrt{2{\tau}_i}}\big)h^+_i\!\!+\!\!\big(1\!\!+\!\!\textrm{erf}\big(\frac{p_i}{\sqrt{2{\tau}_i}}\big)\big)h^-_i\!\!+\!\!2\epsilon\delta({y}_i)}$, ${\cal Z}_{\textrm{ZC}}\!=\!\textrm{erfc}\big(\frac{p_i}{\sqrt{2{\tau}_i}}\big)v^+_i\!\!+\!\!\big(1\!\!+\!\!\textrm{erf}\big(\frac{p_i}{\sqrt{2{\tau}_i}}\big)\big)\delta({y}_i\!\!-\!\!1)$, ${\cal Z}_{\textrm{BSC}}\!=\!\textrm{erfc}\big(\frac{p_i}{\sqrt{2{\tau}_i}}\big)v^+_i\!\!+\!\!\big(1\!\!+\!\!\textrm{erf}\big(\frac{p_i}{\sqrt{2{\tau}_i}}\big)\big)v^-_i$}
	&\\[8pt]
	% \multicolumn{4}{|c|}{${\cal Z}_{\textrm{BSC}}\!=\!\textrm{erfc}\big(\frac{p_i}{\sqrt{2{\tau}_i}}\big)v^+_i\!+\!\big(1\!+\!\textrm{erf}\big(\frac{p_i}{\sqrt{2{\tau}_i}}\big)\big)v^-_i$}
	% &\\[8pt]	
	\hline
	\end{tabular}}
	\vspace{-10px}
\end{table*}

\section{State evolution and potential function}\label{sec:App_SE}

In this appendix we prove Lemma \ref{lemma:fixedpointSE_extPot}. Namely, we show that the stationarity condition
$\partial F_{\text{un}}/\partial E =0$ for the potential function in Definition \ref{def:pot_underlying} implies the 
state evolution equation in Definition \ref{def:SE}. We present a detailed derivation for the underlying uncoupled system.
The proof of Lemma \ref{lemma:fixedpointSE_extPot} for the coupled system follows exactly the same steps.

The calculation is best done by looking at $F_{\text{un}}$ as a function 
of $E$ and $\Sigma(E)^{-2}$, so that  
\begin{align}
\frac{d F_{\text{un}}}{d E} = &  - \frac{1}{2\ln(2) \Sigma(E)^2 } - \frac{1}{R} 
\frac{\partial}{\partial E} \mathbb{E}_Z\Big[\int dy\, \phi(y|Z,E) \log_2 \phi(y|Z,E)\Big]
\nonumber \\ &
- \Big\{\frac{E}{2\ln(2)}  + \frac{\partial}{\partial \Sigma(E)^{-2}}
\mathbb{E}_{\bS,\bZ}\Big[\log_B\int d^B{\bx} \,p_0(\bx) \theta(\bx,\bS,\mathbf{Z},\Sigma(E))\Big]
 \Big\} 
 \frac{d}{dE}\Sigma(E)^{-2}.
 \label{derivativeF}
\end{align}
We first look at the derivative of the bracket $\{\cdots\}$ with respect to $\Sigma^{-2}$. In the next few lines the following notation is used 
for the ``Gibbs'' average
\begin{align*}
 \langle A(\bx)\rangle_{\rm den} = \frac{
 \int d^B{\bx} \,A(\bx) p_0(\bx) \theta(\bx,\bS,\mathbf{Z},\Sigma(E)) }{\int d^B{\bx} \,p_0(\bx) \theta(\bx,\bS,\mathbf{Z},\Sigma(E)) }.
\end{align*}
Using the explicit expression of $\theta(\bx,\bS,\mathbf{Z},\Sigma(E))$ we have
\begin{align*}
\frac{\partial}{\partial \Sigma(E)^{-2}} &
\mathbb{E}_{\bS,\bZ}\Big[\log_B\int d^B{\bx} \,p_0(\bx) \theta(\bx,\bS,\mathbf{Z},\Sigma(E))\Big]
\nonumber \\ &
= 
\frac{\partial}{\partial \Sigma(E)^{-2}}
\mathbb{E}_{\bS,\bZ}\Big[\log_B\int d^B{\bx} \,p_0(\bx) 
e^{-\frac{1}{2}\big(\Vert\bx - \bS\Vert^2\Sigma(E)^{-2}\frac{\ln B}{\ln 2} 
- 2 \bZ\cdot(\bx -\bS)\Sigma(E)^{-1}\sqrt{\frac{\ln B}{\ln 2}}
+ \Vert Z\Vert^2\big)}
\Big]
\nonumber \\ &
= 
- \frac{1}{2\ln 2} \mathbb{E}_{\bS,\bZ}\Big[\langle \Vert\bx - \bS\Vert^2\rangle_{\text{den}}\Big] + \frac{1}{2}\mathbb{E}_{\bS, \bZ}
\Big[\bz \cdot \langle \bx - \bS\rangle_{\text{den}}\Big] \frac{\Sigma(E)}{\sqrt{(\ln B)(\ln 2)} }
\nonumber \\ &
= 
- \frac{1}{2\ln 2} \mathbb{E}_{\bS,\bZ}\Big[\langle \Vert\bx - \bS\Vert^2\rangle_{\text{den}}\Big] + \frac{1}{2}\mathbb{E}_{\bS, \bZ}
\Big[\nabla_{\bZ} \cdot \langle \bx - \bS\rangle_{\text{den}}\Big] \frac{\Sigma(E)}{\sqrt{(\ln B)(\ln 2)} }
\nonumber \\ &
= 
- \frac{1}{2\ln 2} \mathbb{E}_{\bS,\bZ}\Big[\langle \Vert\bx - \bS\Vert^2\rangle_{\text{den}}\Big] + \frac{1}{2\ln 2}\mathbb{E}_{\bS, \bZ}
\Big[\langle \Vert\bx - \bS\Vert^2\rangle_{\text{den}}\Big] 
- \frac{1}{2\ln 2}\mathbb{E}_{\bS, \bZ}
\Big[\Vert\langle \bx - \bS\rangle_{\text{den}}\Vert^2\Big]
\nonumber \\ &
= 
- \frac{1}{2\ln 2}\mathbb{E}_{\bS, \bZ}
\Big[\Vert\langle \bx\rangle_{\text{den}} - \bS \Vert^2\Big]
\nonumber \\ &
= 
- \frac{1}{2\ln 2}\text{mmse}(\Sigma(E)).
\end{align*}
We show below that 
\begin{align}\label{Legendre-relation}
\frac{1}{\Sigma(E)^2 } = - \frac{2}{R}
\frac{\partial}{\partial E} \mathbb{E}_Z\Big[\int dy\, \phi(y|Z,E) \ln \phi(y|Z,E)\Big]
\end{align}
so that \eqref{derivativeF} becomes
\begin{align}\label{integrating-factor-relation}
\frac{d F_{\text{un}}}{d E} = \Big\{\text{mmse}(\Sigma(E)) - E\Big\} 
\frac{1}{2\ln 2}\frac{d}{dE}\Sigma(E)^{-2}
\end{align}
which obviously shows that $d F_{\text{un}}/d E=0$ implies the SE equation $E=T_{\text{un}}(E)$. We point out as a side remark that 
this is the correct ``integrating factor'' which allows to recover the potential function from the SE equation.

It remains to derive \eqref{Legendre-relation}. We will start from the derivative with respect to $E$ in \eqref{Legendre-relation} and 
show that this relation can be transformed into Definition \ref{def:effNoise}, namely
\begin{align}\label{defSigma}
\frac{1}{\Sigma(E)^2} = \frac{1}{R}\int dp\, \frac{e^{-\frac{p^2}{2(1-E)}}}{\sqrt{2\pi (1-E)}}\int dy f(y|p,E) (\partial_p \ln f(y|p,E))^2
\end{align}
where 
\begin{align}\label{def-f}
f(y|p, E) = \int dx P_{\text{out} }(y|x) \frac{e^{-\frac{(x-p)^2}{2E}}}{\sqrt{2\pi E}}.
\end{align}

We first note that $\phi(y|z, E) = f(y|z\sqrt{1-E}, E)$ so the derivative w.r.t $E$ on the right hand side of \eqref{Legendre-relation} becomes
\begin{align}
\frac{\partial}{\partial E} \mathbb{E}_Z\Big[\int dy\, \phi(y|Z,E) \ln \phi(y|Z,E)\Big]
 & = 
\frac{\partial}{\partial E} \int dz\, \frac{e^{-\frac{z^2}{2}}}{\sqrt{2\pi}} \int dy\,  f(y|z\sqrt{1-E}, E) \ln f(y|z\sqrt{1-E}, E)
\nonumber \\ &
= 
\int dz\, \frac{e^{-\frac{z^2}{2}}}{\sqrt{2\pi}}\int dy\,  (1+ \ln f(y|z\sqrt{1-E}, E)) \partial_E f(y|z\sqrt{1-E}, E).
\label{previousstep}
\end{align}
An exercise in differentiation of Gaussians shows\footnote{We thank Christophe Sch\"ulke for pointing out this trick.}
\begin{align*}
\partial_E\Big\{\frac{e^{-\frac{(x- z\sqrt{1-E})^2}{2E}}}{\sqrt{2\pi E}} \Big\}=  \frac{e^{\frac{z^2}{2}}}{2(1-E)}
\partial_z\Big\{e^{-\frac{z^2}{2}}\partial_z \Big\{\frac{e^{-\frac{(x- z\sqrt{1-E})^2}{2E}}}{\sqrt{2\pi E}} \Big\} \Big\}.
\end{align*}
Thus from \eqref{def-f}
\begin{align*}
\partial_E f(y|z\sqrt{1-E}, E) =  \frac{e^{\frac{z^2}{2}}}{2(1-E)}
\partial_z\Big\{e^{-\frac{z^2}{2}}\partial_z f(y|z\sqrt{1-E}, E) \Big\}
\end{align*}
and \eqref{previousstep} becomes
\begin{align*}
\frac{\partial}{\partial E} \mathbb{E}_Z\Big[\int dy\, \phi(y|Z,E) \ln \phi(y|Z,E)\Big]
 & = 
  \frac{1}{2(1-E)}
\int dz\, \int dy\,  (1+ \ln f(y|z\sqrt{1-E}, E))
\nonumber \\ & 
\times 
\partial_z\Big\{\frac{e^{-\frac{z^2}{2}}}{\sqrt{2\pi}}\partial_z f(y|z\sqrt{1-E}, E) \Big\}
\nonumber \\ &
= -
\frac{1}{2(1-E)}
\int dz\, \frac{e^{-\frac{z^2}{2}}}{\sqrt{2\pi}}\int dy\,  
\frac{
\big(\partial_z f(y|z\sqrt{1-E}, E) \big)^2}{f(y|z\sqrt{1-E}, E)}
\nonumber \\ &
= -
\frac{1}{2}
\int dz\, \frac{e^{-\frac{z^2}{2(1-E)}}}{\sqrt{2\pi(1-E)}}\int dy\,  
\frac{
\big(\partial_z f(y|z, E) \big)^2}{f(y|z, E)}
\nonumber \\ &
= -
\frac{1}{2}
\int dz\, \frac{e^{-\frac{z^2}{2(1-E)}}}{\sqrt{2\pi(1-E)}}\int dy\,  
f(y|z, E)
\big(\partial_z \ln f(y|z, E) \big)^2.
\end{align*}
This result explicitly shows that \eqref{Legendre-relation} and \eqref{defSigma} are equivalent as announced.

\section{Bounds on the second derivative of the potential function}\label{sec:App_Bound}
In this appendix, we provide an upper bound on the second derivative 
\begin{align}
\Big[\frac{\partial^2 F_{\text{co}}}{\partial  E_{r}\partial  E_{r'}}\Big]_{\hat{\tbf E}}
 &
= \Big[\frac{\partial^2 }{\partial  E_{r}\partial  E_{r'}}\sum_{r=1}^\Gamma U_{\text{un}}(E_r)\Big]_{\hat{\tbf E}} - 
\frac{\partial^2 }{\partial  E_{r}\partial  E_{r'}}\Big[\sum_{c=1}^\Gamma S_{\text{un}}(\Sigma_c({\tbf E}))\Big]_{\hat{\tbf E}}
\nonumber \\ &
=
\delta_{r, r'}\Big[\frac{\partial^2 U_{\text{un}}(E_r)}{\partial E_r^2}\Big]_{\hat{\tbf E}}
 - \sum_{c=1}^\Gamma \Big[\frac{\partial^2 S_{\text{un}}(\Sigma_c({\tbf E}))}{\partial E_r \partial E_{r^\prime}}\Big]_{\hat{\tbf E}}
 \label{the-crucial-term-to-bound}
\end{align}
of the potential function needed in the proof of Lemma \ref{lemma:quadFormBounded}. We first perform the analysis for general memoryless channels satisfying our two Assumptions \ref{continuity-assumption}, \ref{scaling-assumption}. We then briefly show how to improve the estimate in the special case of the AWGN because of the non vanishing error floor. 

\subsection{General channel}\label{sec:App_Bound_general}

\subsubsection*{Energy term}

Using relation \eqref{Legendre-relation} of Appendix \ref{sec:App_SE} one obtains for the first derivative of the energy term 
\begin{align*}
\frac{\partial U_{\text{un}}(E_r)}{\partial  E_{r}} =  -\frac{E_r}{2 \ln 2} \frac{\partial \Sigma^{-2}}{\partial E_r}.
\end{align*}
Differentiating once more
\begin{align*}
\frac{\partial^2 U_{\text{un}}(E_r)}{\partial  E_{r}^2} = - \frac{1}{2 \ln 2} \frac{\partial \Sigma^{-2}}{\partial E_r} - \frac{E_r}{2 \ln 2} \frac{\partial^2 \Sigma^{-2}}{\partial E_r^2}  .
\end{align*}
Using Assumption \ref{scaling-assumption} we immediately get
\begin{align*}
\frac{\partial^2 U_{\text{un}}(E_r)}{\partial  E_{r}^2} 
\leq 
\frac{C}{2 (\ln 2)R E_r^\beta} + \frac{C E_r}{2 (\ln 2) R E_r^\beta}.
\end{align*}
Now recall that in the proof of Lemma \ref{lemma:quadFormBounded} we have $\hat{E}_r > \bar E_{\rm f} = E_{\rm f} + \epsilon$ where 
$E_{\rm f}$ is the (true) error floor and $\epsilon > 0$. Therefore
\begin{align}
\Big[\frac{\partial^2 U_{\text{un}}}{\partial  E_{r}^2}\Big]_{\hat{\tbf E}}
& \leq 
\frac{C}{2 (\ln 2)R (E_{\rm f} +\epsilon)^\beta} + \frac{C (E_{\rm f} +\epsilon)}{2 (\ln 2) R (E_{\rm f} + \epsilon)^\beta}
\nonumber \\ &
\leq 
\frac{C(2 +\epsilon)}{2 (\ln 2)R (E_{\rm f} +\epsilon)^\beta}.
\label{one-piece-of-bound}
\end{align}
Of course this is the worse possible bound and is valid all the way up to the left boundary of the 
{\it modified} system. As one moves towards the right of the spatially coupled system one could use 
bigger values for $E_r$ and tigthen the bound. This however is not needed to prove Lemma \ref{lemma:quadFormBounded}
as long as $\epsilon >0$. 

\subsubsection*{Entropy term}
For the second derivative of the ``entropy'' term we first apply the 
chain rule
\begin{align*}
\frac{\partial^2 S_{\text{un}}(\Sigma_c({\tbf E}))}{\partial  E_{r}\partial  E_{r'}} 
& 
= 
\frac{\partial}{\partial E_{r'}} \Big( \frac{\partial S_{\text{un}} }{\partial \Sigma_c^{-2}} \frac{\partial \Sigma_c^{-2}}{\partial E_r} \Big) 
\nonumber \\ &
= 
\frac{\partial^2 S_{\text{un}} }{\partial (\Sigma_c^{-2})^2} \frac{\partial \Sigma_c^{-2}}{\partial E_r} \frac{\partial \Sigma_c^{-2}}{\partial E_{r'}} + \frac{\partial S_{\text{un}} }{\partial \Sigma_c^{-2}} \frac{\partial^2 \Sigma_c^{-2}}{\partial E_r \partial E_{r'}}
\nonumber \\ &
=
J_{r,c} J_{r',c}
\frac{\partial^2 S_{\text{un}} }{\partial (\Sigma_c^{-2})^2}  \frac{\partial \Sigma^{-2}}{\partial E_{r}}  \frac{\partial \Sigma^{-2}}{\partial E_{r'}} 
+ 
\delta_{r r^\prime}J_{r,c}
\frac{\partial S_{\text{un}} }{\partial \Sigma_c^{-2}}  \frac{\partial^2 \Sigma^{-2}}{\partial E_r^2},
\end{align*}
where to get the last line we used
\begin{align*}
 \frac{\partial \Sigma_c^{-2}}{\partial E_r} 
 = 
 J_{r,c} \frac{\partial \Sigma^{-2}}{\partial E_r}
 1_{c-w\leq r \leq c+w}, 
 \qquad
  \frac{\partial^2 \Sigma_c^{-2}}{\partial E_r\partial E_{r^\prime}} = \delta_{r r^\prime}J_{r,c} \frac{\partial^2 \Sigma^{-2}}{\partial E_r^2}
  1_{c-w \leq r \leq c+w}
  \end{align*}
which follow directly from the definition of $\Sigma_c^{-2}({\bf E})$.
Recall that by construction $J_{r,c}/\Gamma \leq ({\bar g}/{\underline g})(2w+1)^{-1}$. Recall also Assumption 
\ref{scaling-assumption}. We thus have
\begin{align}
\Big|\frac{\partial^2 S_{\text{un}}(\Sigma_c({\tbf E}))}{\partial  E_{r}\partial  E_{r'}}\Big| 
 &
 \leq 
\frac{\bar g^2}{\underline{g}^2 (2w+1)^2}
\Big|\frac{\partial^2 S_{\text{un}} }{\partial (\Sigma_c^{-2})^2} \Big| \Big|\frac{\partial \Sigma^{-2}}{\partial E_{r}}\Big|  \Big|\frac{\partial \Sigma^{-2}}{\partial E_{r'}}\Big| 1_{c-w \leq r \leq c+w} 1_{c-w \leq r' \leq c+w}
\nonumber \\ &
+ 
\frac{\delta_{r r^\prime} \bar g}{\underline{g} (2w+1)}
\Big|\frac{\partial S_{\text{un}} }{\partial \Sigma_c^{-2}}\Big|  \Big|\frac{\partial^2 \Sigma^{-2}}{\partial E_r^2}\Big|
1_{c-w \leq r \leq c+w}
\nonumber \\ &
\leq 
\frac{\bar g^2 C^2}{\underline{g}^2 (2w+1)^2R^2 E_r^\beta E_{r'}^\beta}
\Big|\frac{\partial^2 S_{\text{un}} }{\partial (\Sigma_c^{-2})^2} \Big| 
1_{c-w \leq r \leq c+w} 1_{c-w \leq r' \leq c+w}
\nonumber \\ &
+ 
\frac{\delta_{r r^\prime} \bar g C}{\underline{g} (2w+1)R E^\beta}
\Big|\frac{\partial S_{\text{un}} }{\partial \Sigma_c^{-2}}\Big| 
1_{c-w \leq r \leq c+w}
\label{second-derivative}
\end{align}
The next step is to compute and estimate the partial derivatives of $S_{\text{un}}$ in this expression. Using Definition
\ref{def:pot_underlying} we find (this involves differentiating under integral signs which can be justified by the ensuing bounds)
\begin{align}
\frac{\partial S_{\text{un}} }{\partial \Sigma_c^{-2}} = \sum_{i=2}^B \mathbb{E}_{\textbf{Z}}\Big[
\Big( \frac{(Z_i -Z_1)\Sigma_c}{2\sqrt{\ln 2 \ln B}} -\frac{1}{\ln 2} \Big) \frac{e_{i} }{1+\sum_{j=2}^B e_{j}}
\Big] 
\label{s1}
\end{align}
\begin{align}
\frac{\partial^2 S_{\text{un}} }{\partial (\Sigma_c^{-2})^2} 
 =  
 &
(\ln B)
\sum_{i=2}^B \mathbb{E}_{\textbf{Z}}\Big[ \Big(\Big( \frac{(Z_i -Z_1)\Sigma_c}{2\sqrt{\ln 2 \ln B}} -\frac{1}{\ln 2} \Big)^2
 - \frac{(Z_i -Z_1)\Sigma_c^{3}}{2\sqrt{\ln 2 \ln B}}\Big)\frac{e_{i}}{1+\sum_{j=2}^B e_{j}}
\nonumber 
\\ 
&
- 
(\ln B)\sum_{i, j=2}^B
 \Big( \frac{(Z_i -Z_1)\Sigma_c}{2\sqrt{\ln 2 \ln B}} -\frac{1}{\ln 2} \Big)
\Big( \frac{(Z_j -Z_1)\Sigma_c}{2\sqrt{\ln 2 \ln B}} -\frac{1}{\ln 2} \Big)  
\frac{e_{i}e_{j} }{(1+\sum_{j=2}^B e_{j})^2} \Big]
\label{s2},
\end{align}
Since $e_i\geq 0$ we have for $2\leq i \leq n$
$$
\frac{e_{i}}{1+\sum_{j=2}^B e_{j}} \leq 1
$$
which easily implies the following bounds for \eqref{s1} and \eqref{s2}
\begin{align}
\Big| \frac{\partial S_{\text{un}} }{\partial \Sigma_c^{-2}} \Big| 
&
\le 
(B-1) \Big(\frac{\Sigma_c}{\sqrt{\pi \ln 2 \ln B}} + \frac{1}{\ln 2}\Big) 
\nonumber \\ &
\leq 
C_1(B)+ C_2(B)\Sigma_c
\label{cc}
\end{align}
\begin{align}
\Big|\frac{\partial S_{\text{un}} }{\partial \Sigma_c^{-2}}\Big|
& \leq  
 (\ln B) (B-1) \Big( \frac{\Sigma^2_{c}}{2 \ln 2 \ln B} + \frac{1}{(\ln 2)^2} + \frac{2 \Sigma_c}{\ln 2 \sqrt{\pi \ln 2 \ln B}} \Big) 
+ (B-1) \frac{\Sigma^3_{c}}{\sqrt{\pi \ln 2 \ln B}}
\nonumber \\ &
+ (\ln B) (B-1)^2 \Big( \frac{(\frac{2\sqrt{3}}{\pi} + \frac{1}{3})\Sigma^2_{c}}{4 \ln 2 \ln B} + \frac{1}{(\ln 2)^2} +  \frac{2 \Sigma_c}{\ln 2 \sqrt{\pi \ln 2 \ln B}} \Big)
\nonumber \\ &
\leq 
C_3(B) + C_4(B)\Sigma_c + C_5(B)\Sigma_c^2 + C_6(B)\Sigma_c^3.
\label{ccc}
\end{align}
where $C_i(B)$, $i=1, \cdots, 6$ are constants that depend only on $B$. Furthermore from the definition of 
$\Sigma_c({\bf E})$ and Assumption \ref{continuity-assumption} we remark that 
$\Sigma_c({\bf E}) \leq \sup_{E\in [0,1]}\Sigma(E)= \Sigma(1)$ so in the bounds \eqref{cc}, \eqref{ccc} we can replace 
$\Sigma_c$ by $\Sigma(1)$. Then using these two bounds the estimate \eqref{second-derivative} becomes
\begin{align*}
\Big|\frac{\partial^2 S_{\text{un}}(\Sigma_c({\tbf E}))}{\partial  E_{r}\partial  E_{r'}}\Big|  
\leq 
&
\frac{\bar g^2 C^2}{\underline{g}^2 (2w+1)^2R^2 E_r^\beta E_{r'}^\beta} 
\Big(C_3(B) + C_4(B)\Sigma(1) + C_5(B)\Sigma^2(1) + C_6(B)\Sigma^3(1)\Big)
\nonumber \\ &
\times 
1_{c-w \leq r \leq c+w} 1_{c-w \leq r' \leq c+w}
\nonumber \\ &
+ 
\frac{\delta_{r r^\prime} \bar g C}{\underline{g} (2w+1)R E_r^\beta} 
\Big(C_1(B) + C_2(B)\Sigma(1)\Big) 1_{c-w \leq r \leq c+w}
\end{align*}
Since 
$$
1_{c-w \leq r \leq c+w} 1_{c-w \leq r' \leq c+w} \leq 1_{r-w \leq c \leq r+w} 1_{\vert r - r'\vert \leq 2w+1}
\quad \text{and}\quad 
1_{c-w \leq r \leq c+w} = 1_{r-w \leq c \leq r+w}
$$ 
when we sum over $c$ we get
\begin{align*}
\sum_{c=1}^\Gamma \frac{\partial^2 S_{\text{un}}(\Sigma_c({\tbf E}))}{\partial  E_{r}\partial  E_{r'}}
\leq &
\frac{\bar g^2 C^2}{\underline{g}^2R^2 (2w+1) E_r^\beta E_{r'}^\beta} 
\Big(C_3(B) + C_4(B)\Sigma(1) + C_5(B)\Sigma^2(1) + C_6(B)\Sigma^3(1)\Big)
1_{\vert r - r' \vert \leq 2w+1}
\nonumber \\ &
+ 
\frac{\delta_{r r^\prime} \bar g C}{\underline{g}R E_r^\beta} 
\Big(C_1(B) + C_2(B)\Sigma(1)\Big) 
\end{align*}
Finally using again $\hat E_r \geq \bar E_{\rm f} = E_{\rm f} +\epsilon$ we obtain
\begin{align}
\Big[\sum_{c=1}^\Gamma \frac{\partial^2 S_{\text{un}}(\Sigma_c({\tbf E}))}{\partial  E_{r}\partial  E_{r'}}\Big]_{\hat{\tbf E}}
\leq &
\frac{\bar g^2 C^2}{\underline{g}^2R^2 (2w+1) (E_{\rm f} +\epsilon)^{2\beta}} 
\nonumber \\ &
\times
\Big(C_3(B) + C_4(B)\Sigma(1) + C_5(B)\Sigma^2(1) + C_6(B)\Sigma^3(1)\Big)
1_{\vert r - r' \vert \leq 2w+1}
\nonumber \\ &
+ 
\frac{\delta_{r r^\prime} \bar g C}{\underline{g}R (E_{\rm f} +\epsilon)^\beta} 
\Big(C_1(B) + C_2(B)\Sigma(1)\Big) 
\label{second-crucial}
\end{align}

\subsubsection*{Final bound} 

Putting \eqref{the-crucial-term-to-bound}, \eqref{one-piece-of-bound} and \eqref{second-crucial} together the triangle inequality implies the important result
\begin{align}
\Big[\frac{\partial^2 F_{\text{co}}}{\partial  E_{r}\partial  E_{r'}}\Big]_{\hat{\tbf E}}
\leq
\delta_{r, r'} \frac{K_1(B, \bar g, \underline g)}{(E_{\rm f} +\epsilon)R} + 1_{\vert r - r' \vert \leq 2w+1} 
\frac{K_2(B, \bar g, \underline g)}{(E_{\rm f} + \epsilon)^{2\beta}R(2w+1)}
\end{align}
for some finite positive $K_1(B, \bar g, \underline g)$ and $K_2(B, \bar g, \underline g)$ independent of $w$ and $\Gamma$. 

\subsection{AWGN Channel}

For the AWGN channel we have an explicit expression for the effective noise, $\Sigma(E)^2 = ({\rm snr}^{-1} +E)R$ which implies
\begin{align}\label{eq:bound_Sigma}
\begin{cases}
 \Sigma^{2}(E_r) &\le \quad R({\rm snr}^{-1} + 1) \\
 \frac{\partial \Sigma^{-2}}{\partial E_r} &\le \quad \frac{\rm{snr}^2}{R} \\
 \frac{\partial^2 \Sigma^{-2}}{\partial E_r^{2}} &\le \quad \frac{\rm{snr}^3}{R}.
\end{cases}
\end{align}
Then using these bounds at the appropriate places in the previous analysis  we get
\begin{align}
\Big[\frac{\partial^2 F_{\text{co}}}{\partial  E_{r}\partial  E_{r'}}\Big]_{\hat{\tbf E}}
\leq
\delta_{r, r'} K_1^\prime(B, \bar g, \underline g){\rm snr}^2 + 1_{\vert r - r' \vert \leq 2w+1} 
\frac{K_2^\prime(B, \bar g, \underline g){\rm snr}^4}{2w+1}
\end{align}
for new constants $K_1^\prime(B, \bar g, \underline g)$, $K_2^\prime(B, \bar g, \underline g)$ (independent of $w$, $\Gamma$).
We can see that the qualitative behaviour of the bound when ${\rm snr}\to +\infty$ is the same than in the case of vanishing 
error floor $E_{\rm f}=0$ and $\epsilon\to 0$.

\section{Potential function and replica calculation}\label{sec:App_Bethe}

The potential functions of the uncoupled and coupled systems, used in this paper, can be viewed as a mathematical tool and we are not really concerned how they are found.
However in practice it is important to have a more or less systematic method which allows to write down ``good'' potential functions. 
There are essentially two ways. One is to ``integrate'' the SE equations as done in \cite{6887298} by using an appropriate ``integrating factor''.
With this method there is some amount of guess involved. For example in the present problem it is not entirely obvious that the correct integrating factor 
is directly related to the Fisher information (as equation \eqref{integrating-factor-relation} in Appendix \ref{sec:App_SE} shows).
The other way is to perform a formal and brute force replica or cavity calculation of the free energy which is then given as 
a variational expression involving the potential function. The disadvantage of such a calculation is that it is painful and maybe also that it is formal, but the 
advantage is that it is quite systematic.
For completeness we give the replica calculation. We stress that the results of the paper do {\it not} rest on this formal calculation and the reader can entirely skip it.

We treat the prototypical case of a spatially coupled compressed-sensing like system where the signal has {\it scalar} components $x_i$, $i=1,\cdots, N$ iid distributed according to a general prior $p_0(x)$. 
The calculation is exactly the same for signals whose components are $B$-dimensional with arbitray priors and sparse superposition codes fall in this class. 
The integration symbol $\mathcal{D}v$ is used for $dv\, e^{-\frac{v^2}{2}}$.

The spatially coupled matrix is made of $\Gamma \times \Gamma$ blocks, 
each with $N/\Gamma$ columns and $ \alpha N/\Gamma$ rows for the blocks part of the $r^{th}$ block-row. The entries 
inside the block $(r,c)$ are i.i.d. with distribution $\mathcal{N}(0,J_{r,c}\Gamma/N)$. Furthermore, we enforce the per block-row 
variance normalization $\sum_{c=1}^{\Gamma} J_{r,c} = 1 \ \forall \ r$. We use the notation $\bx^0$ for the signal and 
define $z_\mu^a \defeq \sum_{c=1}^\Gamma \sum_{i\in c}^{N/\Gamma} F_{\mu i}x_i^a$ where the matrix structure is made explicit. 

The posterior distribution 
is  given by the Bayes rule 
$$
P(\bx|\by) = Z(\by)^{-1}\prod_{i=1}^N p_0(x_i) \prod_{\mu=1}^M P_{\text{out}}(y_\mu|z_\mu)
$$
where $Z(\by)=P(\by)$ is 
the observation dependent normalization, or partition function. The (coupled) free energy ${F}_{\text{co}}$ will be calculated using the replica trick in one of its many 
incarnations
\begin{align}
{F}_{\text{co}} \defeq 
- \lim_{N\to\infty}\lim_{n\to 0} \frac{\partial}{\partial n} \frac{\ln(\mathbb{E}[Z(\by)^n])}{N}, \label{eq:replicatrick}
\end{align}
where $\mathbb{E}$ denotes expectation with respect to the observation $\by(\bF)$ 
which depend on the measurement matrix realization (that will be always implicit). 
We thus need to compute the $n^{th}$ moment of the partition function. For the moment, we consider $n\in\mathbb{N}$ despite that we will let $n\to0$ at the end.

$Z(\by)^n$ can be interpreted as the partition function of $n$ i.i.d. systems, the replicas $a=1, \cdots, n$, each generated independently from the posterior $P(\bx|\by)$
\begin{align}
% Z(\by)&=\int d\bx^0 \prod_{i=1}^N p_0(x_i^0) \prod_{\mu=1}^M P_{\text{out}}(y_\mu|z_\mu^0), \\
% 
Z(\by)^n &= \int \prod_{a=1}^n \left[ d\bx^a \prod_{i=1}^N p_0(x_i^a) \prod_{\mu=1}^M P_{\text{out}}(y_\mu|z_\mu^a)\right],\\
\mathbb{E}[Z(\by)^n] &= \mathbb{E}_{\bF} \int d\by Z(\by)^n P(\by) = \mathbb{E}_{\bF} \int d\by Z(\by)^{n+1} = \mathbb{E}_{\bF} \int d\by\prod_{a=0}^n \[ d\bx^a \prod_{i=1}^N p_0(x_i^a)  \prod_{\mu=1}^M P_{\text{out}}(y_\mu|z_\mu^a)\], \label{eq:eval_Znp1}
\end{align}
where the last equality is implied by $P(\by) = Z(\by)$. This last point is valid only in the Bayes optimal setting and is known to induce
a remarkable set of consequences, among which the correctness of the replica symmetric predictions.

The $\bF$ and $\bx^a$ r.v being i.i.d., we can treat $z_\mu^a$ as a Gaussian random variable by the central limit theorem. Let us compute their distribution. 
As $\bF$ has zero mean, $z_\mu^a$ has zero mean also. Its covariance matrix $\boldsymbol{\tilde{q}}_{r_\mu}$ 
depends on the block-row index $r_\mu \in \{1,\cdots,\Gamma\}$ to which the $\mu^{th}$ measurement index belongs. Similarly, 
$c_i \in \{1,\dots,\Gamma\}$ is the block-column index to which the $i^{th}$ column belongs. We have
\begin{align}\label{qabrmu}
\tilde q^{ab}_{r_\mu} = \mathbb{E}_{\bF}[z_\mu^a z_\mu^b] = \sum_{c,c'=1}^{\Gamma, \Gamma} \sum_{i\in c, j\in c'}^{N/\Gamma,N/\Gamma} \mathbb{E}_{\bF}[F_{\mu i}F_{\mu j}] x_i^a x_j^b = \sum_{c}^{\Gamma} \frac{J_{r_\mu,c}}{N} \sum_{i\in c}^{N/\Gamma} x_i^a x_i^b,
\end{align}
because $\mathbb{E}_{\bF}[F_{\mu i} F_{\mu j}] = \delta_{ij} J_{r_\mu,c_i}/N$ in the present spatial coupling construction. 
We introduce the macroscopic replica overlap matrix, that takes into account the block structure in the signal induced by the matrix structure. Let
\begin{align}
{q}^{ab}_c \defeq \frac{\Gamma}{N} \sum_{i\in c}^{N/\Gamma} x_i^a x_i^b \ \forall \ a, b \in \{0, \cdots,n\}. \label{eq:def_overlap}
\end{align}
Then \eqref{qabrmu}
becomes $\tilde q^{ab}_r =  \sum_{c=1}^{\Gamma} J_{r,c} {q}_c^{ab}$. 

%that assumes that the posterior used for the replicas generation has a "simple" structure in the sense that it does not break into different Gibbs states. More formally, in full generality the posterior could be written as a linear combination of Gibbs measures $P(\bx|\by) = \sum_{u} \lambda_u P_u(\bx|\by)$ with $\sum_{u}\lambda_u = 1$, such that typical (large) samples generated from different Gibbs measures $P_{u}(\bx|\by)$ have an overlap $q_1$, whereas samples generated from the same Gibbs measure have an overlap $q_0$ (i.e. the overlap distribution is non trivial). The replica symmetry says that there actually exists a single $\lambda_u =1$, i.e. a single Gibbs state. It is actually possible to show that in an inference problem which factor graph is locally tree-like or dense and under the Nishimori condition, the replica symmetry is a valid assumption (the replica symmetry breaking phenomenon cannot occur). However, if the Nishimori condition is not verified (i.e. the generative model of the signal does not match the assumed prior and/or the noise channel statistics is unkown), then replica symmetry can happen which complicates a lot the analysis. 

We now introduce the replica symmetric ansatz.
According to this ansatz, the overlap should not depend on the replica index 
$q^{ab}_c = q_c \ \forall \ a\neq b, \, q^{aa}_c = Q_c \ \forall \ a$. 
This implies 
\begin{align}
\tilde q^{ab}_r = \tilde q_r =\sum_{c=1}^\Gamma J_{r,c} q_c \ \forall \ a\neq b, \, \tilde q^{aa}_r = \tilde Q_r = \sum_{c=1}^\Gamma J_{r,c} Q_c \ \forall \ a.
\end{align}
Using the variance  normalization $Q_c = \tilde Q_r$. Then, one can show that in Bayes optimal inference we have  furthermore
$Q_c = \tilde Q_r = \mathbb{E}[S^2] \ \forall \ c, r \in \{1, \cdots,\Gamma\}$,
where $\mathbb{E}[s^2] = \int ds p_0(s) s^2$. In the physics litterature this is often called a ``Nishimori identity''. 

Thus the self overlap $Q_c$ is fixed and the condition (\ref{eq:def_overlap}) for $a=b$ does not need to be enforced. On the other hand, the cross overlap for $a\neq b$ is unknown and so we must keep $\{q_c\}$ as variables. 
%Let us pause for an instant and give an interpretation of (\ref{eq:eval_Znp1}). The signal $\bx^0$ has been generated from $p_0(\bx^0)$ whereas the other replicas are independently drawn from the posterior distribution $P(\bx|\by)$ but due to the Nishimori condition, all replicas play an identical role from the computational point of view. 
Define a distribution of replicated variables at fixed overlap matrices $\{{\boldsymbol{q}}_c, c\in\{1, \cdots,\Gamma\}\}$
\begin{align}
P(\{\bx^a\}|\{{\boldsymbol{q}}_c\})\defeq \frac{1}{\Xi(\{{\boldsymbol{q}}_c\})} \prod_{a=0}^n\Bigg[\prod_{i=1}^N p_0(x_i^a) \prod_{c=1}^{\Gamma} \prod_{b< a}^n \delta\Bigg(\frac{1}{2i\pi}\Bigg[\frac{N}{\Gamma}{q}_c^{ab} - \sum_{i\in c}^{N/\Gamma}x_i^a x_i^b\Bigg]\Bigg)\Bigg], \label{eq:PxagivenQc}
\end{align}
where $\Xi(\{{\boldsymbol{q}}_c\})$ is the associated normalization. The role of the $2i\pi$ appearing in the delta function is purely formal and will become clear later on. Plugging this expression inside (\ref{eq:eval_Znp1}) we get
\begin{align}
\mathbb{E}[Z(\by)^n] &= \mathbb{E}_{\bF} \int d\by \prod_{c=1}^{\Gamma} d\boldsymbol{q}_c P(\{\bx^a\}|\{{\boldsymbol{q}}_c\}) \Xi(\{{\boldsymbol{q}}_c\}) \prod_{a=0}^n \Bigg[d\bx^a \prod_{r=1}^\Gamma \prod_{\mu\in r}^{\alpha N/\Gamma} P_{\text{out}}(y_\mu| z_\mu^a)\Bigg] \label{eq:before_distz}\\
&= \int \prod_{c=1}^{\Gamma} d\boldsymbol{q}_c \Xi(\{{\boldsymbol{q}}_c\}) \int d\by P(\{\bz^a\}|\{{\boldsymbol{q}}_c\}) \prod_{a=0}^n \Bigg[d\bz^a \prod_{r=1}^\Gamma \prod_{\mu\in r}^{\alpha N/\Gamma} P_{\text{out}}(y_\mu| z_\mu^a)\Bigg]. \label{eq:F_averaged}
\end{align}
The second equality is obtained after noticing that the integrand in (\ref{eq:before_distz}) depends on $\{x_i^a\}$ only through $\{z_\mu^a\}$, this allows to replace the 
integration on $\{x_i^a\}$ by an integration on $\{z_\mu^a\}$. As already explained, by the central limit theorem 
\begin{align}
P(\{\bz^a\}|\{{\boldsymbol{q}}_c\}) &= \prod_{\mu=1}^M \mathcal{N}(\bz_\mu|0,\boldsymbol{\tilde{q}}_{r_\mu}) = \prod_{r=1}^{\Gamma} \prod_{\mu \in r}^{\alpha N/\Gamma} \mathcal{N}(\bz_\mu|0,\boldsymbol{\tilde{q}}_{r}) \nonumber\\
&= \prod_{r=1}^{\Gamma} \left[(2\pi)^{n+1} {\rm det}(\boldsymbol{\tilde q}_r)\right]^{-\frac{\alpha  N}{2\Gamma}} \prod_{\mu \in r}^{\alpha N/\Gamma} e^{-\frac{1}{2} \sum_{a,b=0}^{n,n} z_\mu^a [\boldsymbol{\tilde q}_r^{-1}]_{ab} z_\mu^b}.
\end{align}
This is a product of multivariate centered Gaussian distributions, where $\bz_\mu \defeq [z_\mu^a, a\in\{0,\dots,n\}]$, $\bz^a \defeq [z_\mu^a, \mu\in\{1,\dots,M\}]$. 
Recall $\boldsymbol{\tilde{q}}_{r}$ is a function of $\{\boldsymbol{{q}}_c\}$. 
%Let us now make appear the disorder-dependent free entropy, a thermodynamic quantity. As such, it is assumed to be self averaging, i.e. to concentrate around its mean which is the one maximizing the partition function. Let
Let
\begin{align}
\mathbb{E}[Z(\by)^n] &= \int \prod_{c=1}^{\Gamma} d\boldsymbol{q}_c \exp\Big[N\Big(f(\{{\boldsymbol{q}}_c\}) + g(\{\boldsymbol{q}_c\})\Big)\Big], \label{eq:meanZn_beforesaddle}\\
f(\{{\boldsymbol{q}}_c\}) &\defeq \frac{1}{N}\ln\Big[\Xi(\{{\boldsymbol{q}}_c\})\Big], \\
g(\{\boldsymbol{q}_c\}) &\defeq \frac{1}{N}\ln\Bigg[\int d\by P(\{\bz^a\}|\{{\boldsymbol{q}}_c\}) \prod_{a=0}^{n} \Bigg[d\bz^a \prod_{r=1}^\Gamma \prod_{\mu\in r}^{\alpha N/\Gamma} P_{\text{out}}(y_\mu| z_\mu^a)\Bigg]\Bigg]. \label{eq:g_def}
\end{align}
%
% Now we use the saddle point estimation of (\ref{eq:meanZn_beforesaddle}) to obtain the integral up to sub-exponential corrections, 
% which are assumed to be irrelevant in the thermodynamic limit $N\to\infty$ by the concentration property of thermodynamic quantites. 
% Here it is assumed that $f$ and $g$ are both intensive quantities. 
Now we perform a saddle point estimation.
This requires to take the limit $N\to \infty$ 
limit before letting $n\to 0$, and we assume without justification that the final result does 
not depend on the order of limits $n$ and $N$. This gives for the free energy, using (\ref{eq:replicatrick})
% \footnote{The $N\to\infty$ limit has disappeared performing the saddle point estimation, and one 
% should verify a posteriori that indeed $f,g$ do not depend on $N$. As we will see, 
% the dependence of $f$ and $g$ in the number of replicas will be of the 
% form $h(X(q)^{n+1})$ for some $X(q), h$ with $h$ linear. As we have 
% the useful identity $\lim_{n\to 0} \partial_n h(X^{n+1}) = h(X \ln(X) )$ for a 
% linear function $h$ (that will be used for integrations in the next), 
% the $\lim_{n\to 0} \partial_n$ operation commute with the extremization with respect to $q$. 
% Indeed, $\lim_{n\to 0} \partial_n {\rm extr}[h(X(q)^{n+1})] = \lim_{n\to 0} \partial_n h(X(q_*)^{n+1})= 
% h(X(q_*)\ln[X(q_*)]) = {\rm extr} [h(X(q) \ln[X(q)])]$ $= {\rm extr}[\lim_{n\to 0} \partial_n h(X(q)^{n+1}) ] \ \forall \ n$, 
% where $q_*$ is defined by $\partial_q h(X(q))|_{q_*} = 0 \Leftrightarrow \partial_q X(q)|_{q_*}=0$ and we have 
% used that $\partial_q X(q)|_{q_*}=0 \Rightarrow \partial_q [X(q)\ln[X(q)]]|_{q_*}=0$. All 
% this justifies the second equality in (\ref{eq:meanZn_aftersaddle}).}
%
\begin{align}
{F}_{\text{co}} =-\lim_{n\to 0} \frac{\partial}{\partial n} \underset{\{q_c\}}{{\rm extr}} \Big(f(\{{\boldsymbol{q}}_c\}) + g(\{\boldsymbol{q}_c\})\Big) = -\underset{\{q_c\}}{{\rm extr}} \Bigg(\lim_{n\to 0}\frac{\partial f(\{{\boldsymbol{q}}_c\})}{\partial n} + \lim_{n\to 0}\frac{\partial g(\{\boldsymbol{q}_c\})}{\partial n}\Bigg). \label{eq:meanZn_aftersaddle}
\end{align}
Now the replica symmetric ansatz allows to simplify $g$ since $P(\{\bz^a\}|\{{\boldsymbol{q}}_c\})$ becomes 
\begin{align}
P(\{\bz^a\}|\{{\boldsymbol{q}}_c\})=\prod_{r=1}^{\Gamma} \left[(2\pi)^{n+1} {\rm det}(\boldsymbol{\tilde q}_r)\right]^{-\frac{\alpha  N}{2\Gamma}} \prod_{\mu \in r}^{\alpha N/\Gamma} e^{-\frac{C_{1,r}}{2} \sum_{a=0}^{n} (z_\mu^a)^2 -\frac{C_{2,r}}{2} \sum_{a=0, b\neq a}^{n,n} z_\mu^a z_\mu^b}, \label{eq:Pza_coupled}
\end{align}
where $C_{1,r}$ and $C_{2,r}$ depend on $\tilde q_r$ and $\mathbb{E}[s^2]$ as they are obtained from the matrix inversion $\boldsymbol{\tilde{q}}_r^{-1}$. Thanks to the simple structure of $\boldsymbol{\tilde{q}}_r$ under the replica symmetric ansatz, one can easily show that 
\begin{align}
C_{1,r} &= \frac{\mathbb{E}[s^2]+(n-2)\tilde q_r}{\mathbb{E}[s^2](\mathbb{E}[s^2]+(n-2)\tilde q_r)+(1-n)\tilde{q}_r^2} \underset{n\to0}{\to} \frac{\mathbb{E}[s^2] - 2\tilde q_r}{(\mathbb{E}[s^2] - \tilde q_r)^2}, \label{eq:C1}\\
C_{2,r} &= -\frac{\tilde q_r}{\mathbb{E}[s^2](\mathbb{E}[s^2]+(n-2)\tilde q_r)+(1-n)\tilde{q}_r^2} \underset{n\to0}{\to} -\frac{\tilde q_r}{(\mathbb{E}[s^2] - \tilde q_r)^2}. \label{eq:C2}
\end{align}

The replicated variables $\{\bz^a\}$ are correlated through $P(\{\bz^a\}|\{{\boldsymbol{q}}_c\})$. In order to simplify $g$, we decorrelate them by linearizing the exponent of $P(\{\bz^a\}|\{{\boldsymbol{q}}_c\})$ 
using the Gaussian transformation formula for a given $K>0$
\begin{align}
e^{\frac{K}{2} \sum_{a=0, b\neq a}^{n,n} z_\mu^a z_\mu^b} = \int \mathcal{D}\xi_\mu \,  e^{\xi_\mu\sqrt{K} \sum_{a=0}^{n} z_\mu^a -\frac{K}{2} \sum_{a=0}^{n} (z_\mu^a)^2}, \label{eq:stroto}
\end{align}
i.e. the previously correlated $\{z_\mu^a, a\in\{0,\dots,n\}\}$ are now i.i.d. Gaussian variables, but that all interact with a common random Gaussian effective field $\xi_\mu$. Using this with $K=-C_{2,r}$ as we know that $C_{2,r}\le 0$, the integration in $g$ can now be performed starting from (\ref{eq:g_def})
\begin{align}
% g(\{\boldsymbol{q}_c\}) &\defeq \frac{1}{(n+1)N} \ln\Bigg[\int d\by P(\{\bz^a\}|\{{\boldsymbol{q}}_c\}) \prod_{a=0}^{n} \Bigg[d\bz^a \prod_{r=1}^\Gamma \prod_{\mu\in r}^{\alpha_rN/\Gamma} P_{\text{out}}(y_\mu| z_\mu^a)\Bigg]\Bigg]\\
%
g(\{\boldsymbol{q}_c\}) &= \frac{1}{N}\ln\Bigg[\prod_{r=1}^\Gamma \prod_{\mu\in r}^{\alpha N/\Gamma} \int \mathcal{D}\xi_\mu dy_\mu \Bigg(\int dz_\mu \mathcal{N}\Big(z_\mu\Big|m(\xi_\mu,\tilde q_r),V(\mathbb{E}[s^2], \tilde q_r)\Big) P_{\text{out}}(y_\mu| z_\mu)\Bigg)^{n+1}\Bigg] \\
&= \frac{1}{\Gamma}\sum_{r=1}^\Gamma \alpha  \ln\Bigg[\int \mathcal{D}\xi dy \Bigg(\int dz \, \mathcal{N}\Big(z\Big|m(\xi,\tilde q_r),V(\mathbb{E}[s^2], \tilde q_r)\Big) P_{\text{out}}(y|z)\Bigg)^{n+1}\Bigg]\\
&= \frac{1}{\Gamma}\sum_{r=1}^\Gamma \alpha  \ln\Bigg[\int \mathcal{D}\xi dy \Bigg(\int \mathcal{D}z \, P_{\text{out}}\Big(y\Big| m(\xi,\tilde q_r) + z\sqrt{V(\mathbb{E}[s^2], \tilde q_r)}\Big)\Bigg)^{n+1}\Bigg].
\end{align}
As assumed, $g$ does no depend on $N$. Let us compute $m(\xi_\mu , \tilde q_r), V(\mathbb{E}[s^2], \tilde q_r)$. Combining 
(\ref{eq:Pza_coupled}) with (\ref{eq:stroto}), we get that for a $\mu \in r$ and up to a normalization, 
$z_\mu \sim \exp(-z_\mu^2 (C_{1,r} - C_{2,r})/2 + z_\mu \xi_\mu \sqrt{-C_{2,r}})$ which becomes using 
the $n\to 0$ limit of (\ref{eq:C1}), (\ref{eq:C2}) 
$z_\mu \sim \mathcal{N}(z_\mu|\xi_\mu \sqrt{\tilde q_r}, \mathbb{E}[s^2] - \tilde q_r) \exp(\xi_\mu ^2 \tilde q_r / [2(\mathbb{E}[s^2] - \tilde q_r)])$. 
Normalizing $P(z_\mu)$, the term $\exp(\xi_\mu ^2 \tilde q_r / [2(\mathbb{E}[s^2] - \tilde q_r)])$ disappears 
being independent of $z_\mu$ and thus $P(z_\mu) = \mathcal{N}(z_\mu|\xi_\mu \sqrt{\tilde q_r}, \mathbb{E}[s^2] - \tilde q_r)$. Thus $m(\xi_\mu, \tilde q_r) = \xi_\mu \sqrt{\tilde q_r}$, $V(\mathbb{E}[s^2], \tilde q_r) = \mathbb{E}[s^2] - \tilde q_r$. 
Now performing the $\lim_{n\to 0}\partial_n$ operation and using the identity 
\begin{align}
\lim_{n\to 0} \frac{\partial}{\partial n} \ln\left[\int du X(u)^{n+1} \right] = \frac{\int du X(u)\ln(X(u))}{\int dv X(v)}, \label{eq:identity_logxnp1}
\end{align}
we directly obtain 
\begin{align}
\lim_{n\to 0}\frac{\partial g(\{\boldsymbol{q}_c\})}{\partial n} = &\sum_{r=1}^\Gamma \frac{\alpha }{\Gamma}  \int \mathcal{D}\xi dy \mathcal{D}\hat z \, P_{\text{out}}\Big(y\Big|\xi\sqrt{\tilde q_r} + \hat z \sqrt{\mathbb{E}[s^2]- \tilde q_r}\Big) \ln\left[ \int \mathcal{D}z \, P_{\text{out}}\Big(y\Big| \xi\sqrt{\tilde q_r} + z \sqrt{\mathbb{E}[s^2]- \tilde q_r}\Big) \right], \label{g_sipmlified}
\end{align}
where we used the normalization $\int dydu \, P_{\text{out}}(y|u) \mathcal{N}(u|a,b)=1$, such that the denominator in (\ref{eq:identity_logxnp1}) sums to one. Let us now deal with $f(\{{\boldsymbol{q}}_c\})$. We will use the following representation of the delta function $\delta(x)=\int_{\mathbb{R}} d\hat q \exp(2i\pi \hat q x) \Leftrightarrow \delta(x/(2i\pi))=\int_{\mathbb{R}} d\hat q \exp(\hat q x)$ where $\hat q$ can be interpreted as an auxiliary external field.\footnote{We now understand that the presence of the $2i\pi$ in (\ref{eq:PxagivenQc}) is thus just a trick to make the integral real.}

We assume the replica symmetric ansatz for the auxillary fields similarly as for the overlap matrix
$\hat q_c^{ab} = -\hat q_c \ \forall \ a, b\neq a$.
The minus sign is just introduced for convenience. Using again the Gaussian transformation formula, we get  
\begin{align}
f(\{{\boldsymbol{q}}_c\}) &= \frac{1}{N} \ln\Bigg[\prod_{c=1}^{\Gamma} \int d\hat q_c \prod_{a=0}^{n}\prod_{k\in c}^{N/\Gamma} \bigg[p_0(x_k^a) dx_k^a\bigg] e^{- \hat q_c \sum_{a=0,b<a}^{n,n} \big(\frac{N}{\Gamma}{q}_c-\sum_{k\in c}^{N/\Gamma}x_k^a x_k^b\big)} \Bigg] \nonumber\\
&=\frac{1}{N} \sum_{c=1}^\Gamma \ln\Bigg[\int d\hat q_c e^{-\frac{N(n+1)n}{2\Gamma}\hat q_c q_c} \Bigg(\int \prod_{a=0}^n \bigg[dx^a p_0(x^a)\bigg] e^{\frac{\hat q_c}{2} \sum_{a=0,b\neq a}^{n,n} x^a x^b} \Bigg)^{\frac{N}{\Gamma} } \Bigg]\nonumber\\
&=\frac{1}{N} \sum_{c=1}^\Gamma \ln\Bigg[\int d\hat q_c e^{-\frac{N(n+1)n}{2\Gamma}\hat q_c q_c} \Bigg( \int \mathcal{D}\xi \prod_{a=0}^n \bigg[dx^a p_0(x^a)\bigg] e^{-\frac{\hat q_c}{2}\sum_{a=0}^n (x^a)^2+ \xi \sqrt{\hat q_c} \sum_{a=0}^n x^a} \Bigg)^{\frac{N}{\Gamma} }\Bigg]\nonumber\\
% \end{align}
% %
% \begin{align}
&=\frac{1}{\Gamma} \sum_{c=1}^\Gamma \underset{\hat q_c}{{\rm extr}}\Bigg(-\frac{(n+1)n}{2}\hat q_c q_c + \ln\Bigg[\int \mathcal{D}\xi\Bigg( \int dx p_0(x)  e^{-\frac{\hat q_c}{2} x^2+ \xi \sqrt{\hat q_c} x } \Bigg)^{n+1} \Bigg] \Bigg),
\end{align}
where we have assumed that we can treat $\hat q_c$ as a positive variable for the Gaussian transformation transform. This 
will be verified a posteriori at the end of the computation. 
The saddle point method employed for the estimation of the integral over the auxiliary fields is justified similarly as before, as the $N\to \infty$ as already been assumed. Finally, using again (\ref{eq:identity_logxnp1}), we obtain
\begin{align}
\lim_{n\to 0} \frac{\partial f(\{{\boldsymbol{q}}_c\})}{\partial n} = \frac{1}{\Gamma} \sum_{c=1}^\Gamma \underset{\hat q_c}{{\rm extr}}\Bigg(-\frac{\hat q_c q_c }{2} + \int \mathcal{D}\xi ds p_0(s)  e^{-\frac{\hat q_c}{2} s^2+ \xi \sqrt{\hat q_c} s} \ln\Bigg[ \int dx p_0(x)  e^{-\frac{\hat q_c}{2} x^2+ \xi \sqrt{\hat q_c} x } \Bigg] \Bigg). \label{eq:f_beforelastsimpl}
\end{align}

Using (\ref{eq:meanZn_aftersaddle}), (\ref{g_sipmlified}) and this last expression, we get a first version of the replica formula for the 
free energy
\begin{align}
\Gamma {F}_{\text{co}} = \underset{\{q_c, \hat q_c\}}{{\rm extr}} \Bigg\{&-\sum_{r=1}^\Gamma \alpha  \int \mathcal{D}\xi dy \mathcal{D}\hat z \, P_{\text{out}}\Big(y\Big|\xi\sqrt{\tilde q_r} 
+ \hat z \sqrt{\mathbb{E}[s^2]- \tilde q_r}\Big) \ln\Bigg[ \int \mathcal{D}z \, P_{\text{out}}\Big(y\Big| \xi\sqrt{\tilde q_r} + z \sqrt{\mathbb{E}[s^2]- \tilde q_r}\Big) \Bigg] \nonumber
\\&+ \sum_{c=1}^\Gamma \Bigg(\frac{\hat q_c q_c }{2} - \int \mathcal{D}\xi ds p_0(s)  e^{-\frac{\hat q_c}{2} s^2
+ \xi \sqrt{\hat q_c} s} \ln\Bigg[ \int dx p_0(x)  e^{-\frac{\hat q_c}{2} x^2+ \xi \sqrt{\hat q_c} x } \Bigg] \Bigg)\Bigg\}.
\label{extrem}
\end{align}
Recall that in this expression $\tilde q_r =\sum_{c=1}^\Gamma J_{r,c} q_c$.

To make contact with the potential function introduced in this paper we still have to partially solve the extremization problem and reduce \eqref{extrem} to a variational problem 
over one variable. Differentiating the function of $\{q_c, \hat q_c\}$ in \eqref{extrem} with respect to $q_c$ and setting the derivative to zero we find
\begin{align}\label{eq:SE_derivation_1}
\hat q_c = 2 \sum_{r=1}^\Gamma J_{r,c} \alpha  \, \partial_{\tilde q_r} \Bigg( \int dy \mathcal{D}t f (y|t\sqrt{\tilde q_r},\mathbb{E}[s^2] - \tilde q_r) \ln\left[ f(y|t\sqrt{\tilde q_r},\mathbb{E}[s^2] - \tilde q_r) \right]\Bigg),
\end{align}
where 
\begin{align}\label{eq:f_function}
f (y|\mu,\sigma^2) \defeq \int dx \mathcal{N}(x|\mu,\sigma^2) P_{\text{out}} (y | x) = \int \mathcal{D}x P_{\text{out}} (y | x\sigma + \mu ).
\end{align}
One can show the following identity (this has already been shown and used in Appendix \ref{sec:App_SE})
\begin{align}\label{eq:f_functionIdentity}
\partial_{\tilde q_r} f (y|t\sqrt{\tilde q_r},\mathbb{E}[s^2] - \tilde q_r) = \frac{e^{\frac{t^2}{2}}}{2\tilde q_r} \, \partial_t \Bigg( e^{-\frac{t^2}{2}} \, \partial_t \Big(f (y|t\sqrt{\tilde q_r},\mathbb{E}[s^2] - \tilde q_r)\Big) \Bigg).
\end{align}
Hence, (\ref{eq:SE_derivation_1}) can be rewritten as
\begin{align}\label{eq:SE_derivation_2}
\hat q_c &= 2 \sum_{r=1}^\Gamma J_{r,c} \alpha \int dy \mathcal{D}t \Big(1+ \ln\left[ f_{\text{out}} (y|t\sqrt{\tilde q_r},\mathbb{E}[s^2] - \tilde q_r) \right]\Big) \, \partial_{\tilde q_r} f_{\text{out}} (y|t\sqrt{\tilde q_r},\mathbb{E}[s^2] - \tilde q_r) \nonumber \\
&= - \sum_{r=1}^\Gamma J_{r,c} \frac{\alpha}{\tilde q_r} \int dy dt \frac{1}{\sqrt{2\pi}} \Big(1+ \ln\left[ f_{\text{out}} (y|t\sqrt{\tilde q_r},\mathbb{E}[s^2] - \tilde q_r) \right]\Big) \, \partial_t \Bigg( e^{\frac{-t^2}{2}} \, \partial_t \Big(f_{\text{out}} (y|t\sqrt{\tilde q_r},\mathbb{E}[s^2] - \tilde q_r)\Big) \Bigg) \nonumber \\
&= \sum_{r=1}^\Gamma J_{r,c} \frac{\alpha}{\tilde q_r} \int dy \mathcal{D}t \frac{\Big(\partial_t f_{\text{out}} (y|t\sqrt{\tilde q_r},\mathbb{E}[s^2] - 
\tilde q_r)\Big)^2}{f_{\text{out}} (y|t\sqrt{\tilde q_r},\mathbb{E}[s^2] - \tilde q_r)}  
\nonumber \\ &
=
\sum_{r=1}^\Gamma J_{r,c} \alpha \int dy dp dz \frac{\exp\big(-\frac{p^2}{2\tilde q_r}\big)}{\sqrt{2\pi\tilde q_r }} f (y|p,\mathbb{E}[s^2] - \tilde q_r) \big({\partial_p} \ln f (y|p,\mathbb{E}[s^2] - \tilde q_r)\big)^2 \nonumber\\
& = \sum_{r=1}^\Gamma J_{r,c} \alpha \mathbb{E}_{p\vert \tilde q_r} [\mathcal{F}\big(p|\mathbb{E}(s^2) - \tilde q_r\big)].
\end{align}
The final step consists in replacing the stationarity condition \eqref{eq:SE_derivation_2} in \eqref{extrem}. First we remark 
\begin{align}
 \sum_{c=1}^\Gamma \frac{\hat q_c q_c }{2}
 & =
 \frac{1}{2}\sum_{r=1}^\Gamma \sum_{c=1}^\Gamma J_{r,c}q_c \alpha \mathbb{E}_{p\vert \tilde q_r} [\mathcal{F}\big(p|\mathbb{E}(s^2) - \tilde q_r\big)]
 \nonumber \\ &
 = 
 \frac{1}{2}\sum_{r=1}^\Gamma \tilde q_r \alpha \mathbb{E}_{p\vert \tilde q_r} [\mathcal{F}\big(p|\mathbb{E}(s^2) - \tilde q_r\big)]
 \nonumber \\ &
 =
 \frac{1}{2}\sum_{r=1}^\Gamma \tilde q_r \Sigma^{-2}(\mathbb{E}(s^2) - \tilde q_r)
\end{align}
where in the last line we have set
\begin{align}
 \Sigma^{-2}(\mathbb{E}(s^2) - \tilde q_r) = \alpha \mathbb{E}_{p\vert \tilde q_r} [\mathcal{F}\big(p|\mathbb{E}(s^2) - \tilde q_r\big)].
\end{align}
We also set 
\begin{align}
 \Sigma_c^{-2}(\{\tilde q_r\}) = \sum_{r=1}^\Gamma J_{r,c} \alpha \mathbb{E}_{p\vert \tilde q_r} [\mathcal{F}\big(p|\mathbb{E}(s^2) - \tilde q_r\big)]
\end{align}
so that $\hat q_c = \Sigma_c^{-2}(\{\tilde q_r\})$. Then 
\eqref{extrem} becomes
\begin{align}
 & \underset{\{\tilde{q}_r\}}{{\rm extr}} \Bigg\{-\sum_{r=1}^\Gamma \Bigg(\alpha  \int \mathcal{D}\xi dy \mathcal{D}\hat z \, P_{\text{out}}\Big(y\Big|\xi\sqrt{\tilde q_r} 
+ \hat z \sqrt{\mathbb{E}[s^2]- \tilde q_r}\Big) \ln\Bigg[ \int \mathcal{D}z \, P_{\text{out}}\Big(y\Big| \xi\sqrt{\tilde q_r} + z \sqrt{\mathbb{E}[s^2]- \tilde q_r}\Big) \Bigg] \nonumber
\\& +  \frac{1}{2}\tilde q_r\Sigma^{-2}(\tilde{q}_r)\Bigg) - \sum_{c=1}^\Gamma \Bigg( \int \mathcal{D}\xi ds p_0(s)  e^{-\frac{\Sigma_c^{-2}(\{\tilde{q}_r\})}{2} s^2
+ \xi \sqrt{\Sigma_c^{-2}(\{\tilde{q}_r\})} s} \ln\Bigg[ \int dx p_0(x)  e^{-\frac{\hat q_c}{2} x^2+ \xi \sqrt{\Sigma_c^{-2}(\{\tilde{q}_r\})} x } \Bigg] \Bigg)\Bigg\}.
\end{align}
The (courageous) reader can now compare with Definitions \eqref{def:pot-coupled} and \eqref{def:pot_underlying}. The sum over $r$ yields an ``internal energy'' contribution 
 $\sum_r U_{\text{un}}(E_r)$ and the sum over $c$ an ``entropic'' contribution $\sum_c S_{\text{un}}(\Sigma_c({\bf E}))$. To adapt the formula to sparse superposition 
 codes 
 one must replace all scalars by $B$-dimensional vectors, replace $E[s^2]\to 1$, $\alpha\to 1/R$ and set $\tilde q_r \to E[s^2] - E_r = 1-E_r$.
%

\ifCLASSOPTIONcaptionsoff
  \newpage
\fi

\bibliographystyle{IEEEtran}
\bibliography{IEEEabrv,bibliography}

\end{document}